\documentclass{article}
\usepackage{tikz} % Used for figures
\usepackage{longtable}
\usepackage{multirow}
\usepackage{pgfplots}
\usepackage{amsmath, amssymb}
\usepackage{graphicx} % Required for inserting images
\usepackage{optidef} % Prettier optimisation definitions
\usepackage{geometry}
\usepackage{caption}
\usepackage{authblk}

\title{Surrogate-based optimisation of process systems to recover resources from wastewater}
\author[1]{Alex Durkin}
\author[1, 2]{Miao Guo}
\affil[1]{Department of Chemical Engineering, Imperial College London, SW7 2AZ, UK}
\affil[2]{Department of Engineering, King's College London, WC2R 2LS, UK}
\date{}

\begin{document}

\maketitle

\section*{Abstract}

Wastewater systems are transitioning towards integrative process systems to recover multiple resources whilst simultaneously satisfying regulations on final effluent quality. This work contributes to the literature by bringing a systems-thinking approach to resource recovery from wastewater, harnessing surrogate modelling and mathematical optimisation techniques to highlight holistic process systems. A surrogate-based process synthesis methodology was presented to harness high-fidelity data from black box process simulations, embedding first principles models, within a superstructure optimisation framework. Modelling tools were developed to facilitate tailored derivative- free optimisation solutions widely applicable to black box optimisation problems. The optimisation of a process system to recover energy and nutrients from a brewery wastewater reveals significant scope to reduce the environmental impacts of food and beverage production systems. Additionally, the application demonstrates the capabilities of the modelling methodology to highlight optimal processes to recover carbon, nitrogen, and phosphorous resources whilst also accounting for uncertainties inherent to wastewater systems.

\vspace{5mm}
\noindent
\textbf{Keywords}\quad Surrogate modelling, derivative-free optimisation, resource recovery from wastewater

\section*{Acknowledgement}

This work was financially supported by the UK Engineering and Physical Sciences Research  Council (EPSRC) under the DTP CASE-conversion programme ``Systems modelling design for waste resource recovery" [2194316].

\section*{Conflict of interest}

The authors declare no conflicts of interest.

\tableofcontents

\section{Introduction}

\subsection{Process systems engineering for resource recovery}
Process systems engineering (PSE) is currently well poised to harness recent developments in artificial intelligence (AI) \cite{qin_advances_2019} and address the challenges of optimising sustainable systems for resource recovery within a circular economy \cite{pistikopoulos_process_2021}. AI techniques, specifically machine learning (ML) methods \cite{schweidtmann_machine_2021}, can enable the mathematical formulation and simplification of resource recovery process synthesis and design problems \cite{schneider_hybrid_2022}.

The synthesis and design of integrated resource recovery processes, also referred to as biorefineries, is widely applied to the valorisation of organic solids wastes with applications to resource recovery from wastewater relatively lacking \cite{cherubini_toward_2009, gong_sustainable_2015, gong_value-added_2015, kumar_review_2017, pott_wastewater_2018, chen_waste_2018, rajesh_banu_industrial_2020, alibardi_organic_2020, leong_waste_2021}. However, the literature on organic solid waste biorefineries provides a future perspective on the challenges facing integrative process design for resource recovery from wastewater. Specifically, whilst biorefinery concepts provide a computational framework for early-stage integrative process design, more granular challenges arise during implementation, for example: mathematical modelling of resource recovery process units \cite{puchongkawarin_optimization-based_2015}; computational tractability of modelling and optimising integrative superstructures \cite{gong_value-added_2015, alva-argaez_wastewater_1998}; methods to hedge against process and model uncertainties \cite{gong_sustainable_2015}; and holistic decision-making and the corresponding definition of global optimality \cite{gong_sustainable_2015, gong_value-added_2015}.

Computer-aided process synthesis methods can be used to screen alternative process networks without the time and capital expenditures required for pilot studies \cite{yeomans_systematic_1999, chen_recent_2017}. These methods have been widely applied to the design of wastewater treatment plants \cite{ahmetovic_global_2011, bozkurt_mathematical_2015, bozkurt_optimal_2016, castillo_integrated_2016, padron-paez_sustainable_2020, al_stochastic_2020}, whilst applications to integrative processes for resource recovery from wastewater remain relatively unexplored \cite{puchongkawarin_optimization-based_2015, juznic-zonta_smart-plant_2022, durkin_can_2022}. This research gap is constricted by the complexity of modelling resource recovery processes in the context of wastewater systems \cite{puchongkawarin_optimization-based_2015}.

Mathematical modelling is an indispensable tool for designing the resource recovery processes within a circular economy and sustainable future \cite{schneider_hybrid_2022}. Specifically, wastewater treatment plant design has undergone a paradigm shift in the last decades from a dependence on expert knowledge and well-established guidelines, towards sophisticated modelling and simulation technologies \cite{al_stochastic_2020}. This transition was made possible by the development of rigorous mathematical models of wastewater systems, based on first-principles physics, chemistry, and biology, including models for anaerobic digestion \cite{batstone_iwa_2002} and activated sludge processes \cite{henze_activated_2006}. However, these models tend to be highly dimensional and complex, lending to the biological nature of the underlying processes. These models can become particularly cumbersome during equation-orientated integrated process design, wherein the mathematical formulations are coded directly into specialist optimisation software, due to the combinatorial nature of modelling superstructures comprising many possible reaction pathways \cite{alva-argaez_wastewater_1998}. However, the continued development of rigorous models for emerging resource recovery technologies is important as these models form the basis for other modelling approaches (by providing a source of high-fidelity data from process simulations) as well as providing validation for more computationally tractable approximations \cite{solon_resource_2019}.

There exists 3 approaches to address the computational intractability of incorporating rigorous models directly within equation-orientated decision-making frameworks. First, lower-fidelity short-cut models can be formulated as more computationally tractable approximations to rigorous models whilst maintaining their foundations in first-principles \cite{razavi_review_2012}. \cite{bozkurt_mathematical_2015, bozkurt_optimal_2016, castillo_integrated_2016} utilise lower-fidelity general models to optimise municipal wastewater treatment plants, including validation against rigorous simulation-based models. Similarly, there exists significant applications of optimisation methodologies to water networks wherein short-cut models are utilised to represent water and contaminant flows with simple conversion factors \cite{alva-argaez_wastewater_1998, ahmetovic_global_2011}.

Another option is to utilise design of experiments \cite{sacks_design_1989} and process simulation software, embedding rigorous models within sequential modular black boxes along with algorithms to enable solution convergence for complex process networks. Simulation software is widely used to evaluate process design performance within a broad range of applications including prediction, design, operation, sensitivity analysis, and optimisation \cite{razavi_review_2012}. However, the requirement for more accurate process models necessitates a large number of simulation evaluations resulting in excessively high computational costs. Additionally, the underlying code within process simulation software is often embedded as black box models where functional and derivative information is unavailable to the user \cite{amaran_simulation_2016}.

Thirdly, increased data availability and computational efficiency has driven recent advances in data-driven AI and ML models applied to wastewater systems for sustainable decision-making \cite{hadjimichael_machine_2016}, process analysis, operation, and control \cite{corominas_transforming_2018, newhart_data-driven_2019}, prediction, classification, water quality evaluation \cite{zhu_review_2022}, uncertainty analysis, and optimisation \cite{schneider_hybrid_2022}. These data-driven models are effective at modelling nonlinear wastewater systems \cite{zhu_review_2022} but depend highly on the quantity and the quality of data used to train them \cite{schneider_hybrid_2022, dobbelaere_machine_2021}. Additionally, process simulation software can be used to generate large, high-fidelity data from first-principles models with which to train representative ML models as surrogates \cite{bradley_perspectives_2022}. However, a dissociation between traditional engineering specialists and computer science knowledge limits the development and deployment of such data-driven decision-support tools to complex industrial problems \cite{hadjimichael_machine_2016, corominas_transforming_2018, newhart_data-driven_2019, dobbelaere_machine_2021, himmelblau_accounts_2008}.

\subsection{Derivative-free optimisation}
Two branches of AI that are of particular interest in this work are the fields of ML and derivative-free optimisation (DFO). ML models provide a data-driven approach to represent resource recovery processes within larger decision-making frameworks \cite{wei_modeling_2013}. DFO refers to algorithms used to solve black box optimisation (BBO) problems in which the mathematical formulations and derivative information of optimisation objective functions, $f$ (Equation \ref{eq:opt1}), and constraints, $\textbf{g}$ (Equation \ref{eq:opt2}), are unknown or not readily available \cite{conn_introduction_2009, audet_derivative-free_2017}. (This is in contrast to equation-orientated optimisation wherein the mathematical formulations of $f$ and $\textbf{g}$ are explicitly known.) BBO problems arise frequently in simulation-based optimisation applications wherein underlying and unavailable first-principles models are harnessed within decision-making frameworks \cite{amaran_simulation_2016}. This can be achieved by interrogating the black box simulations for high-fidelity data from complex, rigorous underlying models and proceeding to use model-based (or surrogate-based) DFO or direct-search (or sampling-based) DFO approaches \cite{rios_derivative-free_2013}. The former approximates underlying functions (e.g., $f$ replaced by $\hat{f}$) and guides the optimisation search using surrogate (also known as meta- or reduced-order) models whilst the latter sequentially examines data samples for improvements in optimally \cite{bhosekar_advances_2018, alarie_two_2021}. Additionally, Bayesian DFO methods exist at the interface of these two approaches: following a direct-search approach of directly analysing data for optimality whilst simultaneously harnessing features of surrogate modelling components within optimised acquisition functions (such an approach might also be referred to as active ML) \cite{frazier_tutorial_2018}.

\begin{mini!}
    {\textbf{x},\textbf{y}}{f\left( \textbf{x},\textbf{y} \right)}{}{}\label{eq:opt1}
    \addConstraint{\textbf{g}(\textbf{x}, \textbf{y})}{\leq \textbf{0}}{}\label{eq:opt2}
\end{mini!}

DFO algorithms can be further categorised as local or global DFO methods where the former excel at refining current best solutions to obtain local optima using bound tightening algorithms, whilst the latter facilitate exploration for global solutions within the whole search space. Finally, stochastic (or evolutionary) DFO methods differ from deterministic algorithms due to the incorporation of random search steps and heuristics to update characteristics of the entire sample population towards optimality. \cite{kim_surrogate-based_2020}. Rios and Sahinidis \cite{rios_derivative-free_2013} provide an extensive review of DFO algorithms and a comparison of available software implementations.

% Interfacing optimisation programs to process simulation software thereby enables direct evaluations and active ML using high-fidelity input-output data from underlying $f$ and/or $\textbf{g}$ models.

% The training data for surrogate models can come from various sources such as data lakes including sensor data, physical laboratory experiments, or computer experiments using process simulation software to enable the rigorous black box models to be exploited. However, a challenge pertains to the quality of these different data sources; sensor data can be noisy with large uncertainties from the practical operating environment, physical experiments can be unrepresentative of different operational scales, and computer experiments are evaluated through a layer of abstraction due to the underlying black box models.

Surrogate model-based DFO has gained popularity due to the increase in data availability with which to fit surrogate models, improvements in global optimisation, and catalysed by recent interests and developments in ML \cite{alarie_two_2021}. In addition to providing mathematical formulations of the underlying models, surrogate models also address the computational intractability of incorporating rigorous models directly within equation-orientated decision-making frameworks. Surrogate models reduce the computational cost of evaluating expensive model relationships via a reduction in model complexity and/or dimensionality. As such, surrogate modelling is widely used for predictive modelling and feasibility analysis as well as mathematical optimisation \cite{bhosekar_advances_2018}.% Two main emerging methods concern the use of global surrogates versus iteratively updated local surrogates. The former involves fitting a surrogate model representative of the entire design space and performing optimisation to determine the globally optimal solution for this one representation of the underlying data. The latter follows a more Bayesian approach of iteratively updating the surrogate model based on consecutive optimisation solutions highlighting regions for more exploration or exploitation \cite{bhosekar_advances_2018}.

By formulating surrogate models within optimisation problems, their predictive power can be harnessed in solving black box decision-making problems \cite{boukouvala_global_2016, boukouvala_global_2017}. However, a primary challenge lies in writing surrogate model formulations that are tractable within an optimisation problem. That is, translating a predictive formulation of an ML model into a formulation that is compatible with optimisation solver software. Recent literature has also addressed this challenge \cite{schweidtmann_deterministic_2019, maragno_mixed-integer_2021, ceccon_omlt_2022}, harnessing recent advances and popularity of optimisation modelling language Pyomo \cite{bynum_pyomo_2021} and Python-based ML packages, e.g., Scikit-learn \cite{scikit-learn} and PyTorch \cite{paszke_pytorch_2019}. Specifically, these works utilise the object-orientated paradigm to enable the abstraction of surrogate model formulations into independently contained objects which can then be plugged into larger optimisation problems where required. For example, optimisation formulations for neural networks (NNs) exist widely in the literature, owing to both to their popularity as ML models and their highly customisable yet manageable structure \cite{schweidtmann_deterministic_2019, grimstad_relu_2019, tjeng_evaluating_2019}. The weights matrices and bias vectors are optimised during model training and then fixed as optimisation parameters. Since the input to the activation function is a linear weighted sum, it is possible to formulate MILP problems depending on the activation functions used \cite{grimstad_relu_2019, tjeng_evaluating_2019}.

Caballero and Grossmann \cite{caballero_algorithm_2008} present a model-based DFO algorithm using GPs as surrogate models. The authors develop a bounds refinement algorithm to enable convergence upon surrogate-based optima with high confidence of accurate representation of the underlying model. Henao and Maravelias \cite{henao_surrogate-based_2011} use NNs to represent multi-variable mappings alongside explicit constraints within a grey box optimisation framework. In both cases, data for training surrogate models were obtained by using static sampling strategies to generate input samples and evaluating the corresponding outputs by interfacing Matlab with process simulation software. The former formulated NLP problems within the TOMLAB optimisation environment in Matlab \cite{holmstrom_tomlab_2004} and solved these using SNOPT \cite{gill_snopt_2002}. The latter introduced discrete variables to represent a superstructure for optimising process synthesis, and solved the resulting MINLP problem with GAMS \cite{bussieck_general_2004} and the DICOPT solver \cite{grossmann_gamsdicopt_2002}.

Boukouvala and Floudas \cite{boukouvala_argonaut_2017} introduced a DFO framework for constrained grey box problems embedding sampling, bounds refinement, variable selection, surrogate modelling, and global optimisation. A key feature of their work was the incorporation of multiple surrogate functions (linear, quadratic, signomial, radial basis functions (RBFs), and kriging), from which the choice of model to be used was optimised. Beykal et al. \cite{beykal_data-driven_2020} incorporated support vector machines (SVMs) within the DFO framework as supervised classification surrogate models to map the region of numerical infeasibility arising from noisy simulations. Both applications formulated NLP problems which were solved by tuning the solver parameters of ANTIGONE \cite{misener_antigone_2014} to enable local/global solutions.

Despite the rise in popularity of surrogate-based DFO, most of these applications implement NLP wherein optimisation is performed on a continuous search space. However, simulation-based DFO approaches to process synthesis typically result in black box mixed integer nonlinear programming (bb-MINLP) problems \cite{kim_surrogate-based_2020} or constrained DFO (CDFO) problems \cite{boukouvala_global_2016} due to the inclusion of integer programming variables to represent discrete process configurations, for which there has been relatively little research. The challenges facing bb-MINLP and CDFO frameworks are numerous: first, obtaining a representative yet tractable sample set that exists within a discontinuous search space; second, fitting a surrogate model to response surfaces involving continuous and integer decisions; third, formulating multiple continuous surrogates then patching them together at discontinuities can result in complex and less tractable optimisation formulations, particularly as the number of discrete decisions increases \cite{kim_surrogate-based_2020}.

% Despite recent advances in simulation-based DFO, there remains many open challenges constraining the development and deployment of such algorithms and decision-support tools to real-world applications. A key challenge results from fragmented research approaches developing boutique DFO methods for specific applications, with no general one-size-fits-all solution and difficulty comparing methods to determine the best method for a given application. This challenge is compounded by the breadth of DFO applications and variations, including accounting for noisy outputs from stochastic simulations, optimising constrained BBO problems, and optimising both continuous and discrete variables. A second challenge derived from the fragmented research approach is the discontinuation of development after successful application. Recent research advances should be incorporated within existing DFO approaches as well as within commercial simulation software to enable simulation optimisation without necessitating programming experience and interfacing to a programming language. Finally, guaranteeing global solutions within reasonable computational time, particularly for highly dimensional and complex BBO problems, is an ongoing challenge in the optimisation community.

% pasted from chapter 4 intro
The following research gaps were highlighted from a review of the BBO and DFO literature which hinder the deployment of such decision-making frameworks to the process synthesis of resource recovery processes. A key challenge results from fragmented research approaches developing boutique DFO methods for specific applications, with no general one-size-fits-all solution and difficulty comparing methods to determine the best method for a given application. This challenge is compounded by the breadth of DFO applications and variations, including accounting for noisy outputs from stochastic simulations, optimising constrained BBO problems, and optimising both continuous and discrete variables. A second challenge derived from the fragmented research approach is the discontinuation of development after successful application. Recent research advances should be incorporated within existing DFO approaches as well as within commercial simulation software to enable simulation optimisation without necessitating programming experience and interfacing to a programming language. Finally, guaranteeing global solutions within reasonable computational time, particularly for highly dimensional and complex BBO problems, is an ongoing challenge in the optimisation community.

The problems associated with fragmented DFO development can be addressed by harnessing object-orientated programming (OOP) to provide general modelling toolboxes. Such approaches provide flexible foundation models which can be configured and adapted to specific applications. The flexibility of developed DFO methods is thereby increased by addressing the various modelling challenges at the object-level as opposed to the algorithm-level. OOP further lends itself to open-source development and the incorporation of research advances within new or updated modelling objects which can improve the performance of existing DFO configurations. Modelling objects can also be incorporated within tailored DFO implementations in a broad range of commercial simulation software. Finally, OOP for DFO also enables the dissociation between mathematical programming formulations and optimisation solvers, enabling local or global rigorous gradient-based solvers or stochastic metaheuristic solvers to be adopted as required.

A number of object-orientated surrogate modelling and DFO toolboxes have been developed to date, including the SUMO toolbox available in Matlab \cite{gorissen_surrogate_2010} and the surrogate modelling toolbox (SMT) available in Python \cite{bouhlel_python_2019}. The former enables numerous surrogate models to be trained and updated using adaptive sampling methods, with a primary focus on enabling more computationally tractable models for predictive purposes, whilst the latter enables static sampling strategy implementations and different surrogate model formulations with a focus on derivatives for use in gradient-based optimisation. Cozad et al. \cite{cozad_learning_2014} developed a machine learning software able to interface with many black box simulation codes and construct surrogate models from a selection of available basis functions, as well as adaptive sampling implementations, within a no-code interface. The optimisation and machine learning toolkit (OMLT) is a recently developed open-source python package for formulating neural network (NN) and gradient-boosted tree surrogate models within larger optimisation problems \cite{ceccon_omlt_2022}. Audet et al. \cite{audet_nomad_2021} recently updated their popular implementation of the mesh adaptive direct-search (MADS) algorithm for DFO to a new object-orientated architecture to facilitate greater flexibility.

This work contributes an object-orientated DFO modelling suite to address the specific challenges of simulation-based BBO. For the first time, to the best of the authors' knowledge, abstract mathematical formulations of NN and Gaussian process (GP) surrogate models for classification are developed to address uncertainties pertaining to simulation convergence failures. Additional contributions include mathematical programming formulations for GP regression models and their uncertainty predictions enabling the integration of modelling uncertainties into decision-making processes for greater solution interpretability. Further contributions include mathematical programming formulations for adaptive sampling strategies based on GP models and a novel heuristic Delaunay triangulation-based approach. The developed modelling suite packages together the object-orientated surrogate modelling and mathematical programming formulations along with data processing tools to streamline simulation-based DFO. This article proceeds with a review of computer experiments (including adaptive sampling methods) and surrogate modelling before introducing the developed object-orientated DFO methodology. An application is presented to optimise a process system to recover resources from a brewery wastewater.

\subsection{Computer experiments}
Sampling from black box process simulations can be regarded as a set of computer experiments in which the experimental parameters (input data) can be designed in such a way as to maximise the information gained about the underlying system (input-output relationships). Specifically, a computer experiment evaluates mapping of input variables $\textbf{x}$ onto output variables $\textbf{y}$ through some underlying function $f$ as shown in Equation \ref{eq:yfx}. Additionally, black box simulations often return information pertaining to the convergence status of the underlying models. Such information can be processed as binary classification targets, $\textbf{t}$, where a value of 1 represents successful convergence, and a value of 0 means the simulation has failed to converge. Failed convergences can occur due to an infeasible combination of inputs or stochastic/numerical issues within simulator solution algorithms. It is important that the binary convergence targets are used to model feasible operating regions and to discard output data for failed convergences so as not to skew the data.

\begin{equation} \label{eq:yfx}
    \textbf{y}, \textbf{t} = f(\textbf{x})
\end{equation}

Popularised by the Design and Analysis of Computer Experiments (DACE) framework in the late 1980s \cite{sacks_design_1989}, computer experiments involve evaluating a computer simulation for different input designs. Computer experiments enable many input-output relationships to be enumerated without the requirement for expensive pilot studies. However, applications to optimisation problems are often hindered by computationally expensive simulations and lack of derivative and functional information about the underlying complex black box models. Other challenges include handling noisy simulations which are non-deterministic such that a specific input design can converge to a different output and therefore presents a one-to-many input-output mapping.

Sampling strategies have been designed to provide maximum design space coverage with the minimum number of samples. Such sampling strategies are designed to address the trade-off between homogeneous sample coverage (maximising information about underlying input-output relationships) and uniformity of samples (too uniform and the subsequent surrogate modelling can run into issues with correlated data). Sampling strategies are typically adopted in an offline approach where a specific number of samples are evaluated from the black box at once followed by subsequent modelling using all sampled data without any further sampling. This offline approach enables sampling to be removed from the modelling and optimisation workflow, enabling quicker iterations through these stages. Disadvantages include the reliance on the initially sampled data to provide good overall sample space coverage as well as good coverage around globally optimum solutions (exploration and exploitation).

\subsubsection{Variable selection and bounding}
The design of computer experiments begins with the selection of a subset of variables to be sampled from the black box. This thereby defines the selection of variables within the subsequent ML/surrogate modelling/optimisation formulations and can reduce overfitting by avoiding the modelling of unnecessary correlations \cite{bhosekar_advances_2018}. Generally, output variables are determined as those which appear in performance, feasibility, or costing functions within the larger decision-making problem being examined. The subset selection of input variables can include decision variables (design or operating variables) to be determined within a larger decision-making problem or uncertain parameters dependent on practical operating environments. Subset selection methods aim to determine a minimal number of these variables without a significant loss in accurate modelling capabilities, with the added benefit of reducing the surrogate model dimensionality \cite{boukouvala_surrogate-based_2013}. Popular systematic subset selection methods include principal component analysis \cite{jackson_users_1991}, screening techniques \cite{schonlau_screening_2006}, variance-based sensitivity analysis \cite{sobol_global_2001}, and MIP formulations \cite{henao_surrogate-based_2011}. More heuristic methods include performing preliminary experiments to observe variable sensitivities or using expert knowledge to choose relevant variables \cite{forrester_recent_2009}.

Candidate design variables to be selected for resource recovery flowsheet optimisation include volumes of reactors, nominal flowrates, integer variables defining the configuration of the flowsheet, as well as reactor specific design variables such as operating temperatures and recycle rates. Operating variables are variables that can be varied temporally, to alter the operating conditions of the flowsheet in response to externalities deviating from nominal design conditions. During the design stage of flowsheets, lower and upper bounds can be imposed on operating variables to enable the modelling of flowsheet operation and control. It is also possible to treat operating variables as pseudo design variables wherein nominal values are determined. 

The selection of uncertain parameters in practical operating environments considers input flows and compositions which depend on some independent upstream process. Other uncertain parameters that could be selected for modelling include environment temperatures and rainfall, which are particularly important for vessels open to the environment. Selected uncertain parameters can also pertain to the mathematical models, such as model parameters which have been determined via tuning methods, the structure imposed by mathematical models approximating complex systems also introduces uncertainties. Additionally, for simulation-based methods, there is often a degree of uncertainty in the simulation model as well, creating multiple layers of modelling uncertainty. Simulations may also be non-deterministic, necessitating numerical solvers which introduce noise based on different initialisation and convergence criteria.

At the variable selection stage, it is also important to consider in advance the surrogate models that will be used and what input-output relationships to be modelled by each. For example, NNs can be used to correlate many input-output relationships, however, simpler NN structures can be used to represent many input to one output relationships. In this case, it is up to the user to decide whether to have a single complex NN representing many relationships, or many simpler networks each representing just one output variable response. Furthermore, other surrogate models such as Gaussian processes (GPs) are not as flexible as NNs in that they can only effectively model up to 20 input variables per model, so in this case, for highly complex and dimensional systems, principal component analysis can help to reduce the input variable selection.

Data signifying the convergence status of process simulations can also be sampled to determine feasible regions for subsequent classification surrogate modelling and optimisation. A converged sample represents a feasible process design whereas failure of the simulator to converge signifies an infeasible process design. Converged/non-converged samples from black box simulations are typically assigned 0/1 labels, respectively. Another approach is to assign non-converged samples the value of $+\infty$ such that, in the context of minimisation, the optimisation objective function is ensured sub-optimal whilst the feasibility constraints are ensured to violate the form $g(x)\leq 0$ \cite{alarie_two_2021}.

Once input-output variables have been selected, appropriate bounds can be imposed to define the sample space. Typically, lower and upper bounds are imposed on each input dimension to form $m$-vectors of lower and upper bounds $\textbf{x}^L$, $\textbf{x}^U$, which define an $m$-dimensional hypercube \cite{rios_derivative-free_2013}. The $m$-dimensional input vector $\textbf{x}=(x_1, x_2, ..., x_m)$ is then bounded such that $\textbf{x}^L \leq \textbf{x} \leq \textbf{x}^U$. These bounds can be determined based on physical and thermodynamic principles, cost constraints, purchasable equipment, preliminary experiments, or expert knowledge. If the bounds are too tight, the resulting surrogate model will only be valid over a small input space, reducing its predictive capabilities and, in the case of optimisation, failing to capture globally optimal solutions. Conversely, if the bounds are too relaxed, sparse coverage of the large sample space results in regions of high uncertainty in surrogate model predictions. Variable bounds (and even variable selection) can also be adjusted and tightened within optimisation algorithms to iteratively refine the search space towards optimum solutions \cite{caballero_algorithm_2008, boukouvala_argonaut_2017}.

\subsubsection{Sampling}
The goal of sampling strategies is to produce $n$ input-output samples from an $m$-dimensional input space, as the $n \times m$ matrix $X$, such that good sample space coverage is achieved with minimal correlation between samples. Initial sampling in this way is also referred to as static or stationary sampling, as the samples are generated according to some strategy and then evaluated without any new information being used to inform the remaining sample locations as in adaptive sampling. Static sampling strategies include random sampling (also known as Monte Carlo sampling) \cite{mckay_comparison_1979} which can result in non-homogeneous coverage of the sample space leading to large variances in interpolated points from the surrogate model \cite{caballero_algorithm_2008}. On the other extreme, grid sampling enables a uniform projection of samples onto variable axes, but results in subsets of highly correlated samples which share the same value for one or more variables \cite{forrester_recent_2009}. To address this trade-off, quasi-random sampling techniques have been developed, such as Latin hypercube sampling (LHS) \cite{mckay_comparison_1979}, Hammersley \cite{kalagnanam_efficient_1997}, Halton, and Sobol sampling \cite{sobol_distribution_1967}. \cite{garud_design_2017} provide a comprehensive review of static and adaptive sampling strategies in the context of design of experiments.

\begin{figure}[htb]
    \centering
    \includegraphics[width=0.8\textwidth]{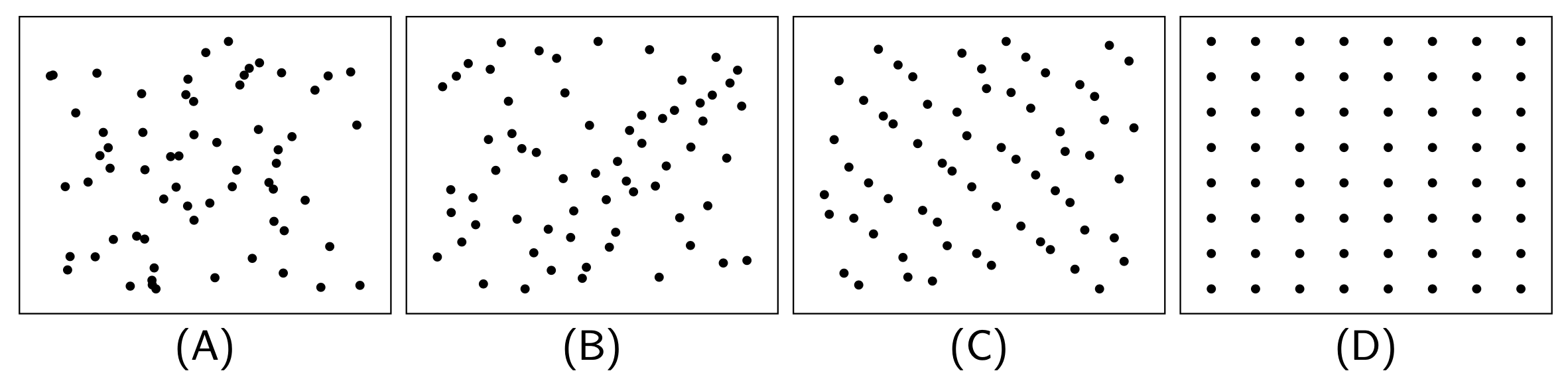}
    \caption[]{Static sampling strategies. (A) Random sampling, (B) Latin hypercube sampling, (C) Sobol sampling, (D) Grid sampling.}
    \label{fig:sampling}
\end{figure}

Figure \ref{fig:sampling} compare the sample space coverage of 64 samples using random sampling, LHS, Sobol sampling, and grid sampling. Random sampling exhibits the least homogeneous space coverage. LHS separates the range of each input variable into $n$ strata where $n$ is the number of samples to be generated. $n$ samples are then positioned such that only one sample is placed in each strata in each dimension thereby ensuring more homogeneous space coverage compared to random sampling. Sobol samples are generated based on underlying quasi-random Sobol sequences resulting in a homogeneity of space coverage between that of LHS and grid sampling. Finally, grid sampling provides the most homogeneous space coverage possible (when $n$ has an integer root for the number of dimensions).

In addition to quasi-random sampling strategies, sampling heuristics have been developed to further optimise the distribution of samples within the search space. For example, LHS samples can be generated such that the correlation between samples is minimised or such that the minimal pairwise distance is maximised or such that the ratio between the maximum pairwise distance and the minimum pairwise distance is minimised \cite{johnson_minimax_1990}. Geometry-based sampling strategies constrain the number of samples that can be generated whilst maintaining their balanced space-filling properties. For example, grid sampling necessitates that the $m$-th root of $n$ is an integer, where $n$ is the total number of grid samples to be generated and $m$ is the number of dimensions of the sample space, such that $\sqrt[m]{n}$ is the number of unique sample values in each dimension. The methods section introduces a heuristic-based method to generate any number of grid samples whilst attempting to maintain homogeneous space coverage. Similarly, Sobol sampling generates an optimal set of balanced samples for $n$ equal to a power of 2.

\subsubsection{Adaptive sampling}

Whilst an increased number of training samples improves the resulting surrogate model accuracy, high sampling requirements contribute significant computational cost for simulation evaluations and subsequent model fitting \cite{bhosekar_advances_2018}. Adaptive sampling techniques aim to minimise expensive sampling requirements by choosing promising locations for sequential samples via acquisition functions which address the trade-off between sample space exploration and exploitation. This online sampling approach builds a surrogate model on sparsely sampled data, which is then updated iteratively by choosing subsequent computer experiments to either explore more of the sample space or exploit potential local optima based on a function. This approach enables better design space exploration and exploitation but at the cost of incorporating expensive black box sampling into the online workflow. In the context of global optimisation, exploration is required to escape local optima, whilst exploitation improves accuracy at available optima. Example acquisition functions include the expected improvement (EI) \cite{jones_efficient_1998} and modified EI \cite{boukouvala_derivative-free_2014} for GP models, postulation as DFO problems \cite{cozad_learning_2014}, and weighted functions with departure functions \cite{garud_smart_2017}. More recently, emerging adaptive sampling frameworks include the use of Delaunay triangulation \cite{Delaunay_1934aa} to partition the sample space \cite{jiang_adaptive_2018, garud_surrogate-based_2019}.

Bayesian optimisation exists at the boundary of surrogate model-based DFO and data-based direct-search DFO. Specifically, Bayesian optimisation determines successive sampling points that are expected to improve the current best solution, thereby following a direct-search approach of directly analysing the samples for optimality. However, Bayesian optimisation also includes a modelling component wherein features of a probabilistic surrogate model (e.g., GP) are used to formulate acquisition functions to inform successive samples, thereby necessitating the iterative generation of surrogate models on new data. In Bayesian optimisation, maximisation of information gained from adaptive samples is achieved using acquisition functions such as the EI function, written below for an optimisation problem to maximise the function represented by the surrogate (Equation \ref{ei}) \cite{jones_efficient_1998}.

\begin{equation} \label{ei}
    \text{EI}(\textbf{x}) = (\hat{y} - y^{(\text{max})} - \xi)\Phi\left(\frac{\hat{y} - y^{(\text{max})} - \xi}{s}\right) + s\phi\left(\frac{\hat{y} - y^{(\text{max})} - \xi}{s}\right)
\end{equation}

where $\text{EI}(\textbf{x})$ is the EI at sample location $\textbf{x}$, $\Phi(\cdot)$ represents the cumulative distribution function, $\phi(\cdot)$ represents the probability distribution function, $\hat{y}$ is the GP predictive mean, $y^{(\text{max})}$ is the current maximum observation, $s$ is the standard deviation, and $\xi$ is a parameter controlling the trade-off between exploration relative to exploitation. The function increases with increasing $\hat{y}$ which corresponds to exploitation around the current maximum. Simultaneously, the function increases with increasing $s$ which corresponds to high uncertainty at sparsely sampled locations thereby enabling exploration. Since both exploration and exploitation are positively correlated with the EI function, both can be addressed simultaneously via maximisation of EI. However, the EI function exhibits multiple local optima that can cause numerical problems \cite{bhosekar_advances_2018}.

It is also possible to use a modified EI function (Equation \ref{ei_mod}) maximising only the second term of the EI function \cite{boukouvala_derivative-free_2014}. This enables for samples with high uncertainty, corresponding to increased $s$, or a predicted value close to the current best solution, represented by the inverse dependence on $(\hat{y} - y^{(\text{max})} - \xi)^2$.

\begin{equation} \label{ei_mod}
    \text{EI}_{\text{mod}}(\textbf{x}) = s\phi\left(\frac{\hat{y} - y^{(\text{max})} - \xi}{s}\right) = \frac{s}{\sqrt{2\pi}}\exp\left({-\frac{(\hat{y} - y^{(\text{max})} - \xi)^2}{2s^2}}\right)
\end{equation}

Another acquisition function is the probability of improvement (PI) shown in Equation \ref{eq:pi}. The PI enables exploration and exploitation to be controlled with the $\xi$ parameter wherein greater values of $\xi$ dampen the influence of samples near to the current best solution. In Equation \ref{eq:pi}, if $\hat{y} - y^{(\text{max})} - \xi \leq 0$ then there is no improvement in the current best solution. However, if $\hat{y} - y^{(\text{max})} - \xi > 0$ then this is the amount by which the function value would improve at that point in the input space. Since GPs are functions sampled from a normal distribution with mean and variance functions, the probability of improvement can therefore be evaluated by sampling from the cumulative distribution function.

\begin{equation} \label{eq:pi}
    \text{PI}(\textbf{x}) = \Phi\left(\frac{\hat{y} - y^{(\text{max})} - \xi}{s}\right)
\end{equation}

Another acquisition function is the upper confidence bound (UCB) shown in Equation \ref{eq:ucb}, where the trade off between exploitation of function performance $\hat{y}$ and exploration in regions of high uncertainty $s$ is clear. Again the tuneable parameter $\xi$ controls how much weight is assigned to exploration compared to exploitation.

\begin{equation} \label{eq:ucb}
    \text{UCB}(\textbf{x}) = \hat{y} + \xi s
\end{equation}

Another option to address the trade-off between exploration and exploitation in adaptive sampling is the bumpiness function used in conjunction with radial basis functions (RBFs). Another is to pose the adaptive sampling problem as a DFO problem to maximise the relative squared error as shown in Equation \ref{dfo} \cite{cozad_learning_2014}.

\begin{equation} \label{dfo}
    \max \left( \frac{y - \hat{y}}{y} \right)^2
\end{equation}

where $\hat{y}$ is the surrogate prediction, $y$ is the true function value, and $x$ are the input variables, typically constrained between lower and upper bounds. Such a function works to boost exploration of the search space, by favouring samples with high uncertainty resulting from lack of prior information. On the other hand, such an approach does not enable exploitation of potential solutions with moderate accuracy until the relative error becomes significant.

Other approaches assign weightings to exploration and exploitation terms within a function representing the trade-off between the two. Exploration is commonly represented by an error term, since sparsely sampled regions will have high uncertainties due to a lack of prior information. For exploitation, a departure function quantifies the impact of a new sample added near an already sampled location (Equation \ref{smart}) \cite{garud_smart_2017}.

\begin{equation} \label{smart}
    \Delta_j (x) = \hat{y} - \hat{y}_j
\end{equation}

where $\Delta_j (x)$ is the impact of sample $j$ on the surrogate function value, $\hat{y}$ is the prediction from the surrogate built using the entire sample set, and $\hat{y}_j$ is the prediction from the surrogate built using all samples except sample $j$.

Delaunay triangulation \cite{Delaunay_1934aa} is a widely employed approach which can be harnessed to partition the input domain to facilitate global exploration and exploitation around local optima. Examples of Delaunay triangulation applications within the optimisation literature include: explorative adaptive sampling at centroids of Delaunay triangulated regions by quantifying the trade-off between sampling within the largest region and sampling within the region with largest estimated modelling error \cite{jiang_adaptive_2018}; Delaunay triangulated region selection balancing exploration and exploitation followed by optimisation, as opposed to heuristic sampling at centroids, of interior sample location \cite{garud_surrogate-based_2019}; objective function uncertainty quantification by interpolating triangulated regions using quadratic functions \cite{beyhaghi_delaunay-based_2016}; sequential sampling at the centroid of each region generated by Delaunay triangulation to improve global surrogate model accuracy \cite{davis_centroid-based_2010}; a direct-search approach adopting a heuristic entropy criterion on Delaunay triangulated vertices to determine promising subsequent sample locations as well as heuristics for sample elimination and thereby triangulation mutations to facilitate global search \cite{wu_triopt_2005}.

\begin{figure}[htb]
    \centering
    \includegraphics[width=5.5in]{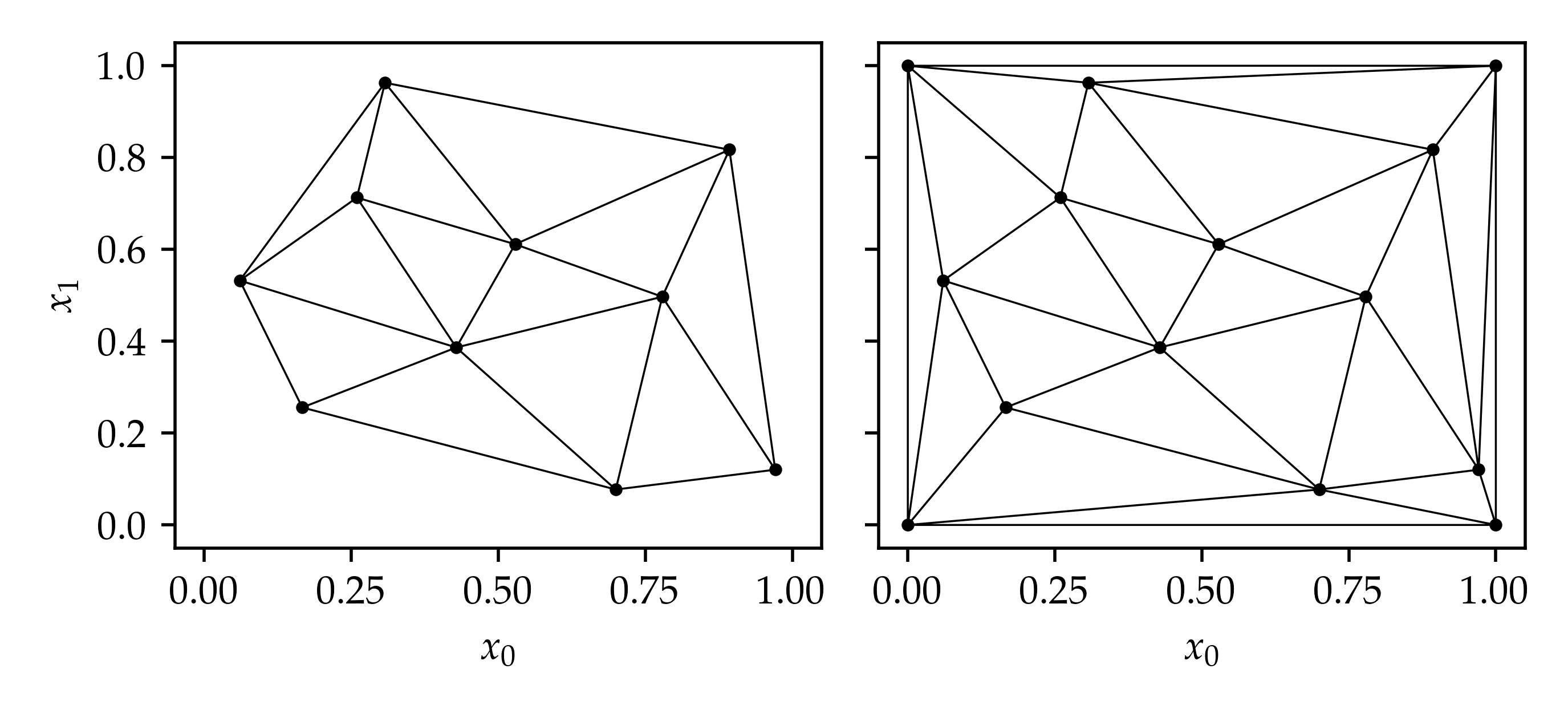}
    \caption[Delaunay triangulation]{Delaunay triangulation within the interior of a convex hull on a point set (left), including vertices of the search space (right).}
    \label{fig:dt}
\end{figure}

Delaunay triangulation provides a method to partition the interior of a convex hull on a point set $X$ into simplex regions $R$. Specifically, a Delaunay triangulation ensures that no point in $X$ exists inside the circumcircle (the unique circle that passes through three vertices of a triangle) of any simplex $R$. Additionally, a Delaunay triangulation is calculated so as to maximise the minimum resulting angle, thereby avoiding sliver triangles. Furthermore, by extending the concepts to circumscribed spheres, it is possible to perform Delaunay tessellation in higher dimensional spaces. Figure \ref{fig:dt} (left) shows the interior of a convex hull of a point set partitioned by Delaunay triangulation whilst Figure \ref{fig:dt} (right) shows the same point set, in addition to the vertices of the search space, partitioned by Delaunay triangulation. To enable global exploration which extrapolates the convex hull of the Delaunay triangulation, the vertices the search space can be included in the Delaunay triangulation. In this way, Delaunay triangulation produces a convex hull as a hypercube over the entire input space. The algorithm can then explore the entire input space including extrapolation between the convex hull of previous samples and the bounds of the input space. Readers are referred to \cite{gartner_computational_2013} for further details on Delaunay triangulation.

\subsection{Surrogate models}

A popular field of PSE research concerns the use of surrogate modelling to represent complex systems within larger decision-making problems. The primary application for surrogate modelling is for predictive purposes wherein some underlying complex model $f$ is substituted by making predictions with a surrogate model $\hat{f}$. Surrogate modelling enables significant savings in computational time and resources and has therefore been utilised for a diverse range of applications \cite{caballero_algorithm_2008, henao_surrogate-based_2011, boukouvala_surrogate-based_2013, cozad_learning_2014, gonzalez-garay_suscape_2018, yondo_review_2018}. To further lower computational cost, reduced-order models can be formulated which also serve to increase the interpretability of the model \cite{cozad_learning_2014}. In addition to generating predictions, the functional representations of surrogate models, $\hat{f}$, can be embedded within larger decision-making problems enabling optimisation and feasibility analysis \cite{bhosekar_advances_2018}.

A surrogate (also known as reduced-order or meta-) model $\hat{f}$ provides a mathematical formulation that approximates some underlying black box function $f$ which often exhibits one of the following properties: highly nonlinear, highly dimensional, unavailable functional form, unavailable derivative information, or no analytical solution. The underlying model is also often computationally expensive to evaluate and packaged with numerical solvers within black box simulation software. Surrogate models enable users to harness these rigorous state of the art black box models with greater computational efficiency. Surrogate models provide readily available functional formulations that are more computationally tractable due to their reduced complexity and/or dimensionality, thereby warranting their widespread use for predictive applications, feasibility analysis, and mathematical optimisation.

In the context of model-based DFO, surrogate models provide tractable mathematical formulations for black box cost functions and constraints. Traditionally, surrogate modelling is used for regression between continuous function outputs and continuous model predictions. For example, \cite{caballero_algorithm_2008} use GPs whilst \cite{henao_surrogate-based_2011} use NNs to represent flows within process flowsheets. Surrogate modelling can also be used to address classification problems between discrete target data. For example, \cite{ibrahim_optimization-based_2018} and \cite{beykal_data-driven_2020} use SVMs to model the boundary between converged and non-converged simulations.

% Figure \ref{fig:sumo} shows a general surrogate modelling framework wherein a single-pass from variable selection and bounding through to model validation represents a global surrogate modelling approach, whilst adaptive sampling and iterative feedback loops are shown by the dotted lines. Global surrogate modelling uses a large amount of sampled data to train a single surrogate model that is accurate over the entire sample space \cite{bhosekar_advances_2018}. One benefit of this approach is that it enables sampling to be taken offline such that the sampled data is saved and the simulator shutdown whilst the modelling proceeds separately. The disadvantage of this approach is the reliance on the initially sampled data to provide good coverage over the entire sample space and, in the case of optimisation, around possible optima. The more iterative approaches of adaptive sampling and variable bounds tightening, keeps the first two stages online such that the simulator is iteratively evaluated at each pass based on criteria from the model validation stage. This approach enables better exploration of the sample space and exploitation around potential optima, but at the cost of incorporating expensive black box sampling into the online workflow \cite{beykal_data-driven_2020}.

% \begin{figure}[htb]
%     \centering
%     \begin{tikzpicture}
%         \node[rounded corners, draw=black, align=center, text width=4cm, minimum height=3cm]{Remove this figure?};
%     \end{tikzpicture}
%     \caption[]{Caption.}
%     \label{fig:sumo}
% \end{figure}

A (global) surrogate model approximates some complex underlying model (over the entire design space). Simple regression techniques such as linear or polynomial regression are the most computationally tractable surrogate models but can fail in accurately modelling nonlinearities in the underlying model. Conversely, artificial NNs or interpolation techniques such as GPs offer better accuracy albeit with less tractable formulations.

\subsubsection{Neural networks}\label{sec:nn}
NNs are popular surrogate and ML models widely used in many different fields including image recognition and speech processing to novel drug discovery \cite{lecun_deep_2015}. With the rise of ML and surrogate modelling, NNs have also found applications in modelling chemical processes \cite{thompson_modeling_1994} due to their high accuracy and ability to represent multiple input-output relationships \cite{himmelblau_accounts_2008}. An NN consists of layers of nodes which map input variables $\textbf{x}$ onto predictive output variables $\hat{\textbf{y}}$. A node receives inputs from nodes in the previous layer, then transmits the output of a function evaluation to nodes in the subsequent layer. Each NN contains an input layer and an output layer, which transmit input variables $\textbf{x}$ and predictive output variables $\hat{\textbf{y}}$, respectively. The input layer, where the input data is passed into the NN has a number of nodes equal to the number of input variables. Similarly, the output layer, from which output variable predictions are made, has a number of nodes equal to the number of output variables that the NN is designed to predict. In this way, only the input layer and output layer are evaluated by the user, whilst the function evaluations enabling nonlinear modelling are carried out in hidden layers – so called because the user does not directly observe them as they exist in between the input and output layer.

\begin{figure}[htb]
    \centering
    \includegraphics[width=0.8\textwidth]{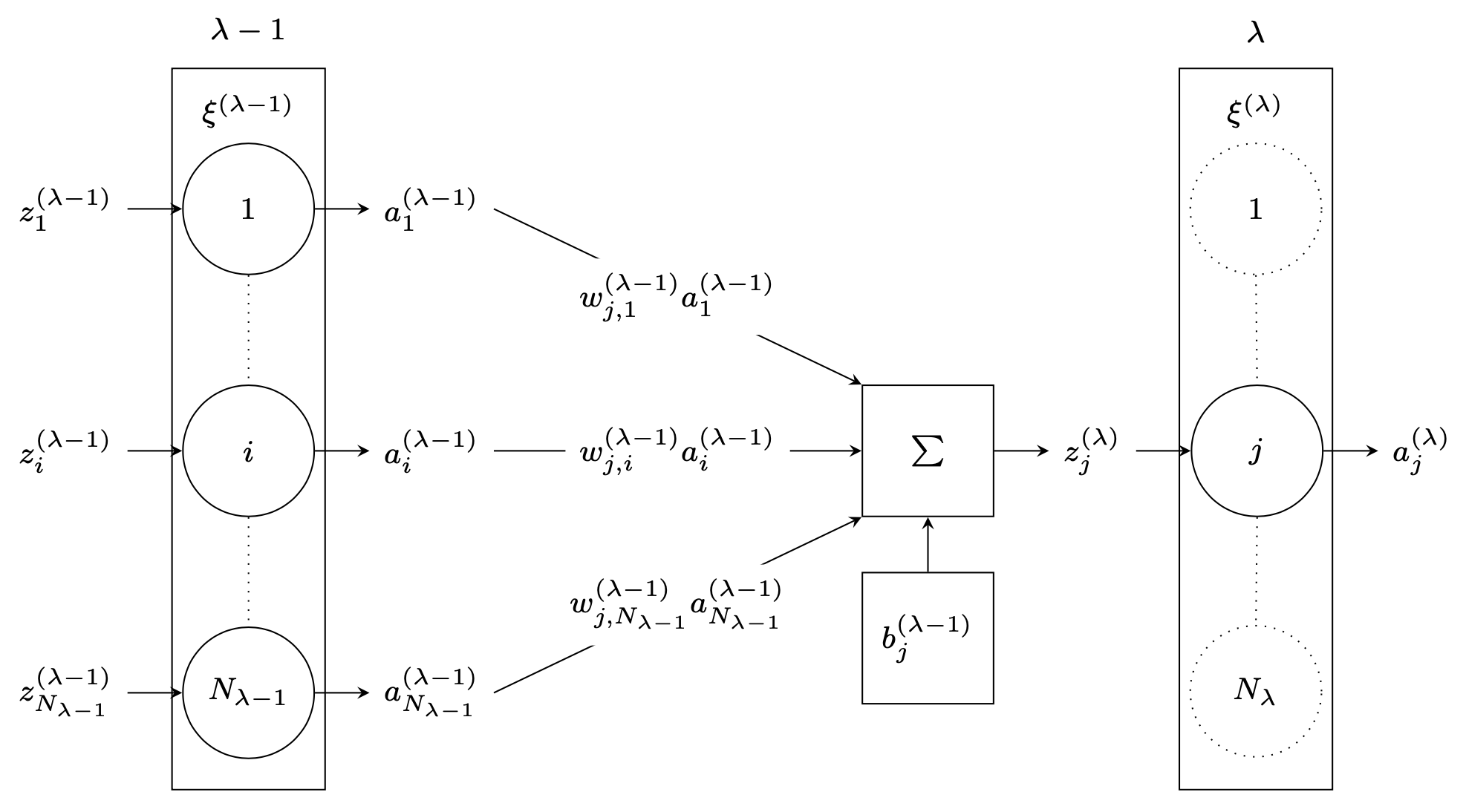}
    \caption[]{Neural network node mapping outputs from previous layer $z_i^{(\lambda-1)}$ for nodes $i=1, ..., N_{\lambda-1}$ onto the $j^{th}$ node in layer $\lambda$. $w_{j,i}^{(\lambda-1)}$ is the element from the weight matrix mapping the $i^{\text{th}}$ output from the previous layer onto the $j^{\text{th}}$ node. $b_{j}^{(\lambda-1)}$ is the element from the bias vector applied to the weighted sum for the $j^{\text{th}}$ node. $\xi^{(\lambda)}$ is the activation function for layer $\lambda$.}
    \label{fig:nn1}
\end{figure}

Hidden layers behave similarly to the output layer but make predictions of intermediate features instead of output variables. These intermediate features do not have to be specified by the user or even evaluated at any time; during training, the NN automatically determines how the intermediate features map input variables onto the output variables. Figure \ref{fig:nn1} shows the calculation of the intermediate output value $a_j^{(\lambda)}$ for the $j^{\text{th}}$ node within an intermediate NN layer $\lambda$. Within a general NN structure, layer $\lambda$ consists of $N_\lambda$ nodes thereby defining a vector of intermediate layer outputs $\textbf{a}^{(\lambda)}$. The outputs from the previous layer, $a^{(\lambda-1)}_i$, are first multiplied by the relevant elements from the matrix of weights $w^{(\lambda-1)}_{j,i}$ and summed along with a bias parameter $b^{(\lambda-1)}_j$ to yield the subsequent node inputs, $z^{(\lambda)}_j$. The result is a linear function with weights and bias parameters which can be tuned during training (Equation \ref{eq:nn1}). Node inputs are then mapped onto node outputs via an activation function $\xi^{(\lambda)}$ (Equation \ref{eq:nn2}). Equations \ref{eq:nn-in} and \ref{eq:nn-out} enable inputs to be passed to the network, and outputs to be interpreted from the network, respectively.

\begin{subequations}
    \begin{align}
        \textbf{z}^{(\lambda)} ={}  & \; W^{(\lambda-1)} \textbf{a}^{(\lambda-1)} + \textbf{b}^{(\lambda-1)} & \text{for} \; \lambda = 2,..., \Lambda \label{eq:nn1} \\
        \textbf{a}^{(\lambda)} ={}  & \; \xi^{(\lambda)} \left( \textbf{z}^{(\lambda)} \right) & \text{for} \; \lambda = 1,..., \Lambda \label{eq:nn2} \\
        \textbf{z}^{(1)} ={}  & \; W^{(0)} \textbf{x} + \textbf{b}^{(0)} \label{eq:nn-in} \\
        \hat{\textbf{y}} ={} & \;  W^{(\Lambda)} \textbf{a}^{(\Lambda)} + \textbf{b}^{(\Lambda)} \label{eq:nn-out}
    \end{align}
\end{subequations}

The structure of nodes and layers within NNs define the type of network being used. One type of NN structure is a feedforward NN (or multilayer perceptron), in which layers are structured sequentially and data flow through the network from input layer, through the hidden layers, to the output layer. Specifically, a special type of feedforward NN is a fully dense feedforward network in which each node is connected to every other node in the adjacent layers. This generality of fully dense NNs makes them widely applicable to a broad range of problems independent of the input data \cite{himmelblau_accounts_2008}. Another type of structure are recurrent NNs which include feedback connections where layer outputs are fed back into the NN.

\begin{figure}[htb]
    \centering
    \includegraphics[width=0.8\textwidth]{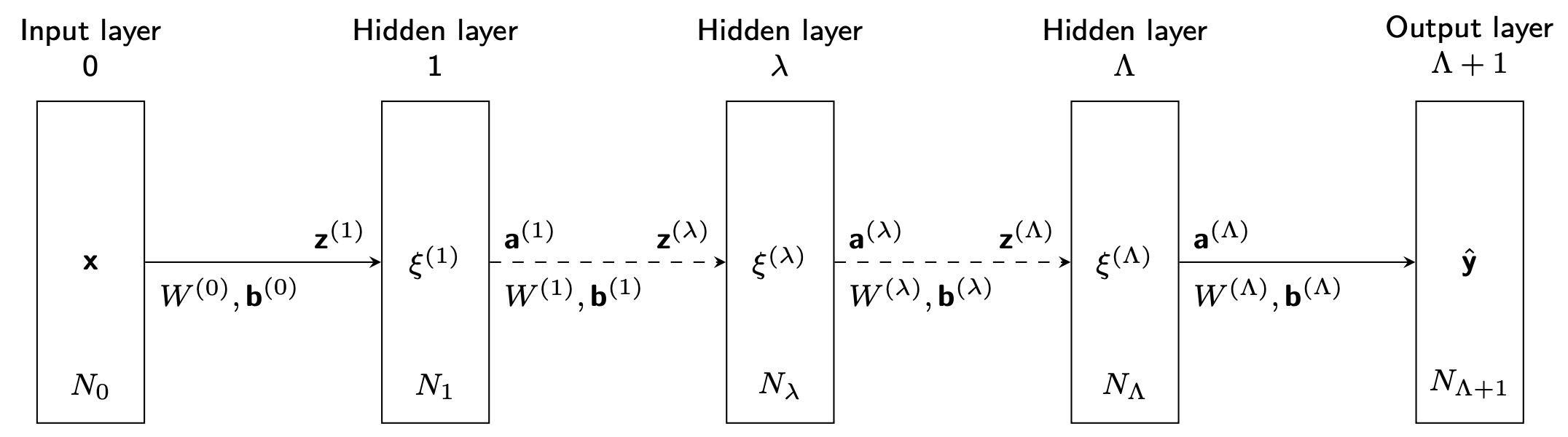}
    \caption[]{Feedforward neural network structure to map inputs $\textbf{x}$ onto predictions $\hat{\textbf{y}}$. The network consists of an input layer with $N_0$ nodes, an output layer with $N_{\Lambda+1}$ nodes, and $\Lambda$ hidden layers with $N_{\lambda}$ nodes for $\lambda=1,..., \Lambda$. Each hidden layer has an activation function $\xi^{(\lambda)}$ for $\lambda=1,..., \Lambda$. Layer inputs $\textbf{z}^{(\lambda)}$ are calculated by multiplying the previous layer outputs $\textbf{a}_{\lambda-1}$ by a weights matrix $W^{(\lambda-1)}$ and added to a bias vector $\textbf{b}^{(\lambda-1)}$ Finally, layer outputs are determined by passing the inputs through an activation function $\xi^{(\lambda)}$.}
    \label{fig:nn}
\end{figure}

Figure \ref{fig:nn} shows the mapping of inputs $\textbf{x}$ onto predictions $\hat{\textbf{y}}$ for a general fully-dense feedforward NN with $\Lambda + 1$ layers. The convention for naming NNs is to count the number of hidden layers plus the output layer whilst not counting the input layer. Here, the input layer is thereby assigned $\lambda=0$ whilst the final hidden layer is denoted by $\Lambda$ such that the output layer is layer $\Lambda+1$. Predictions from the output layer are interpretable by the user and so the notation for the outputs from the output layer is refined to $\hat{\textbf{y}}$. Since the outputs from the input layer are the input variables themselves, Equation \ref{eq:nn-in} is used to determine the outputs from the first hidden layer. Intermediate outputs from the hidden layers are evaluated as shown in Figure \ref{fig:nn1} and Equations \ref{eq:nn1} and \ref{eq:nn2}. Finally, predictive outputs from the output layer are evaluated using Equation \ref{eq:nn-out}.

Activation functions enable users to introduce nonlinear modelling capabilities to the NN as well as incorporate expert knowledge about the system characteristics. Activation functions include the rectified linear unit (ReLU) model which is a piecewise linear function and the default choice for feedforward NN activation functions wherein negative inputs return 0 enabling deactivation of node connections, otherwise positive inputs are directly returned as outputs. Smooth activation functions include the Sigmoid function and the Tanh function. The former squashes outputs between 0 and 1 and has the property that very large magnitude inputs have very small gradients which ensures that small deviations at high values have less importance than deviations in inputs around 0. The latter squashes outputs between -1 and 1 thereby exhibiting the property that inputs of 0 return outputs of 0 as well as have the largest gradient. Other activation functions include the Softplus function which is a smooth equivalent to the ReLU function and which has the property that the derivative of the function is equivalent the Sigmoid function. The Hardsigmoid and Hardtanh functions are piecewise linear approximations of the Sigmoid and Tanh functions, respectively. There also exists variations such as the Leaky ReLU or ReLU6 models which implement the ReLU model but with a very shallow linear model for negative inputs or crop the positive outputs at 6, respectively. Finally, a linear (or identity) activation function can be used if the relationship being modelled is suspected to be linear. Some of these activation functions are detailed in Table \ref{tab:activationfuncs} \cite{nwankpa_activation_2018}.

\begin{table}[htb]
    \centering
    \caption[Neural network activation functions]{Neural network activation functions.}
    \begin{tabular}{l c p{4.9cm} l l}
        \hline
        Name & Plot & Function, $\xi(x)$ & Derivative, $\xi'(x)$ & Range\\
        \hline
        \\
        Linear & \raisebox{-.5\height}{\begin{tikzpicture}
            \begin{axis}[height=3cm, width=0.2\textwidth, yticklabels={,,}, xticklabels={,,}, grid=both, major grid style={line width=.1pt,draw=gray!10}, axis line style={line width=.2pt,draw=gray!50}, axis lines*=middle, enlargelimits={abs=0.1}, ticks=none]
            \addplot[] table [x=x,y=x,col sep=comma] {data/tanh.csv};
        \end{axis}
        \end{tikzpicture}} & $x$ & 1 & $[-\infty, \infty]$ \\[10mm]
        Tanh & \raisebox{-.5\height}{\begin{tikzpicture}
            \begin{axis}[height=3cm, width=0.2\textwidth, yticklabels={,,}, xticklabels={,,}, grid=both, major grid style={line width=.1pt,draw=gray!10}, axis line style={line width=.2pt,draw=gray!50}, axis lines*=middle, enlargelimits={abs=0.1}, ticks=none]
            \addplot[] table [x=x,y=y,col sep=comma] {data/tanh.csv};
        \end{axis}
        \end{tikzpicture}} & $\frac{e^x - e^{-x}}{e^x + e^{-x}} = \frac{e^{2x} - 1}{e^{2x} + 1} \newline = 1 - \frac{2}{e^{2x} + 1} = \frac{1 - e^{-2x}}{1 + e^{-2x}}$ & $1-\xi(x)^2$ & $[-1, 1]$ \\[10mm]
        Sigmoid & \raisebox{-.5\height}{\begin{tikzpicture}
            \begin{axis}[height=3cm, width=0.2\textwidth, yticklabels={,,}, xticklabels={,,}, grid=both, major grid style={line width=.1pt,draw=gray!10}, axis line style={line width=.2pt,draw=gray!50}, axis lines*=middle, enlargelimits={abs=0.1}, ticks=none]
            \addplot[] table [x=x,y=y,col sep=comma] {data/sigmoid.csv};
        \end{axis}
        \end{tikzpicture}} & $\frac{1}{1 + e^{-x}}$ & $\xi(x)\left(1-\xi(x)\right)$ & $[0, 1]$ \\[10mm]
        Softplus & \raisebox{-.5\height}{\begin{tikzpicture}
            \begin{axis}[height=3cm, width=0.2\textwidth, yticklabels={,,}, xticklabels={,,}, grid=both, major grid style={line width=.1pt,draw=gray!10}, axis line style={line width=.2pt,draw=gray!50}, axis lines*=middle, enlargelimits={abs=0.1}, ticks=none]
            \addplot[] table [x=x,y=y,col sep=comma] {data/softplus.csv};
        \end{axis}
        \end{tikzpicture}} & $\frac{1}{\beta} \log\left( 1 + \exp\left(\beta x\right) \right)$ & $\frac{1}{1 + e^{-\beta x}}$ & $[0, \infty]$ \\[10mm]
        ReLU & \raisebox{-.5\height}{\begin{tikzpicture}
            \begin{axis}[height=3cm, width=0.2\textwidth, yticklabels={,,}, xticklabels={,,}, grid=both, major grid style={line width=.1pt,draw=gray!10}, axis line style={line width=.2pt,draw=gray!50}, axis lines*=middle, enlargelimits={abs=0.1}, ticks=none]
            \addplot[] table [x=x,y=y,col sep=comma] {data/relu.csv};
        \end{axis}
        \end{tikzpicture}} & $\max(0, x) = \begin{cases}
            0 & \text{if} \; x \leq 0 \\
            x & \text{otherwise}
        \end{cases}$ & $\begin{cases}
            0 & \text{if} \; x \leq 0 \\
            1 & \text{otherwise}
        \end{cases}$ & $[0, \infty]$ \\[10mm]
        Hardsigmoid & \raisebox{-.5\height}{\begin{tikzpicture}
            \begin{axis}[height=3cm, width=0.2\textwidth, yticklabels={,,}, xticklabels={,,}, grid=both, major grid style={line width=.1pt,draw=gray!10}, axis line style={line width=.2pt,draw=gray!50}, axis lines*=middle, enlargelimits={abs=0.1}, ticks=none]
            \addplot[] table [x=x,y=y,col sep=comma] {data/hardsigmoid.csv};
        \end{axis}
        \end{tikzpicture}} & $\begin{cases}
            0 & \text{if} \; x \leq -3 \\
            1 & \text{if} \; x \geq +3 \\
            \frac{x}{6} + \frac{1}{2} & \text{otherwise}
        \end{cases}$ & $\begin{cases}
            0 & \text{if} \; |x| \geq 3 \\
            \frac{1}{6} & \text{otherwise}
        \end{cases}$ & $[0, 1]$ \\[10mm]
        \hline
    \end{tabular}
    \label{tab:activationfuncs}
\end{table}

NNs for classification can be implemented by interpreting predictions from the output through a logistic function to map the NN outputs (logits) onto interpretable probabilities, $p(t=1)$. In this way, NNs for classification still make predictions in $\mathbb{R}$ and trained using specialised loss functions for discrete target data such as binary cross-entropy (BCE) loss on the probability predictions. It is also possible to use BCE combined with a sigmoid function on the logits. Training classification NNs on the continuous logit predictions and then squashing predictions though a logistic function enables increased numerical stability compared to training on probabilities or class predictions directly \cite{paszke_pytorch_2019}.

\subsubsection{Gaussian processes}\label{sec:gps}
Traditionally developed in the field of geostatistics in the 1950s \cite{krige_statistical_1951}, GPs gained popularity after their use in design of experiments \cite{sacks_design_1989}. Today, GPs have gained much traction as surrogate models due to the relatively simple computations required for supervised ML and Bayesian inference \cite{rasmussen_gaussian_2006}. GPs are a statistical modelling method wherein an underlying function is approximated by a probability distribution over functions enabling interpolating predictions and an estimate of uncertainty in these predictions to be evaluated simultaneously \cite{Davis2007}. The GP predictive mean function and covariance function can be written as shown by Equation \ref{eq:posterior-mean} and Equation \ref{eq:posterior-cov}, respectively.

\begin{equation} \label{eq:posterior-mean}
    \begin{aligned}
        \hat{f} ={} & \boldsymbol{\mu} + \textbf{k}^\intercal K^{-1} \left( \textbf{y} - \boldsymbol{\mu} \right) & \\
        ={} & \textbf{k}^\intercal K^{-1} \textbf{y} & & \text{for} \; \boldsymbol{\mu} = \textbf{0} \\
        ={} & \sum_{i=1}^n \alpha_i k(X_i, \textbf{x}) & & \text{where} \; \boldsymbol{\alpha} = K^{-1} \textbf{y}
    \end{aligned}
\end{equation}

\begin{equation} \label{eq:posterior-cov}
    \mathbb{V}\left[ \hat{f} \right] = \sigma_f^2 - \textbf{k}^T K^{-1} \textbf{k}
\end{equation}

The mean function (Equation \ref{eq:posterior-mean}) can be used for predictive purposes where $n$ is the number of training samples, $\boldsymbol{\alpha}$ is an $n$-vector of fitted parameters dependent only on the training data, and $k(X_i, \textbf{x})$ is the GP kernel function evaluated between training sample $X_i$ and new input vector \textbf{x}. An attractive property of GPs is that their foundation in probabilistic distributions enables the covariance in predictions to be evaluated using Equation \ref{eq:posterior-cov} where $\sigma_f^2$ is a kernel parameter representing the function variance, \textbf{k} is equivalent to $k(X, \textbf{x})$, and $K$ is the GP kernel function evaluated between the training data equivalent to $k(X, X)$.

The GP kernel parameters can be optimised to maximise some measure of the function fit. Specifically, maximum likelihood estimation (MLE) can be used to maximise the log marginal likelihood function (or minimise the negative log marginal likelihood) shown in Equation \ref{eq:lml}, where the first term is the data-fit based on observed data, the middle term is a complexity penalty, and the third term is a normalisation constant.

\begin{equation} \label{eq:lml}
    \min \frac{1}{2} \textbf{y}^T K^{-1}\textbf{y} + \frac{1}{2} \log \left| K \right| + \frac{n}{2} \log 2 \pi
\end{equation}

The GP kernel function enables users to incorporate expert knowledge into the model, for example favouring smooth, periodic (used to model function shapes which repeat themselves over some periodic dimension such as energy demand over time), or noisy functions. A primary challenge in modelling with GPs is the selection of the kernel function, akin to the arbitrary specification of NN structure and activation function. This work includes linear and polynomial kernels as shown by Equation \ref{eq:covpoly}, where $\sigma_0^2$ is a kernel parameter optimised during fitting. The polynomial order, $\omega$, is an integer that is specified and fixed by prior to model fitting where the linear kernel is specified with $\omega=1$. GP regression with linear and polynomial kernels is equivalent to Bayesian linear and polynomial regression, respectively. Although Bayesian linear and polynomial regression can be implemented more efficiently using specialised software packages (e.g., Scikit-learn \cite{scikit-learn}), linear and polynomial GP kernels are included here for demonstrative purposes. The linear kernel is unique among GP kernels in that it is non-stationary.

\begin{equation} \label{eq:covpoly}
    k(\textbf{x},\textbf{x}^{\prime}) = \sigma_f^2 \left( \sigma_0^2 + \textbf{x} \cdot \textbf{x}^\prime \right)^\omega
\end{equation}

Practically, the most common covariance function is the squared exponential covariance function (Equation \ref{eq:sqexp1}) since it is universal, exhibiting a flexible approximation to many underlying functions and providing some resolve to the challenge of kernel selection. It is also possible to integrate the squared exponential kernel function against most necessary functions and it is infinitely differentiable, making it effective within optimisation frameworks \cite{rasmussen_gaussian_2006}.

\begin{equation} \label{eq:sqexp1}
    k(\textbf{x},\textbf{x}^{\prime}) = \sigma_f^2 \exp \left( -\frac{1}{2l^2} \sum_{j=1}^m \left| x_j - x^{\prime}_j \right|^2 \right)
\end{equation}

In Equation \ref{eq:sqexp1}, $\sigma_f^2$ and $l$ are parameters representing the positive function variance and characteristic length scale, respectively. Greater values of $\sigma_f^2$ result in greater variances over the input domain. The length scale $l$ represents the sensitivity of the function, where higher values ensure that even two points far away are correlated resulting in smoother, lower frequency mean predictions, whilst lower values result in higher frequency function changes. Note that the length scale is raised to the exponent 2 in Equation \ref{eq:sqexp1} as it has been taken outside of the squared distance term, although it could equally be written as $\left| \frac{x_j - x^{\prime}_j}{l} \right|^2$. The covariance function can be evaluated between two points $\textbf{x}, \textbf{x}^{\prime}$ in $m$-dimensional space, such that $x_j,x^{\prime}_j$ represent coordinates for $j=1, ..., m$.

GPs can also be utilised to address (binary) classification problems, based on the fundamental idea: place a GP prior over a latent function $u(\textbf{x})$ and then squash this through the logistic sigmoid function to obtain a prior with outputs that are constrained between 0 and 1 and therefore interpretable as probabilities. These predictive probabilities can be evaluated using Equation \ref{eq:probit},

\begin{equation} \label{eq:probit}
    p(t=1) = \sigma \left( \textbf{k}^T (\textbf{t}-\sigma(u)) \left( 1 + \frac{\pi}{8} \left( \sigma_f^2 - \textbf{k}^T \left( W^{-1} + K \right)^{-1} \textbf{k} \right) \right)^{-1/2} \right)
\end{equation}

where $\sigma(\cdot)$ is the sigmoid function, $\textbf{t}$ are the training targets, and $W$ is a matrix parameter dependent on the latent function values. The derivation of this formulation requires the Laplace approximation to approximate non-Gaussian distributions as Gaussian \cite{rasmussen_gaussian_2006} as well as the inverse probit approximation to map the outputs onto interpretable probabilities \cite{bishop_pattern_2006}. To fit the GPC models, the negative log marginal likelihood in minimised. Evaluating this function once again requires the Laplace approximation such that the parameter optimisation objective function can be written as Equation \ref{eq:nll}, where $\hat{u}$ is the latent posterior mode.

\begin{equation} \label{eq:nll}
    \min - \textbf{t}^T \hat{u} + \frac{1}{2} \hat{u}^T K^{-1} \hat{u} + \frac{1}{2} \log \left| K \right| + \frac{1}{2} \log \left| W + K^{-1} \right| + \sum_{i=1}^n \log \left( 1 + \exp(\hat{u}_i) \right)
\end{equation}

\section{Object-orientated derivative-free optimisation development}

\begin{figure}[htb]
    \centering
    \includegraphics[width=0.8\textwidth]{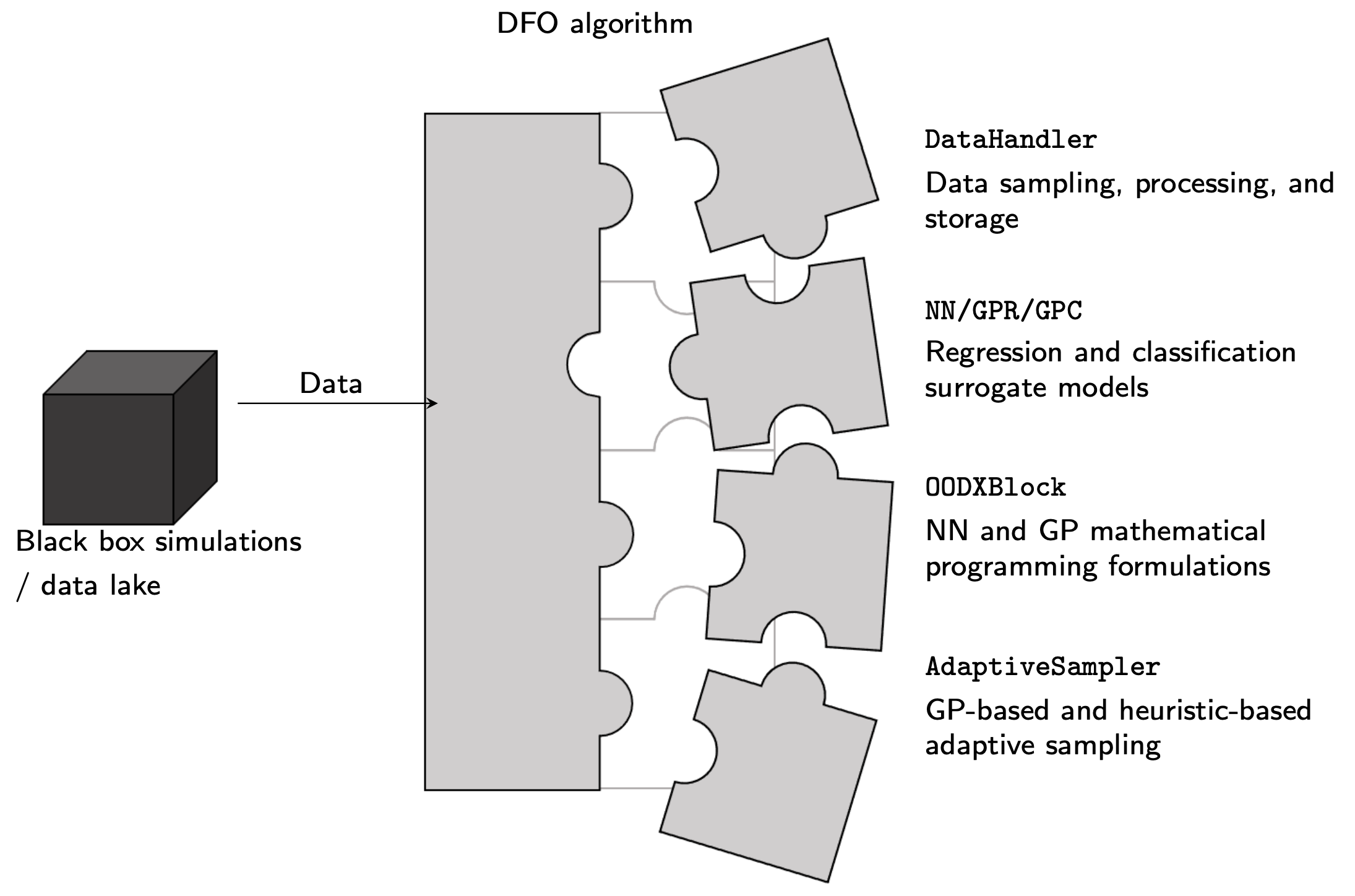}
    \caption[]{Object-orientated derivative-free optimisation (OODX) modelling objects: \texttt{DataHandler, NN, GPR, GPC, OODXBlock, AdaptiveSampler}. The modelling objects are used to construct a DFO algorithm harnessing data from black box simulations and/or data lakes.}
    \label{fig:method}
\end{figure}

This section presents a methodology for object-orientated derivative-free optimisation (OODX). Specifically, the presented method is available as an open-source Python package \cite{atdurkin-gh}, harnessing the capabilities of Pyomo and Python's rich set of machine learning, data analysis, visualisation, and optimisation libraries. Figure \ref{fig:method} depicts the modelling capabilities of OODX wherein the modelling objects are used to construct a DFO algorithm harnessing data from black box simulations and/or data lakes (built with data from online sensors, computer simulations/code evaluations, or even physical experiments).

The OODX method currently contains six modelling objects which enable entire DFO workflows to be developed and customised to specific applications. The \texttt{DataHandler} object enables data generation with sampling strategies, data processing, and data storage during DFO algorithm iterations. Three surrogate modelling objects enable (1) \texttt{NN} - NNs for regression and classification problems, (2) \texttt{GPR} - GP regression (GPR) models, and (3) \texttt{GPC} - GP classification (GPC) models. The \texttt{OODXBlock} object enables abstracted Pyomo formulations to be instantiated to represent trained \texttt{NN}/\texttt{GPR}/\texttt{GPC} objects within larger mathematical programming formulations. Finally, the \texttt{AdaptiveSampler} object enables adaptive sampling Pyomo formulations to maximise exploration of the sample space and exploitation of incumbent optimal samples.

In addition to using the OODX modelling objects within tailored DFO solutions to BBO problems, the objects can also be utilised individually or in any combination required. For example, the machine learning applications for predictive purposes might only require the \texttt{DataHandler} and \texttt{NN} objects. As another example, a classification model might be trained on existing data using the \texttt{GPC} object before plugging a representative feasibility constraint into an existing Pyomo formulation using the \texttt{OODXBlock} object.

The OODX modelling objects can be used to construct both surrogate model-based DFO algorithms and data-driven direct-search DFO methods. The former might be constructed as shown in Figure \ref{fig:method} wherein design of experiments is used to generate data from black box simulations for training surrogate models (supervised machine learning), before a mathematical programming formulation of the surrogate model is plugged into a larger decision-making problem and solved iteratively with adaptive sampling (active machine learning). On the other hand, the latter might only utilise data-driven explorative and exploitative adaptive sampling formulations to guide the search towards a global optimum.

This section proceeds by introducing the foundational concepts for each of the six objects within OODX. Specifically, the static sampling strategies and data processing methods implemented in the \texttt{DataHandler} object are presented. Similarly, the methods underpinning the \texttt{NN}, \texttt{GPR}, and \texttt{GPC} surrogate modelling objects are highlighted. Likewise, the abstracted surrogate model mathematical programming formulations used in the \texttt{OODXBlock} object are detailed. To conclude this section, the the adaptive sampling formulations available within the \texttt{AdaptiveSampler} object are presented.

\subsection{Data sampling, processing, and storage}

The \texttt{DataHandler} object enables users to perform key data processing methods for machine learning frameworks such as: static sampling strategies for input sample generation; splitting data into training and testing sets; data scaling using standardisation; replicate scaling methods to standardise new data into the same modelling space; and inverse scaling methods to return outputs from the modelling space back to the original space. The \texttt{DataHandler} object also enables data storage as a collection of attributes within a single object instance. Storing data in this way ensures that the data are not accidentally manipulated or overwritten whilst also enabling access for later analysis such as iterative model validation on testing data.
 
Static sampling strategies enable a specified number of input samples to be generated within a search space specified with lower and upper bounds on each input dimension. Sampling strategies available in the \texttt{DataHandler} object are random sampling, Latin hypercube sampling (LHS), Sobol sampling, and grid sampling. Random sampling was implemented by scaling random numbers generated between 0 and 1 into the specified search space. LHS and Sobol sampling were implemented using the implementations available within Scikit-optimize version 0.9.0, where the LHS method utilises the \emph{maximin} criterion to maximise the minimum Euclidean distance between the generated samples. The grid sampling method implemented in this work begins by calculating the value of $\sqrt[m]{n}$ where $n$ is the number of samples to be generated in $m$-dimensional space. The result of this evaluation provides the number of discrete values to be sampled at in each dimension, and is subsequently rounded up to ensure that this value, $n^{\text{grid}}$, is an integer such that $n^{\text{grid}} \geq n$. $n^{\text{grid}}$ evenly spaced samples, between the lower and upper bound of each dimension inclusive, are then generated. The resulting $n^{\text{grid}}$ samples are then shuffled and the first $n$ samples are used as the grid samples. This method thereby enables $n$ grid samples to be generated even where $n$ is not a factor of $m$ but the characteristics of grid sampling are maintained. Additionally, this method ensures that the $\left(n^{\text{grid}} - n\right)$ samples are removed randomly so that the homogeneous space coverage is maintained as much as possible.

The generated input samples were evaluated by black box models to produce output data for subsequent surrogate model fitting. Since the evaluation of output samples is highly dependent on different black box software, necessitating customised programming scripts, the \texttt{DataHandler} object enables users to directly store output data offline from their black box sampling scripts. Similarly, the \texttt{DataHandler} is flexible to enable storage of existing input-output data, from computer experiments or physical experiments. Here, input-output data (including any binary classification target data), existing in the original space, are stored along with the lower and upper bounds on the search space.

Following the generation of input-output data, the \texttt{DataHandler} object enables the data to be randomly split into training and testing sets for fitting and validating surrogate models, respectively. This implementation uses Scikit-learn version 1.0.2 and enables the specification of the fraction of the data to be reserved in the testing set (with a default value of 0.3). Training and testing inputs, outputs, and binary classification targets are stored as attributes along with the raw data.

The \texttt{DataHandler} object enables input-output data to be scaled via standardisation as shown by Equation \ref{eq:standardisation}, where $\textbf{x}$ is the data to be scaled, $\bar{x}$ is the mean, $\sigma_x$ is the standard deviation, and $\tilde{\textbf{x}}$ is the standardised data. Standardisation scales the data so that is follows a normal distribution with a mean of 0 and a standard deviation of 1.

\begin{equation} \label{eq:standardisation}
    \tilde{\textbf{x}} = \frac{\textbf{x} - \bar{x}}{\sigma_x}
\end{equation}

\begin{figure}[htb]
    \centering
    \includegraphics[width=\textwidth]{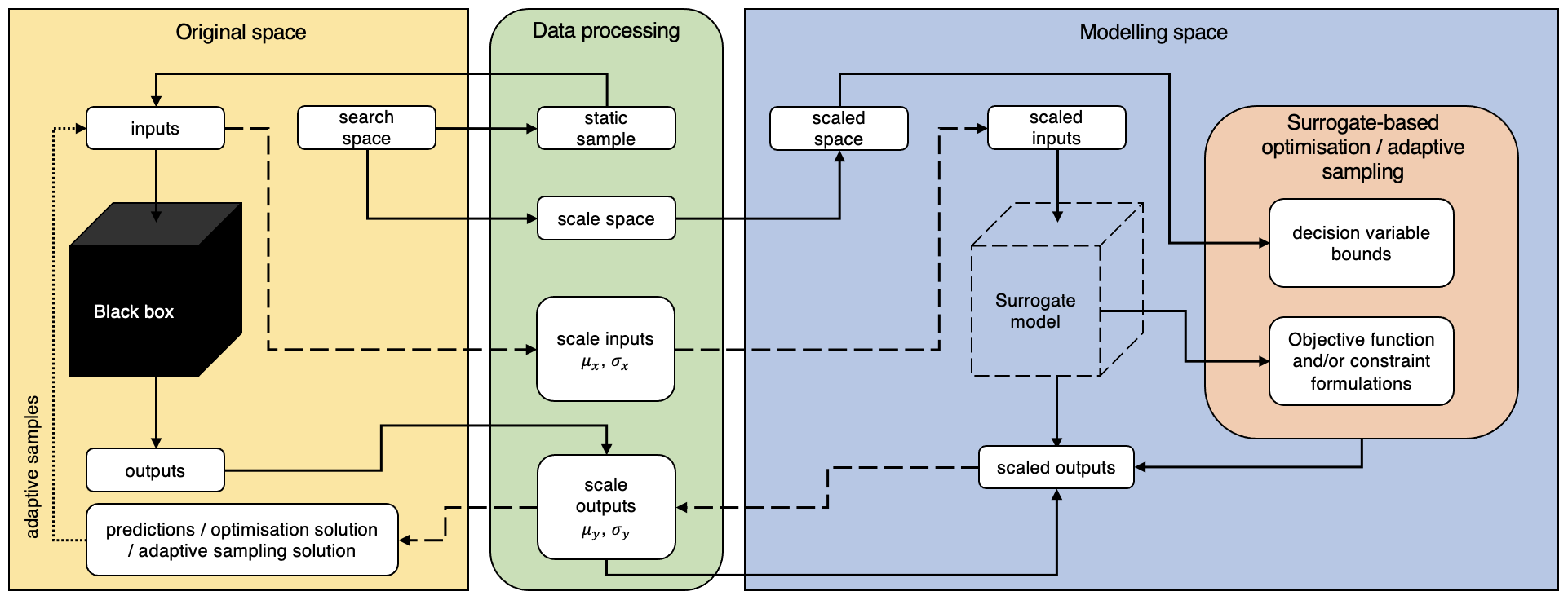}
    \caption{Data processing overview for the surrogate-based optimisation methodology. A search space is first posed (from within the original space) and static sampling methods are used to generate inputs to black box evaluations to determine corresponding outputs. Inputs and outputs are scaled to the modelling space for surrogate model training. The search space is also scaled using the statistical moments of the inputs ($\mu_x$, $\sigma_x$) to provide a scaled search space for use as optimisation decision variable bounds. Trained surrogate models are embedded in optimisation and/or adaptive sampling formulations which provide solutions in the scaled modelling space. The outputs are therefore scaled back to the original space (using $\mu_y$, $\sigma_y$) for interpretation or adaptive sampling from the black box (dotted line). Dashed lines show surrogate modelling predictions beginning from new inputs which are scaled prior to evaluation by a trained surrogate model to provide outputs which are then scaled back to the original space for interpretation.}
    \label{fig:data}
\end{figure}

The standardisation method automatically operates on the data stored within the OODX \texttt{DataHandler} object, providing standardised sets of raw inputs/outputs, training and testing inputs/outputs, as well as standardised bounds on the search space. The standardised data are stored as attributes of the \texttt{DataHandler} object instance along with the mean and standard deviation values used for scaling. Storing the mean and standard deviation enables the replication of the standardisation methods on new input-output data as well as inverse scaling methods for mapping data back to the original space (Figure \ref{fig:data}). Figure \ref{fig:data} shows an overview of the data processing tools for the surrogate-based optimisation methodology.

For example, in Figure \ref{fig:data} a typical workflow begins with input data generated from a specified search space using static sampling strategies. The corresponding outputs are then obtained by evaluating these inputs using the black box. Prior to training a surrogate model, scaling methods were used to map the original input-output data into the modelling space. Standardisation into the modelling space ensures that the input-output data are scaled over the samples, for each dimension, to avoid overfitting to dimensions with larger magnitudes. Without this scaling, the larger magnitude dimensions would have greater influence during model fitting. Scaled input-output data can then be used to train a surrogate model to enable the mapping from scaled scaled inputs to predictions in the modelling space. New inputs in the original space are then scaled into the modelling space , mapped onto surrogate model predictions (bottom right), and finally mapped back to the original space (dashed dataflow). In this way, the black box input-output mapping can be achieved with the appropriate scaling, surrogate model predictions, and inverse scaling.

The standardisation methods are based on a set of assumptions which systematise the methods and ensure that they are flexible to different configurations of stored data. For example, the methods automatically adapt based on whether training and testing sets have been generated, or whether binary classification targets representing sample feasibility have been stored. The assumptions used to systematise the standardisation, along with further details on all the \texttt{DataHandler} instance attributes and methods, are shown in Table \ref{tab:datahandler}.

\subsection{Neural networks}

Section \ref{sec:nn} introduced some background on NNs. The \texttt{NN} object within OODX enables building, training, and evaluation of NN models using PyTorch version 2.0 \cite{paszke_pytorch_2019}. Specifically, the customisation capabilities enable the number of layers ($\Lambda$), the number of nodes in each layer ($N_{\lambda}$), and the activation functions ($\xi^{(\lambda)}$) to be defined for fully-dense feedforward NNs (Figure \ref{fig:nn}). By default, the \texttt{NN} object is trained using mini-batch gradient descent (Table \ref{tab:nn-training}) with a mean squared error loss function for regression NNs. The optimised weights and biases are saved within trained \texttt{NN} objects to enable explicit mathematical formulations of NNs to be realised from generalised formulations. The general formulations, owing to the generality of fully-dense feedforward NN structures, were formulated for linear layers (Equation \ref{eq:nn1}) passed through activation functions (Equation \ref{eq:nn2}) and constructed layer by layer from inputs (Equation \ref{eq:nn-in}) to outputs (Equation \ref{eq:nn-out}). Predictions from explicit mathematical formulations were validated against predictions made directly from the PyTorch-based \texttt{NN} object. Further details on all the \texttt{NN} instance attributes and methods, are shown in Table \ref{tab:nn}.

\subsection{Gaussian processes}

Section \ref{sec:gps} introduced some background on GPs for both regression and classification. GP regression (GPR) in OODX uses the \texttt{GPR} object which was based on the GPR implementation in Scikit-learn version 1.0.2 \cite{scikit-learn} with additional functionality for deriving explicit mathematical formulations for use in DFO. The \texttt{GPR} object implements linear/polynomial and squared exponential kernel functions which are optimised via maximum likelihood estimation (MLE). Predictions and standard deviations in these predictions can be evaluated directly from the underlying Scikit-learn implementation or via an explicit mathematical formulation constructed using the saved optimised parameters. Predictions from the mathematical formulation were validated against predictions from the Scikit-learn object. Further details on all the \texttt{GPR} instance attributes and methods, are shown in Table \ref{tab:gpr}.

\subsection{Classification models}

The OODX \texttt{NN} object can also be adapted for classification surrogate modelling by utilising specialised training functions such as the binary cross entropy loss. Further details regarding NNs for classification were presented at the end of Section \ref{sec:nn} whilst further details on using the OODX \texttt{NN} object for classification can be found in Table \ref{tab:nn}.

The OODX \texttt{GPC} object was coded entirely in NumPy version 1.21.2 to enable the GPC model to be constructed in such a way that enabled an explicit mathematical formulation of the predictive function to be formulated for compatibility with Pyomo. Despite the existence of a GPC implementation in Scikit-learn, this object uses a statistical approximation which complicates formulating an equivalent analytical formulation. Accordingly, the \texttt{GPC} object used the squared exponential kernel function (Equation \ref{eq:sqexp1}) and the probit approximation \cite{bishop_pattern_2006} to enable the explicit mathematical formulation shown in Equation \ref{eq:probit}. The MLE formulation to fit the GPC model was also implemented NumPy using Equation \ref{eq:nll}. For machine learning applications, it was sufficient to make predictions using the linear algebraic notation that the underlying model formulation lends itself to, however, for mathematical optimisation applications, is was necessary to write an equivalent formulation incorporating summations over parameter/variable indices. Predictions from the explicit Pyomo-compatible formulation were validated against the linear algebra predictive function. Further details on GPC were presented at the end of Section \ref{sec:gps} whilst all of the \texttt{GPC} instance attributes and methods are shown in Table \ref{tab:gpc}.

% \subsection{Regularisation}

\subsection{Mathematical programming formulations}

Abstracted Pyomo formulations enable hierarchical optimisation modelling by adding modelling components as attributes to object instances \cite{bynum_pyomo_2021}. Such object instances can then be added to larger to decision-making problems to enforce the surrogate model formulations in a plug-and-play framework. The \texttt{OODXBlock} object harnesses the object-orientated capabilities of Pyomo to contain the relevant sets, parameters, variables, and constraints for general surrogate model formulations which can be used to build surrogate-based DFO algorithms for solving BBO problems. The general \texttt{OODXBlock} formulations can be instantiated as formulations representative of specific OODX \texttt{NN}, \texttt{GPR} and \texttt{GPC} models. Equality constraints then connect the block inputs and outputs to corresponding variables in the larger optimisation formulations. The hierarchical optimisation modelling structure means that the larger optimisation formulations can access the surrogate modelling parameters as well as enforcing the constraints.

\subsubsection{Neural network formulations}

Fully dense feedforward NNs with $\lambda=0, ..., \Lambda+1$ layers (where $\lambda=0$ is the input layer, $\lambda=1, ..., \Lambda$ are the hidden layers, and $\lambda=\Lambda+1$ is the output layer) and $N_\lambda$ nodes in each layer were formulated using Equations \ref{eq:nn1}, \ref{eq:nn-in}, and \ref{eq:nn-out}. These equations define the NN structure and mapping of inputs to predictive outputs. The equivalent optimisation formulation for these Equations (Table \ref{tab:algorithm-nngen}) enforce the NN structure and optimised weights and biases thereby enforcing an exact formulation of the linear mappings wthin the NN. This means that a prediction from the \texttt{NN} object at a given input returns the exact same value as a prediction enforced within optimisation problems using this \texttt{OODXBlock} formulation. At this stage, prior to the definition of the activation functions and the required mapping from node inputs $\textbf{z}$ onto activated node outputs $\textbf{a}$, the mathematical program in Table \ref{tab:algorithm-nngen} is a linear programming (LP) formulation.

\begin{table}[htb]
    \centering
    \caption[General feedforward NN LP formulation]{Optimisation formulation for general fully dense feedforward neural network structure as linear program. Note that this formulation cannot be solved for any meaningful solution without defining the activation function constraints.}
    \begin{tabular}{l l}
        \hline
        \multicolumn{2}{l}{Sets} \\
        \hline
        $\lambda \in \{0,..., \Lambda+1\}$ & set of layers \\
        $N_\lambda$ & set of nodes in each layer \\
        \hline
        \multicolumn{2}{l}{Parameters} \\
        \hline
        $w^{(\lambda)}_{i,i'}$ & weights indexed by $\lambda = 0,..., \Lambda$, $i\in N_{\lambda+1}$, $i' \in N_{\lambda}$ \\
        $b^{(\lambda)}_{i}$ & biases indexed by $\lambda = 0,..., \Lambda$, $i\in N_{\lambda+1}$ \\
        \hline
        \multicolumn{2}{l}{Variables} \\
        \hline
        $x_i$ & network inputs indexed by $i \in N_0$ \\
        $\hat{y}_i$ & network outputs indexed by $i \in N_{\Lambda+1}$ \\
        $z^{(\lambda)}_i$ & node inputs indexed by $\lambda = 1,..., \Lambda$, $i \in N_{\lambda}$ \\
        $a^{(\lambda)}_i$ & activated outputs indexed by $\lambda = 1,..., \Lambda$, $i \in N_{\lambda}$ \\
        \hline
        \multicolumn{2}{l}{Equations} \\
        \hline
        \multicolumn{2}{l}{
            \textbf{for} $i$ \textbf{in} $N_1$ \textbf{do}
        }\\
        \multicolumn{2}{l}{
            \quad $z^{(1)}_i = \sum\limits_{i'\in N_0} w^{(0)}_{i,i'} x_{i'} + b^{(0)}_i$
        } input layer \\
        \\
        \multicolumn{2}{l}{
            \textbf{for} $\lambda=2$ \textbf{to} $\Lambda$ \textbf{do}
        }\\
        \multicolumn{2}{l}{
            \quad \textbf{for} $i$ \textbf{in} $N_\lambda$ \textbf{do}
        }\\
        \multicolumn{2}{l}{
            \qquad $z^{(\lambda)}_i = \sum\limits_{i'\in N_{\lambda-1}} w^{(\lambda-1)}_{i,i'} a^{(\lambda-1)}_{i'} + b^{(\lambda-1)}_i$
        } hidden layer inputs \\
        \\
        \multicolumn{2}{l}{
            \textbf{for} $i$ \textbf{in} $N_{\Lambda+1}$ \textbf{do}
        }\\
        \multicolumn{2}{l}{
            \quad $\hat{y}_i = \sum\limits_{i'\in N_\Lambda} w^{(\Lambda)}_{i,i'} a^{(\Lambda)}_{i'} + b^{(\Lambda)}_i$
        } output layer \\
        \hline
    \end{tabular}
    \label{tab:algorithm-nngen}
\end{table}

In the LP formulation in Table \ref{tab:algorithm-nngen}, $a_i^{(\lambda)}$ represents the activated outputs from layer $\lambda$ for nodes $i=1, ..., N_\lambda$, defined over the hidden layers $\lambda=1, ..., \Lambda$. To complete the NN-based formulations, activation function constraints were defined to map node inputs $z_i^{(\lambda)}$ onto the activated output variables. A linear activation function multiplies the layer outputs by 1, retaining the LP formulation for the NN model. Whilst the hyperbolic tangent function is commonly used as an activation function, the trigonometric expression, tanh, is typically unsupported by optimisation solvers. The four explicit algebraic formulations of tanh containing exponential terms (Table \ref{tab:activationfuncs}) have been implemented and their performance analysed in the literature \cite{schweidtmann_deterministic_2019}. Here, the tanh formulation in Table \ref{tab:3acts} was implemented due to the single exponential term dependent on variable inputs. Similar formulations containing exponential terms were formulated for sigmoid and softplus activation functions (Table \ref{tab:3acts}).

The activation function formulations in Table \ref{tab:3acts} can be combined with the LP in Table \ref{tab:algorithm-nngen} depending on the activation function used within a trained NN, and be embedded with larger optimisation problems. Formulations for tanh, sigmoid, and softplus functions thereby provide nonlinear programming (NLP) formulations due to the nonlinearities introduced by these activation function constraints. The 4 activation functions in Table \ref{tab:3acts} were formulated in Pyomo using the LP/NLP formulations and then used to form complete Pyomo-based LP/NLP formulations by combining with the general NN structural formulation in Table \ref{tab:algorithm-nngen}. In practice, the appropriate constraints are automatically activated depending on the NN object passed as an argument to the \texttt{OODXBlock} object.

\begin{longtable}{l l}
    \caption[NN activation function LP/NLP formulations]{Optimisation formulations for neural network activation functions: linear (LP), tanh (NLP), sigmoid (NLP), softplus (NLP), ReLU (MILP), and hardsigmoid (MILP).} \label{tab:3acts} \\

    \hline
    \multicolumn{2}{l}{Sets} \\
    \hline
    \endfirsthead

    \multicolumn{2}{c}%
    {{\tablename\ \thetable{} -- continued from previous page}} \\
    \hline
    \endhead

    \hline \multicolumn{2}{r}{{Table continued on next page}} \\
    \endfoot

    \hline
    \endlastfoot
    
    $\lambda \in \{1,..., \Lambda\}$ & set of hidden layers \\
    $N_\lambda$ & set of nodes in each layer \\
    \hline
    \multicolumn{2}{l}{Parameters} \\
    \hline
    $M$ & arbitrary large positive scalar for big-M formulations \\
    \hline
    \multicolumn{2}{l}{Variables} \\
    \hline
    $z^{(\lambda)}_i$ & node inputs indexed by $\lambda = 1,..., \Lambda$, $i \in N_{\lambda}$ \\
    $a^{(\lambda)}_i$ & activated outputs indexed by $\lambda = 1,..., \Lambda$, $i \in N_{\lambda}$ \\
    $p^{(\lambda)}_i$ & binary variable indexed by $\lambda = 1,..., \Lambda$, $i \in N_{\lambda}$ \\
    $q^{(\lambda)}_i$ & binary variable indexed by $\lambda = 1,..., \Lambda$, $i \in N_{\lambda}$ \\
    \hline
    \multicolumn{2}{l}{Equations} \\
    \hline
    \multicolumn{2}{l}{
        \textbf{for} $\lambda=1$ \textbf{to} $\Lambda$ \textbf{do}
    }\\
    \multicolumn{2}{l}{
        \quad \textbf{for} $i$ \textbf{in} $N_\lambda$ \textbf{do}
    }\\
    \multicolumn{2}{l}{
        \qquad $a^{(\lambda)}_i = z^{(\lambda)}_i$
    } linear LP \\
    \hline
    \multicolumn{2}{l}{
        \textbf{for} $\lambda=1$ \textbf{to} $\Lambda$ \textbf{do}
    }\\
    \multicolumn{2}{l}{
        \quad \textbf{for} $i$ \textbf{in} $N_\lambda$ \textbf{do}
    }\\
    \multicolumn{2}{l}{
        \qquad $a^{(\lambda)}_i = 1 - \frac{2}{\exp\left(2z_i^{(\lambda)}\right) + 1}$
    } tanh NLP \\
    \hline
    \multicolumn{2}{l}{
        \textbf{for} $\lambda=1$ \textbf{to} $\Lambda$ \textbf{do}
    }\\
    \multicolumn{2}{l}{
        \quad \textbf{for} $i$ \textbf{in} $N_\lambda$ \textbf{do}
    }\\
    \multicolumn{2}{l}{
        \qquad $a^{(\lambda)}_i = \frac{1}{1 + \exp\left(-z^{(\lambda)}_i\right) }$
    } sigmoid NLP \\
    \hline
    \multicolumn{2}{l}{
        \textbf{for} $\lambda=1$ \textbf{to} $\Lambda$ \textbf{do}
    }\\
    \multicolumn{2}{l}{
        \quad \textbf{for} $i$ \textbf{in} $N_\lambda$ \textbf{do}
    }\\
    \multicolumn{2}{l}{
        \qquad $a^{(\lambda)}_i = \log\left( 1 + \exp\left( z^{(\lambda)}_i \right) \right)$
    } softplus NLP \\
    \hline
    \multicolumn{2}{l}{
        \textbf{for} $\lambda=1$ \textbf{to} $\Lambda$ \textbf{do}
    }\\
    \multicolumn{2}{l}{
        \quad \textbf{for} $i$ \textbf{in} $N_\lambda$ \textbf{do}
    }\\
    \multicolumn{2}{l}{
        \qquad $a^{(\lambda)}_i \geq 0$
    } \\
    \multicolumn{2}{l}{
        \qquad $a^{(\lambda)}_i \geq z^{(\lambda)}_i$
    } \\
    \multicolumn{2}{l}{
        \qquad $a^{(\lambda)}_i \leq M p^{(\lambda)}_i$
    } \\
    \multicolumn{2}{l}{
        \qquad $a^{(\lambda)}_i \leq z^{(\lambda)}_i + M \left(1 - p^{(\lambda)}_i\right)$
    } ReLU MILP \\
    \hline
    \multicolumn{2}{l}{
        \textbf{for} $\lambda=1$ \textbf{to} $\Lambda$ \textbf{do}
    }\\
    \multicolumn{2}{l}{
        \quad \textbf{for} $i$ \textbf{in} $N_\lambda$ \textbf{do}
    }\\
    \multicolumn{2}{l}{
        \qquad $a^{(\lambda)}_i \leq p^{(\lambda)}_i$
    } \\
    \multicolumn{2}{l}{
        \qquad $a^{(\lambda)}_i \geq \frac{1}{6} z^{(\lambda)}_i + \frac{1}{2} - M\left(1 - p^{(\lambda)}_i + q^{(\lambda)}_i\right)$
    } \\
    \multicolumn{2}{l}{
        \qquad $a^{(\lambda)}_i \leq \frac{1}{6} z^{(\lambda)}_i + \frac{1}{2} + M\left(1 - p^{(\lambda)}_i + q^{(\lambda)}_i\right)$
    } \\
    \multicolumn{2}{l}{
        \qquad $a^{(\lambda)}_i \geq q^{(\lambda)}_i$
    } \\
    \multicolumn{2}{l}{
        \qquad $z^{(\lambda)}_i \leq M p^{(\lambda)}_i - 3$
    } \\
    \multicolumn{2}{l}{
        \qquad $z^{(\lambda)}_i \geq M\left(p^{(\lambda)}_i - 1\right) - 3$
    } \\
    \multicolumn{2}{l}{
        \qquad $z^{(\lambda)}_i \leq M q^{(\lambda)}_i + 3$
    } \\
    \multicolumn{2}{l}{
        \qquad $z^{(\lambda)}_i \geq M \left(q^{(\lambda)}_i - 1\right) + 3$
    } hardsigmoid MILP \\
\end{longtable}

Mixed-integer linear programs (MILPs) can also be formulated for NNs utilising piecewise linear activation functions such as ReLU and hardsigmoid. Equations \ref{eq:nn-relu1} - \ref{eq:nn-relu4} enforce the MILP formulation for ReLU activation functions on layer $\lambda$ for nodes $i=1, ..., N_\lambda$. This formulation utilises a binary activation variable $p_i^{(\lambda)}$ that determines the location of $z_i^{(\lambda)}$ in the ReLU function input domain and whether it is positive or negative. Constraints are enforced and relaxed when required using the big-M formulation to enforce the following properties:

\begin{itemize}
    \item $a_i^{(\lambda)}$ is always greater than or equal to 0 (Equation \ref{eq:nn-relu1}),
    \item if $z_i^{(\lambda)}$ is negative, then $p_i^{(\lambda)}$ is constrained to 0 by Equation \ref{eq:nn-relu4}, and $a_i^{(\lambda)}$ is constrained equal to 0 by Equation \ref{eq:nn-relu3},
    \item if $z_i^{(\lambda)}$ is positive, then $p_i^{(\lambda)}$ is constrained to 1 in order to satisfy Equations \ref{eq:nn-relu2} and \ref{eq:nn-relu3}, and $a_i^{(\lambda)}$ is constrained equal to $z_i^{(\lambda)}$ by Equations \ref{eq:nn-relu2} and \ref{eq:nn-relu4}.
\end{itemize}

\begin{subequations}
    \begin{align}
        a_i^{(\lambda)} \geq \; & 0 & & \text{for } i = 1, ..., N_\lambda \quad \text{for } \lambda = 1, ..., \Lambda \label{eq:nn-relu1} 
        \\
        a_i^{(\lambda)} \geq \; & z_i^{(\lambda)} & & \text{for } i = 1, ..., N_\lambda \quad \text{for } \lambda = 1, ..., \Lambda \label{eq:nn-relu2} 
        \\
        a_i^{(\lambda)} \leq \; & M p_i^{(\lambda)} & & \text{for } i = 1, ..., N_\lambda \quad \text{for } \lambda = 1, ..., \Lambda \label{eq:nn-relu3} 
        \\
        a_i^{(\lambda)} \leq \; & z_i^{(\lambda)} + M \left( 1 - p_i^{(\lambda)} \right) & & \text{for } i = 1, ..., N_\lambda \quad \text{for } \lambda = 1, ..., \Lambda \label{eq:nn-relu4}
    \end{align}
\end{subequations}

Equations \ref{eq:nn-hs1} - \ref{eq:nn-hs8} enforce the MILP formulation for hardsigmoid activation functions on layer $\lambda$ for nodes $i=1, ..., N_\lambda$. This formulation utilises 2 binary activation variables $p_{i}^{(\lambda)}$ and $q_{i}^{(\lambda)}$ to determine the location of $z_i^{(\lambda)}$ in the function input domain and whether it is less than -3, greater than +3, or between -3 and +3. The big-M formulation is used again here to enforce the following properties:

\begin{itemize}
    \item if $z_i^{(\lambda)} \leq -3$, $p_{i}^{(\lambda)}$ is constrained to 0 by Equation \ref{eq:nn-hs2}, $q_{i}^{(\lambda)}$ is constrained to 0 by Equation \ref{eq:nn-hs4}, and $a_i^{(\lambda)}$ is constrained equal to 0 by Equations \ref{eq:nn-hs5} and \ref{eq:nn-hs8},
    \item if $z_i^{(\lambda)} \geq +3$, $p_{i}^{(\lambda)}$ is constrained to 1 by Equation \ref{eq:nn-hs1}, $q_{i}^{(\lambda)}$ is constrained to 1 by Equation \ref{eq:nn-hs3}, and $a_i^{(\lambda)}$ is constrained equal to 1 by Equations \ref{eq:nn-hs5} and \ref{eq:nn-hs8},
    \item if $-3 \leq z_i^{(\lambda)} \leq +3$, $p_{i}^{(\lambda)}$ is constrained to 1 by Equation \ref{eq:nn-hs1}, $q_{i}^{(\lambda)}$ is constrained to 0 by Equation \ref{eq:nn-hs4}, and $a_i^{(\lambda)}$ is constrained equal to $\frac{1}{6}z_i^{(\lambda)} + \frac{1}{2}$ by Equations \ref{eq:nn-hs6} and \ref{eq:nn-hs7}.
\end{itemize}

\begin{subequations}
    \begin{align}
        z_i^{(\lambda)} \leq \; & M p_{i}^{(\lambda)} -3 & & \text{for } i = 1, ..., N_\lambda \quad \text{for } \lambda = 1, ..., \Lambda \label{eq:nn-hs1} 
        \\
        z_i^{(\lambda)} \geq \; & M \left( p_{i}^{(\lambda)} - 1 \right) -3 & & \text{for } i = 1, ..., N_\lambda \quad \text{for } \lambda = 1, ..., \Lambda \label{eq:nn-hs2} 
        \\
        z_i^{(\lambda)} \leq \; & M q_{i}^{(\lambda)} +3 & & \text{for } i = 1, ..., N_\lambda \quad \text{for } \lambda = 1, ..., \Lambda \label{eq:nn-hs3} 
        \\
        z_i^{(\lambda)} \geq \; & M \left( q_{i}^{(\lambda)} - 1 \right) +3 & & \text{for } i = 1, ..., N_\lambda \quad \text{for } \lambda = 1, ..., \Lambda \label{eq:nn-hs4}
        \\
        a_i^{(\lambda)} \leq \; & p_{i}^{(\lambda)} & & \text{for } i = 1, ..., N_\lambda \quad \text{for } \lambda = 1, ..., \Lambda \label{eq:nn-hs5} 
        \\
        a_i^{(\lambda)} \geq \; & \frac{1}{6} z_i^{(\lambda)} + \frac{1}{2} - M \left( 1 - p_{i}^{(\lambda)} + q_{i}^{(\lambda)} \right) & & \text{for } i = 1, ..., N_\lambda \quad \text{for } \lambda = 1, ..., \Lambda \label{eq:nn-hs6} 
        \\
        a_i^{(\lambda)} \leq \; & \frac{1}{6} z_i^{(\lambda)} + \frac{1}{2} + M \left( 1 - p_{i}^{(\lambda)} + q_{i}^{(\lambda)} \right) & & \text{for } i = 1, ..., N_\lambda \quad \text{for } \lambda = 1, ..., \Lambda \label{eq:nn-hs7} 
        \\
        a_i^{(\lambda)} \geq \; & p_{i}^{(\lambda)} & & \text{for } i = 1, ..., N_\lambda \quad \text{for } \lambda = 1, ..., \Lambda \label{eq:nn-hs8} 
    \end{align}
\end{subequations}

The Pyomo formulations for the 2 MILP NN activation functions (ReLU and hardsigmoid) are available within the \texttt{OODXBlock} object as shown in Table \ref{tab:3acts}. The ReLU and hardsigmoid formulations in Table \ref{tab:3acts} are used to form a complete Pyomo-based MILP formulations by combining with the general NN structural formulation in Table \ref{tab:algorithm-nngen}.

\subsubsection{Gaussian process formulations}

GPR surrogate models were formulated exactly using the mathematical programming formulations shown in Table \ref{tab:gprmp}. Specifically, the GPR predictive mean function (Equation \ref{eq:posterior-mean}) was combined with explicit formulations for the squared exponential (Equation \ref{eq:sqexp1}) and polynomial (Equation \ref{eq:covpoly}) kernel functions to yield Pyomo compatible NLP formulations (translating matrix and vector notation into formulations readable by optimisation solvers). NLP formulations for trained GPR models with optimised parameters are automatically generated from the general formulations using the \texttt{OODXBlock} object. These NLP formulations thereby provide exact representations of the GPR predictive mean function which was validated by comparing predictions from the explicit formulations and \texttt{GPR} object instances. 

\begin{longtable}{l l}
    \caption[GPR formulations for different kernel function]{Optimisation formulation for Gaussian process regression (GPR) predictions with squared exponential and polynomial (including linear with $\omega=1$) kernel functions.}\label{tab:gprmp} \\

    \hline
    \multicolumn{2}{l}{Sets} \\
    \hline
    \endfirsthead

    \multicolumn{2}{c}%
    {{\tablename\ \thetable{} -- continued from previous page}} \\
    \hline
    \endhead

    \hline \multicolumn{2}{r}{{Table continued on next page}} \\
    \endfoot

    \hline
    \endlastfoot

    $n$ & set over the number of training samples \\
    $m$ & set over the number of input dimensions \\
    \hline
    \multicolumn{2}{l}{Parameters} \\
    \hline
    $x^{(\text{train})}_{i,j}$ & training inputs with dimensions $i\in n$, $j \in m$ \\
    $l$ & scalar length scale \\
    $\sigma_f^2$ & scalar process variance \\
    $\sigma_0^2$ & scalar inhomogenity variance \\
    $\omega$ & scalar integer polynomial order \\
    $\alpha_i$ & linear predictor weights with dimensions $i\in n$ \\
    \hline
    \multicolumn{2}{l}{Variables} \\
    \hline
    $x_j$ & model inputs with dimension $j \in m$ \\
    $\hat{y}$ & model predictive output \\
    \hline
    \multicolumn{2}{l}{Equations} \\
    \hline
    \multicolumn{2}{l}{
        $\hat{y} = \sum\limits_{i \in n} \alpha_i \sigma_f^2 \exp\left( -\sum\limits_{j \in m} \frac{1}{2 l^2} \left( x_j - x^{(\text{train})}_{i,j} \right)^2 \right)$
    } squared exponential \\
    \\
    \multicolumn{2}{l}{
        $\hat{y} = \sum\limits_{i \in n} \alpha_i \sigma_f^2 \left( \sigma_0^2 + \sum\limits_{j \in m} x^{(\text{train})}_{i,j} x_j \right)^\omega$
    } polynomial \\
\end{longtable}

The uncertainty in GPR models can also be formulated such that optimisation solvers can implement them in larger decision making problems. Equation \ref{eq:posterior-cov} shows the variance in GPR predictions at a new input vector $\textbf{x}$. Since $\sigma_f^2$ and $K^{-1}$ are parameters dependent only on training data and fixed prior to optimisation, the only dependence on new inputs is via $\textbf{k}$. Equation \ref{eq:gprstd-con} therefore shows the optimisation formulation for the varying $\textbf{k}^T K^{-1}\textbf{k}$ part of the variance formulation, where care must be taken to reformulate optimisation problems accordingly (i.e. to maximise the uncertainty, Equation \ref{eq:gprstd-con} must be minimised). Note that, where the uncertainty requires formulating, Equation \ref{eq:gprstd-con} must be subtracted from $\sigma_f^2$ to provide the variance, square-rooted to provide the standard deviation, then multiplied by the correct constant to obtain the necessary confidence interval (e.g., $1.96\sigma$ approximates the 95\,\% confidence interval).

\begin{equation} \label{eq:gprstd-con}
    \upsilon^{(proxy)} = \sum_{i=1}^n\left( k_i \sum_{i^\prime=1}^n \left( K^{-1}_{i, i^\prime} k_{i^\prime} \right)\right)
\end{equation}

The Pyomo-based formulation of the GPR uncertainty proxy (Equation \ref{eq:gprstd-con}) is available in the \texttt{OODXBlock} object for squared exponential, linear, and polynomial kernel function as shown in Table \ref{tab:gprump}.

\begin{longtable}{l l}
    \caption[GPR uncertainty proxy formulations]{Optimisation formulations for Gaussian process regression (GPR) uncertainty proxy with squared exponential and polynomial (including linear with $\omega=1$) kernel functions.}\label{tab:gprump} \\

    \hline
    \multicolumn{2}{l}{Sets} \\
    \hline
    \endfirsthead

    \multicolumn{2}{c}%
    {{\tablename\ \thetable{} -- continued from previous page}} \\
    \hline
    \endhead

    \hline \multicolumn{2}{r}{{Table continued on next page}} \\
    \endfoot

    \hline
    \endlastfoot

    $n$ & set over the number of training samples \\
    $m$ & set over the number of input dimensions \\
    \hline
    \multicolumn{2}{l}{Parameters} \\
    \hline
    $x^{(\text{train})}_{i,j}$ & training inputs with dimensions $i\in n$, $j \in m$ \\
    $l$ & scalar length scale \\
    $\sigma_f^2$ & scalar process variance \\
    $\sigma_0^2$ & scalar inhomogenity variance \\
    $\omega$ & scalar integer polynomial order \\
    $K^{-1}_{i, j}$ & inverse covariance matrix with dimensions $i, j \in n$ \\
    \hline
    \multicolumn{2}{l}{Variables} \\
    \hline
    $x_j$ & model inputs with dimension $j \in m$ \\
    $\upsilon^{(proxy)}$ & model uncertainty proxy output \\
    \hline
    \multicolumn{2}{l}{Equations} \\
    \hline
    \multicolumn{2}{l}{
        $\begin{aligned}
            \upsilon^{(proxy)} ={} & \; \sum\limits_{i \in n} \left( \sigma_f^2 \exp\left( -\sum\limits_{j \in m} \frac{1}{2 l^2} \left( x_j - x^{(\text{train})}_{i,j} \right)^2 \right)\right. \\ 
            & \;\times \left.\sum\limits_{k \in n} K^{-1}_{i, k} \sigma_f^2 \exp\left( -\sum\limits_{j \in m} \frac{1}{2 l^2} \left( x_j - x^{(\text{train})}_{i,j} \right)^2 \right) \right) \\
        \end{aligned}$
    } \\
    \multicolumn{2}{r}{squared exponential} \\
    \\
    \multicolumn{2}{l}{
        $\begin{aligned}
            \upsilon^{(proxy)} ={} & \; \sum\limits_{i \in n} \left( \sigma_f^2 \left( \sigma_0^2 + \sum\limits_{j \in m} x^{(\text{train})}_{i,j} x_j \right)^\omega\right. \\ 
            & \;\times \left.\sum\limits_{k \in n} K^{-1}_{i, k} \sigma_f^2 \left( \sigma_0^2 + \sum\limits_{j \in m} x^{(\text{train})}_{i,j} x_j \right)^\omega \right) \\
        \end{aligned}$
    } \\
    \multicolumn{2}{r}{polynomial} \\
\end{longtable}

In addition to GPR-based optimisation, a GPC model with given training inputs and saved optimal model parameters can be enforced with the NLP constraint shown in Equation \ref{eq:gpc-con} (where $k_i$ has been defined to clean up the notation). Once again, this constraint is equivalent to Equation \ref{eq:probit}, where the kernel function has been formulated explicitly and matrix/vector operations have been translated into summations over indices to enable interpretation by optimisation solvers.

\begin{equation} \label{eq:gpc-con}
    \begin{aligned}
        p(t=1) = \left( 1 + \exp \left[ - \frac{ \sum_{i=1}^n \left( \delta_i k_i \right) }{\left( 1 + \frac{\pi}{8} \left[ \sigma_f^2 - \sum_{i=1}^n \left( k_i \sum_{k=1}^n \left[ k_i P_{i,k}^{-1} \right] \right) \right] \right)^{1/2}} \right] \right)^{-1}
        \\
        \text{where } k_i = \sigma_f^2 \exp \left( - \sum_{j=1}^m \frac{1}{2l^2} \left( x_j - x^{(\text{train})}_{i, j} \right)^2 \right)
    \end{aligned}
\end{equation}

The Pyomo formulation of Equation \ref{eq:gpc-con} exists in the \texttt{OODXBlock} object with the corresponding NLP formulation shown in Table \ref{tab:gpcmp}.

\begin{longtable}{l l}
    \caption[GPC NLP formulation]{Optimisation formulation for Gaussian process classification (GPC) with squared exponential kernel function probability predictions.}\label{tab:gpcmp} \\

    \hline
    \multicolumn{2}{l}{Sets} \\
    \hline
    \endfirsthead

    \multicolumn{2}{c}%
    {{\tablename\ \thetable{} -- continued from previous page}} \\
    \hline
    \endhead

    \hline \multicolumn{2}{r}{{Table continued on next page}} \\
    \endfoot

    \hline
    \endlastfoot

    $n$ & set over the number of training samples \\
    $m$ & set over the number of input dimensions \\
    \hline
    \multicolumn{2}{l}{Parameters} \\
    \hline
    $x^{(\text{train})}_{i,j}$ & training inputs with dimensions $i\in n$, $j \in m$ \\
    $l$ & scalar length scale \\
    $\sigma_f^2$ & scalar process variance \\
    $\delta_{i}$ & aggregate parameter with dimension $i \in n$ \\
    $P^{-1}_{i, j}$ & aggregate parameter with dimensions $i, j \in n$ \\
    \hline
    \multicolumn{2}{l}{Variables} \\
    \hline
    $x_j$ & model inputs with dimension $j \in m$ \\
    $p(t=1)$ & model probability prediction \\
    \hline
    \multicolumn{2}{l}{Equations} \\
    \hline
    \multicolumn{2}{l}{
        $\begin{aligned}
            p(t=1) ={} & \; \left( 1 + \exp \left[ - \sum_{i=1}^n \left( \delta_i \sigma_f^2 \exp \left( - \sum_{j=1}^m \frac{1}{2l^2} \left( x_j - x^{(\text{train})}_{i, j} \right)^2 \right) \right) \right.\right.\\
            & \; \times\left( 1 + \frac{\pi}{8} \left[ \sigma_f^2 - \sum_{i=1}^n \left( \sigma_f^2 \exp \left( - \sum_{j=1}^m \frac{1}{2l^2} \left( x_j - x^{(\text{train})}_{i, j} \right)^2 \right)\right.\right.\right.\\ 
            & \; \times \left.\left.\left.\left.\left. \sum_{k=1}^n \left[ \sigma_f^2 \exp \left( - \sum_{j=1}^m \frac{1}{2l^2} \left( x_j - x^{(\text{train})}_{i, j} \right)^2 \right) P_{i,k}^{-1} \right] \right) \right] \right)^{-1/2} \right] \right)^{-1} \\ 
        \end{aligned}$
    } \\
    \multicolumn{2}{r}{GPC probability predictions} \\
\end{longtable}

\subsection{Adaptive sampling}

The OODX \texttt{AdaptiveSampler} object provides GP-based and heuristic-based adaptive sampling Pyomo formulation objects. Specifically, exploration and exploitation are enabled using GP-based adaptive sampling formulations to maximise GPR uncertainty and the modified expected improvement (EI) function, respectively. Adaptive sampling is also enabled by using Delaunay triangulation to partition the search space into regions, then using heuristics to choose which region to place samples so as to achieve exploration and exploitation equivalent to the GP-based methods. The resulting GP-based formulations were NLPs whilst the heuristic-based formulations were implemented as MILP formulations.

The 4 adaptive sampling formulations were implemented in Pyomo version 6.2 enabling the surrogate model block formulations to be harnessed with \texttt{OODXBlock}. For the GP-based adaptive sampling formulation, GPR predictive and uncertainty proxy formulations were utilised within the adaptive sampling formulations. Additionally, the adaptive sampling formulations are flexible, unconstrained, foundational models which enable classification surrogate formulations to be plugged-in to enable adaptive sampling with online feasibility constraints. Upon addition of such classification-based constraints, the foundational adaptive sampling NLP and MILP formulations can become mixed integer nonlinear programming (MINLP) formulations depending on the classification model formulations used. Finally, by implementing the adaptive sampling formulations using Pyomo, it is possible to use any appropriate compatible solver to find local or global solutions as required.

\subsubsection{Gaussian process-based adaptive sampling}

The OODX \texttt{AdaptiveSampler} provides a Pyomo formulation for adaptive sampling to maximise the uncertainty of a GPR surrogate model. Specifically, the uncertainty proxy, $\upsilon^{(\text{proxy})}$, of a trained instance of a \texttt{GPR} model is minimised by using the \texttt{OODXBlock} formulation from Table \ref{tab:gprump} in combination with the objective function in Equation \ref{eq:gpuas} to provide a complete NLP problem which can be solved for adaptive samples.

\begin{equation}\label{eq:gpuas}
    \min_{\textbf{x}} \quad \upsilon^{(\text{proxy})}(\textbf{x})\\
\end{equation}

The OODX \texttt{AdaptiveSampler} provides a Pyomo formulation to maximise the modified EI for explorative and exploitative adaptive sampling of GPR models. Specifically, the objective function in Equation \ref{eq:meias} was used in combination with \texttt{OODXBlock} formulations for GPR predictions y(\textbf{x}) (Table \ref{tab:gprmp}) and uncertainty proxy $\upsilon^{(\text{proxy})}(\textbf{x})$ (Table \ref{tab:gprump}) to complete an NLP problem which can be solved for adaptive samples.

\begin{equation}\label{eq:meias}
    \max_{\textbf{x}} \quad \left(\frac{\sigma_f^2 - \upsilon^{(\text{proxy})}(\textbf{x})}{2\pi}\right)^{1/2} \exp\left( - \frac{\left(\hat{y}(\textbf{x}) - y^{(\text{max})} - \xi\right)^2}{2\left( \sigma_f^2 - \upsilon^{(\text{proxy})}(\textbf{x}) \right)} \right)\\
\end{equation}

\subsubsection{Heuristic-based adaptive sampling}

The OODX \texttt{AdaptiveSampler} object contains exploration and exploitation Pyomo-based MILP formulations incorporating heuristic-based adaptive sampling using Delaunay triangulation implementation within SciPy version 1.8.0 \cite{2020SciPy-NMeth}. The MILP formulation for the unconstrained heuristic-based adaptive sampling is presented in Table \ref{tab:as3}. These adaptive sampling formulations are refereed to as heuristic-based due to the rule-based decision-making of subsequent samples as opposed to the statistics informed adaptive sampling enabled by GP surrogate models, i.e. rigorous maximisation of GPR uncertainty (Equation \ref{eq:gpuas}) or rigorous maximisation of the GP-based modified EI function (Equation \ref{eq:meias}). Specifically, the heuristic-based adaptive sampling MILP chooses the largest region within the input space partitioned by Delaunay triangulation. This is implemented by performing Delaunay triangulation between samples in the input space, calculating the size of each region, assigning a binary variable to each distinct region, and selecting the binary variable corresponding to the region with the largest size. An adaptive sample is subsequently placed at the centroid of the largest simplex, thereby exploring the most sparsely sampled region. The benefit of such heuristic-based adaptive sampling is quicker iterations involving solving MILP formulations as opposed to complex surrogate-based NLP problems. The formulation in Table \ref{tab:as3} provides a heuristic-based MILP formulation for efficient \emph{exploration} of the search space using the method described. 

\begin{table}[htb]
    \centering
    \caption[Heuristic-based adaptive sampling formulation]{Heuristic-based adaptive sampling MILP formulation using regions from Delaunay triangulation of the search space.}
    \begin{tabular}{l l}
        \hline
        \multicolumn{2}{l}{Sets} \\
        \hline
        $R$ & set of regions \\
        $m$ & set over the number of input dimensions \\
        \hline
        \multicolumn{2}{l}{Parameters} \\
        \hline
        $S_i$ & sizes of regions with dimensions $i \in R$ \\
        $C_{i,j}$ & centroid coordinates with dimensions $i \in R$, $j \in m$ \\
        \hline
        \multicolumn{2}{l}{Variables} \\
        \hline
        $x_j$ & decision variable with dimensions $j \in m$ \\
        $z_i$ & binary variable with dimensions $i \in R$ \\
        \hline
        \multicolumn{2}{l}{Equations} \\
        \hline
        \multicolumn{2}{l}{
            \textbf{for} $j$ \textbf{in} $m$ \textbf{do}
        }\\
        \multicolumn{2}{l}{
            \quad $x_j = \sum\limits_{i\in R} C_{i, j} z_i$
        } possible $\textbf{x}$ values bound to centroids \\
        \\
        \multicolumn{2}{l}{
            $\sum\limits_{i \in R} z_{i} = 1$
        } exactly one region selected \\
        \\
        \hline
        \multicolumn{2}{l}{Objective} \\
        \hline
        \multicolumn{2}{l}{
            $\max$ \quad $\sum\limits_{i\in R} S_i z_i$
        } maximise size \\
        \hline
    \end{tabular}
    \label{tab:as3}
\end{table}

The formulation in Table \ref{tab:as3} can be adapted by embedding \texttt{OODXBlock} formulations for classification models and plugging in feasibility constraints into the adaptive sampling MILP. In doing so, depending on whether the classifier formulation is MILP or NLP, the resulting adaptive sampling formulation with be MILP or MINLP, respectively. It is also possible to adjust the formulation in Table \ref{tab:as3} to enable heuristic-based \emph{exploitation} of the current most promising sample with regards to surrogate-based optimisation. This Delaunay triangulation-based exploitation adaptive sampling MILP was inspired by the GP-based EI function. The exploitation adaptive sampling with Delaunay triangulation thereby includes the influence of the current best sample. Accordingly, the formulation in Table \ref{tab:as3} can be updated to consider only the regions for which the current best sample is a vertex. In this way, the set $R$ as well as the $S$ and $C$ parameters are configured differently along with the binary selector variable vector, $\textbf{z}$. Otherwise, the MILP formulation remains the same, although now the optimisation will determine the largest region about the current best sample.

\subsubsection{Bounds adjustments}

The OODX \texttt{AdaptiveSampler} object can be used to adjust the bounds of the adaptive sampling search space to enable refinement around potential optima and/or feasible regions. Practically, the bounds on the input variables to the GP-based NLP formulations are enforced such that adaptive sampling solutions are constrained within the specified bounds. With regards to the Delaunay triangulation-based MILP formulations, the set of centroids which constrains the input variables to a discrete set of locations, is updated to only include centroids within the specified bounds. Additionally, the vertices of the search space hypercube can be included within the Delaunay triangulation. The \texttt{AdaptiveSampler} object therefore enables box bounds to be adjusted within bounds tightening algorithmic implementations whilst enabling the bounds tightening rules to be tailored to specific applications. In this way, it is possible to enable bounds tightening, relaxation, and/or translation in any direction and dimension as is necessitated for the desired exploitation/exploration of the search space.

\section{Application: resource recovery from brewery wastewater}

\subsection{Case study introduction}
% pasted from chapter 5 intro
This section presents a an application of the developed methodology to the optimisation of a process system to recover resources from a brewery wastewater. Brewery wastewater was chosen due to the high concentrations of chemical oxygen demand (COD), total nitrogen (TN), and total phosphorous (TP), typical among food and beverage processing industries which have been highlighted as a focus for sustainable resource recovery \cite{simate_treatment_2011, asgharnejad_comprehensive_2021}. Typical brewery wastewater characteristics are shown in Table \ref{tab:bww} \cite{rao_ph_2007}.

\begin{table}[htb]
    \centering
    \caption[Typical brewery wastewater characteristics]{Typical brewery wastewater characteristics \cite{rao_ph_2007}.}
    \begin{tabular}{l l}
        \hline
        Parameter & Value \\
        \hline
        pH & 3--12 \\
        Temperature ($^\circ$C) & 14--18 \\
        COD (g\,m$^{-3}$) & 2,000--6,000 \\
        BOD (g\,m$^{-3}$) & 1,200--3,600 \\
        COD:BOD & 1.667 \\
        VFA (g\,m$^{-3}$) & 1,000--2,500 \\
        Phosphates as PO$_4$ (g\,m$^{-3}$) & 10--50 \\
        TN (g\,m$^{-3}$) & 25--80 \\
        TS (g\,m$^{-3}$) & 5,100--8,750 \\
        TSS (g\,m$^{-3}$) & 2,900--3,000 \\
        TDS (g\,m$^{-3}$) & 2,020--5,940 \\
        \hline
    \end{tabular}
    \label{tab:bww}
\end{table}

A simulation-based superstructure optimisation methodology was thereby developed to optimise process systems to recover carbon, nitrogen (N), and phosphorous (P) resources from a representative brewery wastewater. The resource recovery superstructure was modelled in a state of the art wastewater treatment process simulation software. This enabled computer experiments to provide high-fidelity training data for surrogate models \cite{bradley_perspectives_2022}. The superstructure itself embedded high-rate anaerobic digestion for the recovery for biogas as well as biological nutrient (N and P) recovery (BNR) processes.

BNR has gained popularity over physico-chemical processes due to the realised performance at reduced costs arising from the non-requirement for additional dosing chemicals and subsequent separation of precipitates \cite{hasan_recent_2021}. However, challenges associated with BNR include modelling and controlling biological systems particularly in uncertain practical environments and colder climates \cite{hasan_recent_2021}. BNR enables the concentration of N and P nutrients into a solid digestate by-product which can be utilised as a fertiliser \cite{wrap_compost_2016}. Additionally, the recovery of nutrient-rich fertiliser from wastewater enables the substitution of traditional N- and P-fixating production processes thereby alleviating the related environmental issues associated with human interference in the natural biogeochemical cycles. Furthermore, the removal of nutrients from wastewater effluents reduces the emissions to receiving water bodies thereby alleviating the associated impacts of water toxicity and eutrophication \cite{puchongkawarin_optimization-based_2015}.

\begin{figure}[htb]
\centering
    \includegraphics[width=0.8\textwidth]{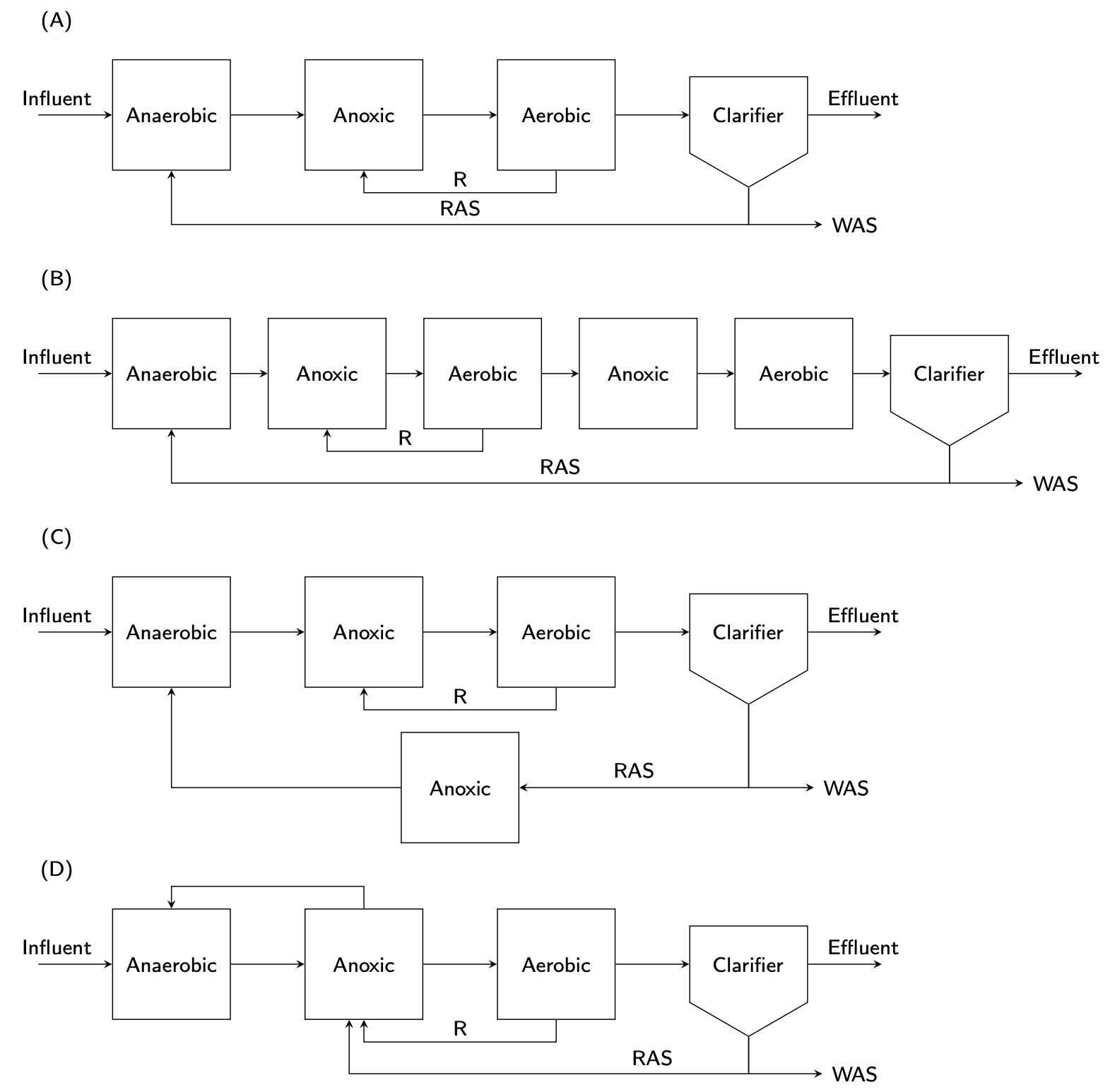}
    \caption[]{Biological nutrient recovery processes. (A) A$^2$O process, (B) Bardenpho process, (C) Johannesburg process, (D) UCT process. R: recycle from aerobic to anoxic zone, RAS: return activated sludge, WAS: waste activated sludge.}
    \label{fig:bnr}
\end{figure}

BNR process configurations include the anaerobic-anoxic-oxic (A$^2$O) process, the Bardenpho process, the Johannesburg process, and the University of Cape Town (UCT) process. The A$^2$O process (Figure \ref{fig:bnr}A) facilitates both biological N and P recovery by releasing ammonia within an anaerobic zone whilst simultaneously facilitating polyhydroxyalkanoate (PHA) storage by phosphorous accumulating organisms (PAOs). Following the anaerobic zone, an anoxic zone reduces nitrates, recycled from a final aerobic zone, to nitrogen. The aerobic zone simultaneously oxidises nitrogenous compounds to nitrates and facilitates the uptake of phosphates by PAOs. The Bardenpho process (Figure \ref{fig:bnr}B) operates similarly to the A$^2$O process with an additional anoxic-aerobic step between the original aerobic stage and the final clarifier. The Bardenpho process thereby provides superior nutrient recovery compared to the A$^2$O process at the expense of additional capital costs for additional reactors and increased operating costs for the second aeration tank. The Johannesburg process (Figure \ref{fig:bnr}C) is equivalent to the A$^2$O process with an additional fermenter on the return activated sludge (RAS) from the clarifier to the anaerobic stage. This facilitates strict anaerobic conditions within the first stage by enabling denitrification of any nitrates in the RAS. Additionally, increased volatile fatty acid (VFA) production in the side stream enables enhanced PHA storage by PAOs in the anaerobic stage. The UCT process (Figure \ref{fig:bnr}D) adopts a unique recycle configuration with the RAS recycled to the anoxic stage and with an additional recycle from the anoxic stage to the anaerobic stage. There also exists a modified UCT version which utilises a multistage anoxic reactor configuration to improve the denitrification capacity by recycling nitrates to latter stages of the anoxic zones whilst the mixed liquor recycle is withdrawn from earlier stages of the anoxic zone \cite{demir_comparison_2021}. Finally, the literature contains some works focusing on integrative BNR configurations, for example: integrating high-rate anaerobic digestion within the Bardenpho process \cite{li_uasb-modified_2020}; combining the A$^2$O process with chemical P precipitation \cite{tomei_holistic_2020}; and the optimisation of the Bardenpho process and moving bed biofilm reactor using response surface methodology \cite{ashrafi_optimising_2019}.

Surrogate models were formulated to represent different BNR process configurations within the simulated superstructure. Specifically, GPR models and classification NNs were modelled and formulated as mathematical programming formulations using the OODX methodology presented \cite{atdurkin-gh}. However, the application of the OODX methodology to the superstructure optimisation of resource recovery process necessitated several challenges to be addressed. Firstly, this chapter builds on the previous one by formulating the surrogate models within a superstructure optimisation problem including binary variables for the selection of discrete resource recovery pathways within the superstructure. As such, this work develops a bb-MINLP framework \cite{kim_surrogate-based_2020}. In addition to the challenges associated with formulating and solving bb-MINLP problems, this chapter also consider the competing decision criteria pertaining to the recovery of different resources and economic cost. This challenge was addressed by multi-objective optimisation to highlight different non-dominated solutions on the Pareto frontier.

Finally, optimisation under uncertainty methods were used to optimise resource recovery process design whilst accounting for uncertainties in the brewery wastewater composition. Specifically, the flow and composition of wastewaters are inherently stochastic, with loading shocks potentially leading to reduced process performance in the best case, and process infeasibility in the worst case. As for the objective function, economic and environmental performance criteria are often the focus when designing systems to recover resources from wastewater, but process operability to ensure that the designed performance is realisable within practical operating environments is of equal importance. 

To address the challenges in this application, superstructure optimisation methodology embedding surrogate process unit models trained and formulated using the OODX package was developed and implemented to optimise the recovery of carbon and nutrient resources from a brewery wastewater. The well-established commercial wastewater treatment process simulation software GPS-X version 8.0 was utilised as a source of high-fidelity data from first-principles black box models using design of experiments \cite{bradley_perspectives_2022}. These computer experiments provided high-fidelity data of resource recovery pathways from a superstructure embedding high-rate anaerobic digestion (AD) via an upflow anaerobic sludge blanket (UASB) reactor, 4 different BNR pathways, and side-stream AD including return flows. Binary selection variables enabled the optimisation of resource recovery pathway whilst the volume of the UASB was optimised as a continuous process design variable. Additionally, the uncertainty in the brewery wastewater COD was accounted for by varying this parameter during the computer experiments between the lower and upper brewery wastewater COD compositions cited in the literature (2,000--6,000 g\,m$^{-3}$).

\subsection{Methods}

\subsubsection{Wastewater characterisation}

The characterisation of the brewery effluent was in accordance with the typical brewery wastewater characteristics from Table \ref{tab:bww} and is shown in Table \ref{tab:thisbww}. A wastewater production of 1,000\,m$^{3}\,d^{-1}$ was considered for demonstrative purposes, although it should be noted that typical brewery effluent loads vary greatly based on the size of the brewery and its production capacity. As such, 1,000\,m$^{3}$\,d$^{-1}$ is an approximating overestimate of typical brewery wastewater production to demonstrate the modelling capabilities and it should be noted that the methodology could also be applied to other wastewater compositions and different production capacities.

\begin{table}[htb]
    \centering
    \caption[Brewery wastewater characterisation]{Brewery wastewater characterisation implemented in this study.}
    \begin{tabular}{l l}
        \hline
        Parameter & Value \\
        \hline
        Flow (m$^3$\,d$^{-1}$) & 1,000 \\
        COD (g$\,$m$^{-3}$) & 4,000 \\
        BOD (g$\,$m$^{-3}$) & 2,400 \\
        COD:BOD & 1.66 \\
        VFA (g$\,$m$^{-3}$) & 2,133 \\
        TN (g$\,$m$^{-3}$) & 80 \\
        Ammonia nitrogen (g$\,$m$^{-3}$) & 50 \\
        TP (g$\,$m$^{-3}$) & 30 \\
        Ortho-phosphate (g$\,$m$^{-3}$) & 20 \\
        \hline
    \end{tabular}
    \label{tab:thisbww}
\end{table}

The wastewater characterisation also accounted for uncertainties in the COD composition. Specifically, for the optimisation under uncertainty case study, the COD was considered to vary between the lower and upper bounds cited in literature (2,000--6,000 g\,m$^{-3}$) \cite{simate_treatment_2011}. The influent COD was thereby modelled as following a normal distribution with a mean of 4,000 g\,m$^{-3}$ and a standard deviation of 1,000 g\,m$^{-3}$. This probability distribution thereby ensures that 95\,\% of the inlfuent COD realisations fall within 2 standard deviations of the mean ($4,000 \pm 2,000$ g\,m$^{-3}$).

% \begin{figure}[htb]
% \centering
%     \begin{tikzpicture}
%         \node[rounded corners, draw=black, align=center, text width=4cm, minimum height=3cm]{Placeholder};
%     \end{tikzpicture}
%     \caption[]{Caption.}
%     \label{fig:pdf}
% \end{figure}

\subsubsection{Superstructure postulation}
\begin{figure}[htb]
\centering
    \includegraphics[width=0.8\textwidth]{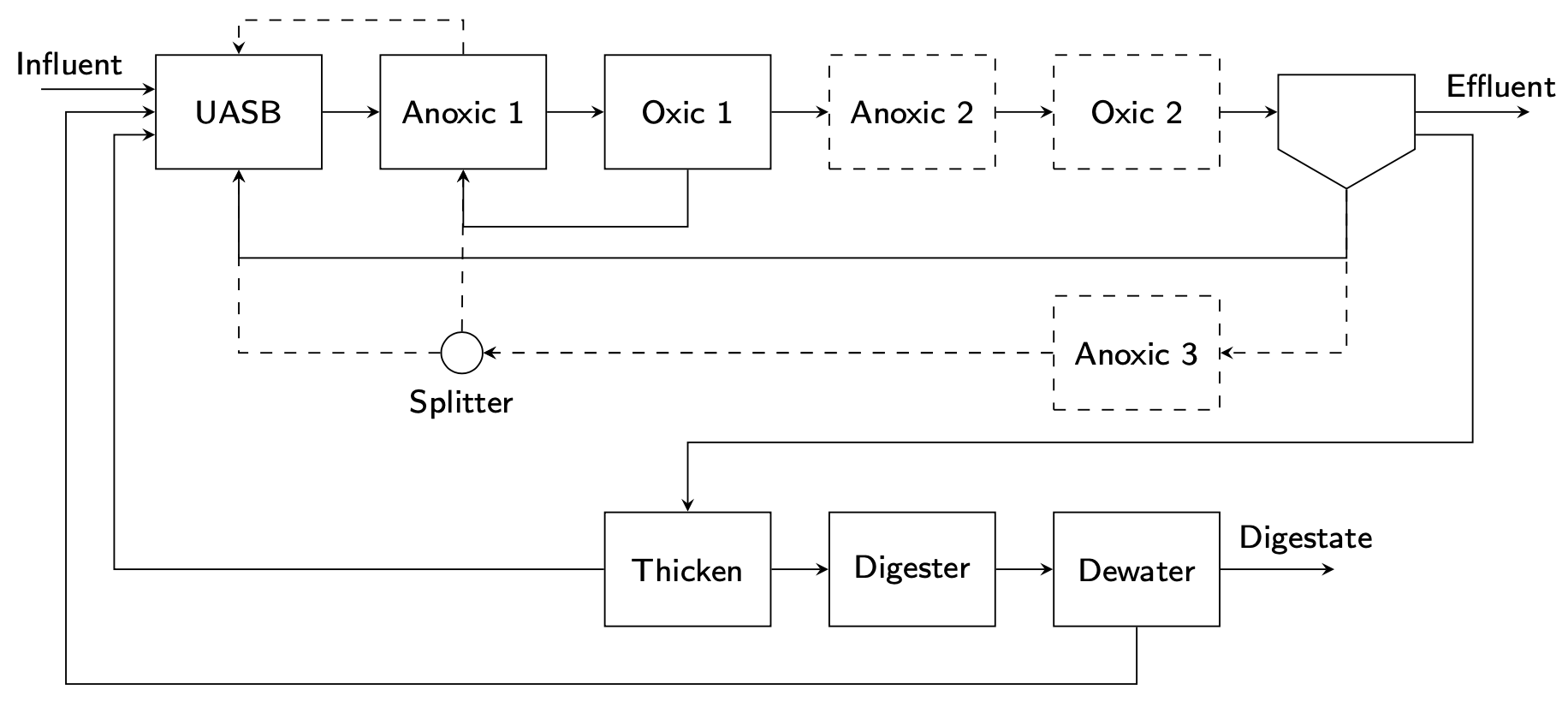}
    \caption[]{Carbon, nitrogen, phosphorous recovery superstructure embedding A$^2$O, Bardenpho, Johannesburg, and UCT biological nutrient recovery (BNR) processes. Solid lines depict reactors and streams present in all BNR configurations whilst dashed lines show reactors and steams which are only activate for Bardenpho, Johannesburg, or UCT process dependent on the configurations detailed in Table \ref{tab:configs}. UASB: upflow anaerobic sludge blanket reactor.}
    \label{fig:ss1}
\end{figure}

A superstructure comprising biological resource recovery pathways was postulated (Figure \ref{fig:ss1}) and modelled within the GPS-X wastewater treatment process simulation software. Specifically, the UASB serves the double purpose as an anaerobic zone within BNR process configurations whilst also facilitating high-rate anaerobic digestion to recover biogas. The superstructure then embeds 4 different BNR configurations including A$^2$O, Bardenpho, Johannesburg, and UCT processes. The waste activated sludge from a secondary clarification unit is then sent to a conventional anaerobic digestion process, including prior thickening and post dewatering, from which the digestate is recovered as a rich source of N and P nutrients.

The parameterisation to model each BNR pathway within the superstructure is given in Table \ref{tab:configs}. Specifically, the A$^2$O process is modelled by constraining the volumes of the second anoxic and aerobic reactors (Anoxic 2 and Oxic 2, respectively) to zero. Additionally, the third anoxic reactor (Anoxic 3) is constrained to have zero volume and the return activated sludge is recycled to the UASB. Finally, the recycle from the first anoxic reactor (Anoxic 1) to the UASB is not utilised. The Bardenpho process is the same as the A$^2$O process configuration except that the Bardenpho process utilises Anoxic 2 and Oxic 2. The difference between the A$^2$O configuration and the Johannesburg process is the activation of Anoxic 3 within the RAS recycle. The UCT process is the same as the A$^2$O process except that the RAS is recycled to Anoxic 1 instead of directly the UASB and there is a recycle from Anoxic 1 to the UASB.

\begin{table}[htb]
    \centering
    \caption[BNR configuration parameterisation]{BNR configuration parameterisation within the superstructure (Figure \ref{fig:ss1}).}
    \begin{tabular}{l c c c c}
        \hline
        Parameter & A$^2$O & Bardenpho & Johannesburg & UCT \\
        \hline
        Anoxic 1 volume (m$^3$) & 400 & 400 & 400 & 400 \\
        Oxic 1 volume (m$^3$) & 1,000 & 1,000 & 1,000 & 1,000 \\
        Anoxic 2 volume (m$^3$) & 0 & 400 & 0 & 0 \\
        Oxic 2 volume (m$^3$) & 0 & 200 & 0 & 0 \\
        Clarifier volume (m$^3$) & 600 & 600 & 600 & 600 \\
        Anoxic 3 volume (m$^3$) & 0 & 0 & 200 & 0 \\
        Splitter fraction to UASB & 1 & 1 & 1 & 0 \\
        Anoxic 1 to UASB recycle (m$^3$\,d$^{-1}$) & 0 & 0 & 0 & 4,000 \\
        Thickener underflow (m$^3$\,d$^{-1}$) & 6 & 6 & 6 & 6 \\
        Digester volume (m$^3$) & 300 & 300 & 300 & 300 \\
        Dewatered digestate flow (m$^3$\,d$^{-1}$) & 1.5 & 1.5 & 1.5 & 1.5 \\
        \hline
    \end{tabular}
    \label{tab:configs}
\end{table}

\subsubsection{Computer experiments}

Computer experiments enabled high-fidelity data from GPS-X to be harnessed for surrogate modelling training and to represent underlying black box models within the superstructure optimisation framework. GPS-X version 8.0 was interfaced using Python scripts to iterate over a set of input samples, generated using static sampling strategies, and evaluate and save the corresponding output data. The computer experiments were designed by specifying the input-output variables to be sampled, the lower, upper bounds on the input space, and generating a set of well-spaced static input samples.

The volume of the UASB was selected as a continuous input variable due to its influence on COD conversion to recovered biogas as well as the dependence of downstream BNR processes on the anaerobic zone. This critical design parameter therefore has an influence on COD, N, and P recovery providing a good candidate for multi-objective optimisation \cite{li_uasb-modified_2020}. Lower and upper bounds of 50\,m$^3$ and 500\,m$^3$ respectively, were imposed on the UASB volume based on preliminary experiments and  to provide a reasonable range for the optimisation search.

The other input variable to computer experiments was the concentration of COD in the brewery wastewater, thereby defining a 2-dimensional search space with the variable representing the volume of the UASB. The influent COD was varied in the computer experiment so as to gain an understanding of the process system performance in response to uncertain variations in wastewater quality within practical operating environments. The lower and upper bounds of 2,000 g\,m$^{-3}$ and 6,000 g\,m$^{-3}$ were imposed on the influent COD in accordance with the typical range cited for brewery wastewaters \cite{simate_treatment_2011}.

The output variables sampled from computer experiments included the final effluent COD, TN, and TP concentrations which were used to ensure effluent quality constraints were met. Biogas production, specifically the volumetric flow of methane produced, from both the UASB and the digester were sampled to assess the total biogas recovery performance of the flowsheet. The mass of the recovered digestate was sampled along with TN and TP composition data for the digestate so as to determine the N and P quality in kg\,t$^{-1}$. The aeration requirements of the 2 aerobic zones were evaluated to determine the total aeration requirement of the process as a proxy for operational cost of which the aeration constitutes a primary factor. Finally, binary data signifying convergence of the simulator (failed convergences were assigned a label of 0 whilst successful convergences were labelled 1) was sampled for modelling feasibility constraints using classification surrogate models. 

The input and output variables are summarised in Table \ref{tab:vars and bounds} along with their corresponding  indexes used in the mathematical notation for surrogate model and mathematical programming formulations. Specifically, the variables were assigned incremental integer indexes with indexes $u=\{1, 2\}$ assigned to the input variables and indexes $v=\{3, 4, ..., 9\}$ assigned to the output variables.

\begin{table}[htb]
    \centering
    \caption[Variable selection and bounds]{Input and output variable selection and input space bounds.}
    \begin{tabular}{l l l}
        \hline
        \multicolumn{3}{c}{Input variables} \\
        \hline
        $u$ & Variable & Bounds \\
        \hline 
        1 & UASB volume & $[50, 500]$ \\
        2 & Influent COD & $[2000, 6000]$ \\
        \hline
        \multicolumn{3}{c}{Output variables} \\
        \hline
        $v$ & \multicolumn{2}{l}{Variable}\\
        \hline
        3 & \multicolumn{2}{l}{Effluent COD}\\
        4 & \multicolumn{2}{l}{Effluent TN}\\
        5 & \multicolumn{2}{l}{Effluent TP}\\
        6 & \multicolumn{2}{l}{Biogas recovered}\\
        7 & \multicolumn{2}{l}{Digestate nutrient quality}\\
        8 & \multicolumn{2}{l}{Aeration cost}\\
        9 & \multicolumn{2}{l}{Convergence}\\
        \hline
    \end{tabular}
    \label{tab:vars and bounds}
\end{table}

Sobol sampling was adopted to provide good initial search space coverage in a small number of samples \cite{sobol_global_2001}. Sampling for black box superstructure optimisation necessitates a strategy to sample uniformly over the discrete variables as well as the continuous search space \cite{kim_surrogate-based_2020}. This was achieved by generating 32 samples over the 2-dimensional input space and then evaluating these 32 samples for each of the 4 discrete BNR configurations within the superstructure resulting in 128 total simulation interrogations.

\begin{figure}[htb]
\centering
    \includegraphics[width=0.8\textwidth]{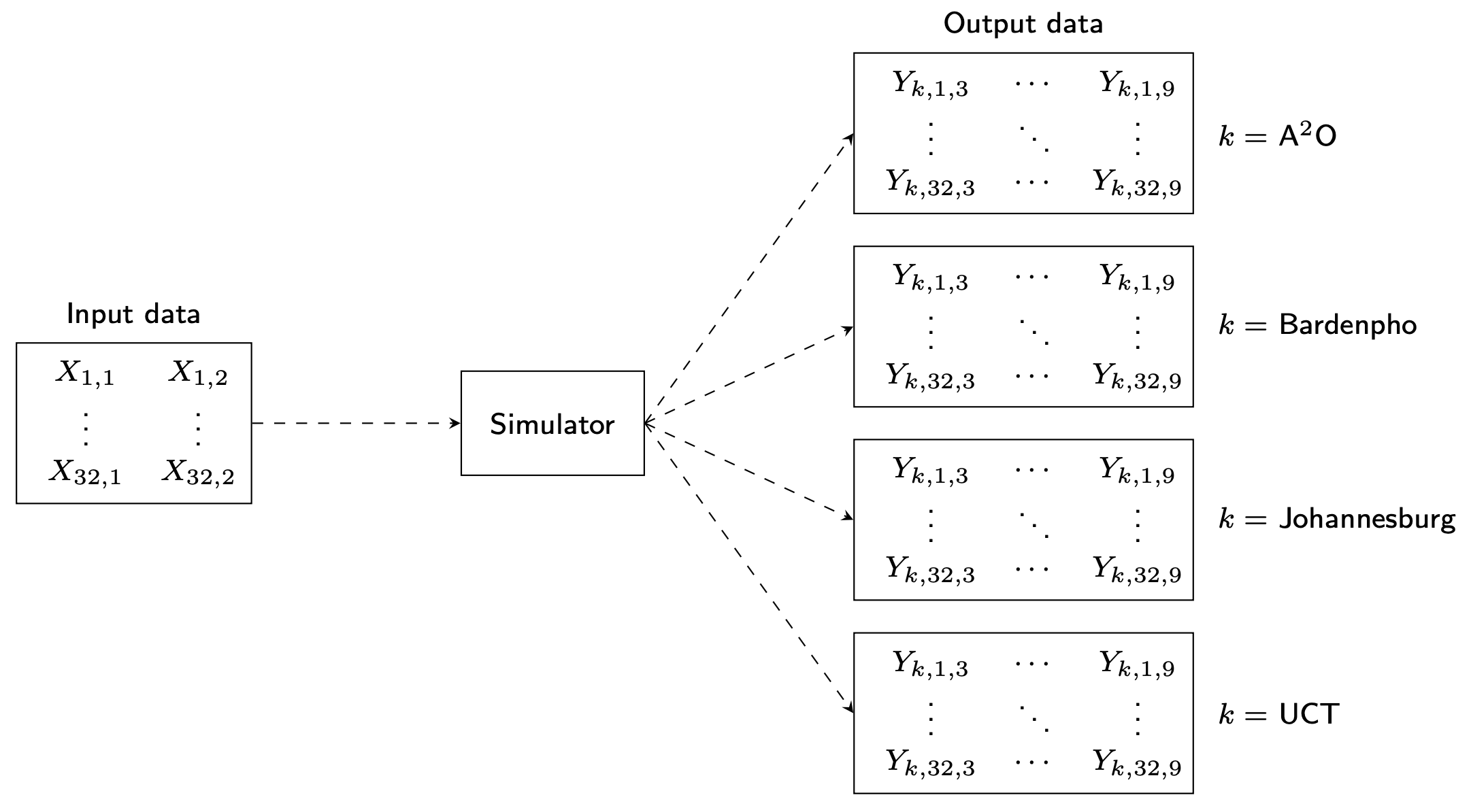}
    \caption[]{Sampling input-output mappings for discrete BNR configurations.}
    \label{fig:sampling1}
\end{figure}

The Sobol sampling thereby obtains the mappings of input data onto output data as shown by Figure \ref{fig:sampling1} where the following notation was used:

\begin{itemize}
    \item $k$ indexes the set of BNR configurations \{A$^2$O, Bardenpho, Johannesburg, UCT\}
    \item $n$ parameter denoting the number of samples (equal to 32 here)
    \item $u$ denotes the input variable index \{1, 2\}
    \item $v$ denotes the output variable index \{3, 4, ..., 9\}
    \item $i$ indexes the set of samples \{1, 2, ..., n\}
    \item $X_{i, u}$ matrix of input samples
    \item $Y_{k, i, v}$ matrix of output samples for each BNR configuration
    % \item $\mu_u$, $\mu_v$ mean input and output values
    % \item $\sigma_u$, $\sigma_v$ standard deviation of inputs and outputs
\end{itemize}

\subsubsection{Surrogate modelling}

Prior to training the surrogate models, the input-output data were standardised and split into training and testing data. Standardisation of the data ensured that the surrogate models were not overfit to variables with greater magnitude. The training data was subsequently used to train the surrogate models whilst the testing data was reserved for validating the model performance. The mean absolute error was used as a validation metric. In this work, 25\,\% of the data was reserved for testing (equating to 8 samples per BNR configuration) and the remaining 75\,\% (24 samples) were used for model training. The same train-test-split was used for each discrete BNR configuration so as to facilitate an equal comparison due to the same uncertainties introduced due to the train-test-split. Regression models were fitted to converged training data only so as not to skew the predictions over the converged, feasible region.

GPR surrogate models with squared-exponential kernel functions were formulated to map the 2 continuous input variables onto each continuous output variable for each discrete BNR configuration. In total, 6 continuous output variables (including effluent COD, TN, and TP concentrations, total biogas recovery, nutrient quality of the digestate, and total aeration requirement) and 4 BNR configurations thereby necessitated 24 GPR models. Predictions from GPR models can thereby be written as $\hat{y}_{k,v} = \text{GPR}_{k,v}\left(\textbf{x}\right)$ where $k$ is the BNR configuration and $v=3, ..., 8$ corresponding to the 6 continuous output variables indices shown in Table \ref{tab:vars and bounds} and \textbf{x} is a 2-dimensional input vector. GPR models were validated based on the mean absolute error (MAE) and mean absolute percentage error (MAPE).

\begin{figure}[htb]
    \centering
    \includegraphics[width=0.8\textwidth]{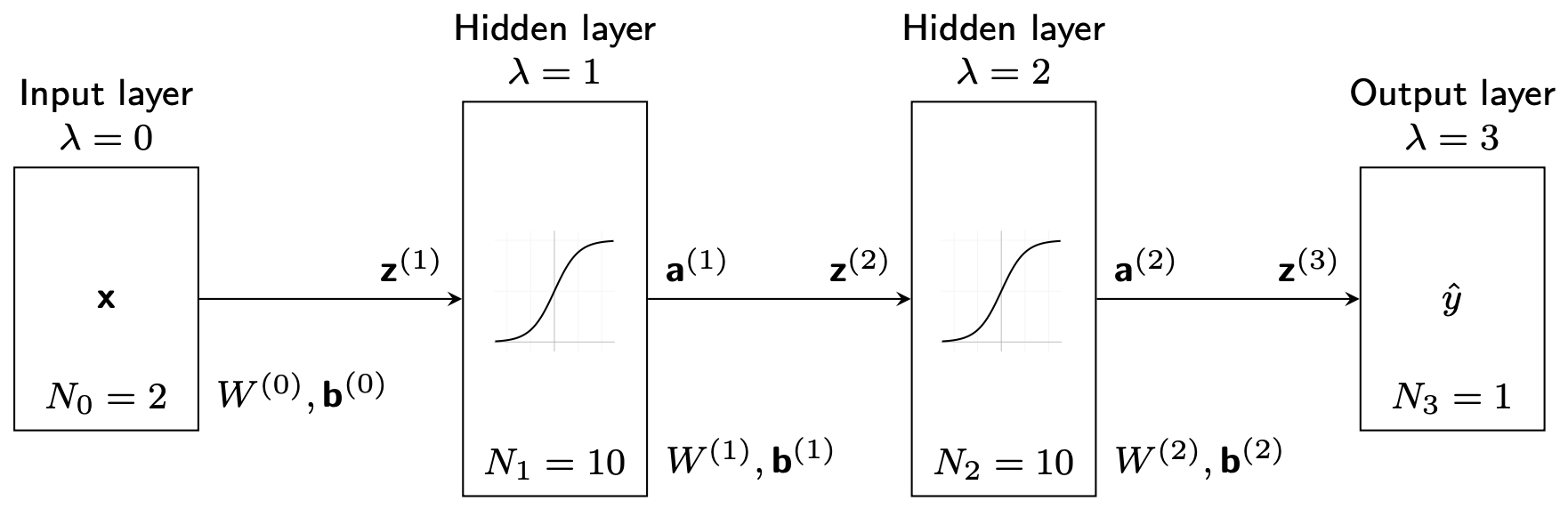}
    \caption[]{Feedforward neural network structure to map inputs $\textbf{x}$ onto predictions $\hat{\textbf{y}}$. The network consists of an input layer with $N_0$ nodes, an output layer with $N_{\Lambda+1}$ nodes, and $\Lambda$ hidden layers with $N_{\lambda}$ nodes for $\lambda=1,..., \Lambda$. Each hidden layer has an activation function $\xi^{(\lambda)}$ for $\lambda=1,..., \Lambda$. Layer inputs $\textbf{z}^{(\lambda)}$ are calculated by multiplying the previous layer outputs $\textbf{a}_{\lambda-1}$ by a weights matrix $W^{(\lambda-1)}$ and added to a bias vector $\textbf{b}^{(\lambda-1)}$ Finally, layer outputs are determined by passing the inputs through an activation function $\xi^{(\lambda)}$.}
    \label{fig:nnapp}
\end{figure}

Classification NNs with 2 hidden layers, each with 10 nodes and sigmoid activation functions (Figure \ref{fig:nnapp}), were trained on the relationships between the 2 continuous input variables and binary convergence targets (output variable assigned index 9 in Table \ref{tab:vars and bounds}). Trained for each discrete BNR pathway, there were a total of 4 classification NNs within this decision-making methodology. To maintain the notation, the predictions from the classification models can be written as $\hat{y}_{k,9}=\text{NN}_k\left(\textbf{x}\right)$ where $k$ is the BNR configuration and the 9 index corresponds to the convergence output mapping in Table \ref{tab:vars and bounds} and \textbf{x} is a 2-dimensional input vector. The classification NNs were validated based on the precision and recall scores which represent the ability of the classifier to avoid false positives and to correctly label all the positives, respectively.

\subsubsection{Optimisation formulation}

The trained GPR models were formulated within the mathematical optimisation problem using the NLP constraint shown in Equation \ref{eq:mapping}. In total there were 24 GPR constraints $\hat{\bar{y}}_{k,v}$ for each BNR configuration (A$^2$O, Bardenpho, Johannesburg, UCT), $k$ and each continuous output variable $v$. The predictions of these GPR constraints exist in the standardised modelling space thereby necessitating inverse scaling back to the original space. The GPR parameters which were optimised in the surrogate modelling stage ($\alpha_i$, $\sigma_f$, $l$) as well as the standardised training data, $\bar{X}_{i,u}$ (where $i$ indexes the number of training samples $n$ and $u$ indexes the dimensionality of the input vector $m=2$), were implemented as optimisation parameters for each BNR configuration $k$ and continuous output variable $v$. Inputs to the GPR NLP constraints were the optimisation decision variable, $\bar{x}_1$ and $\bar{x}_2$, which were bounded by the standardised lower, upper bounds of the UASB volume and influent COD, respectively.

\begin{equation}\label{eq:mapping}
    \hat{\bar{y}}_{k,v} ={} \left.\sum\limits_{i \in n} \alpha_i \sigma_f^2 \exp\left( -\sum\limits_{u \in m} \frac{1}{2 l^2} \left( \bar{x}_u - \bar{X}_{i,u} \right)^2 \right)\right|_{k,v}
\end{equation}

The modelling outputs from each GPR constraint were scaled back to original space by multiplying by the relevant standard deviation, $\sigma_v$, and adding the relevant mean, $\mu_v$, saved during the data standardisation which was facilitated by using the OODX data handling object. The inverse scaled surrogate modelling outputs were then multiplied by a binary variable, $\gamma_k$, used to activate/deactivate each BNR pathway within the superstructure optimisation model. Coupled with a ``choose exactly one" constraint which constrains the sum of these binary variables to equal 1 (Equation \ref{eq:choose1}), the summation over the product between these binary selection variables and the inverse scaled modelling outputs enabled MINLP formulations representing the superstructure. Equations \ref{eq:c1}, \ref{eq:c2}, and \ref{eq:c3} show the resulting constraints to enforce effluent quality constraints on the COD, TN, and TP below limits of 50\,g\,m$^{-3}$, 10\,g\,m$^{-3}$, and 5\,g\,m$^{-3}$, respectively.

\begin{equation}\label{eq:choose1}
    \sum_k \gamma_k = 1
\end{equation}

\begin{equation}\label{eq:c1}
    \sum_{k} \gamma_k \left(\hat{\bar{y}}_{k,3} \sigma_{3} + \mu_{3} \right) \leq 50
\end{equation}

\begin{equation}\label{eq:c2}
    \sum_{k} \gamma_k \left(\hat{\bar{y}}_{k,4} \sigma_{4} + \mu_{4} \right) \leq 10
\end{equation}

\begin{equation}\label{eq:c3}
    \sum_{k} \gamma_k \left(\hat{\bar{y}}_{k,5} \sigma_{5} + \mu_{5} \right) \leq 5
\end{equation}

Constraints representing the classification NNs mapped the scaled optimisation input variables $\bar{x}_u$ onto predicted probabilities of feasibility for each BNR configuration. The NNs with 2 hidden layers, each with 10 nodes and sigmoid activation functions, were formulated as the following set of NLP constraints, defined for each BNR configuration $k$, where $\hat{y}_{k,7}$ is a scalar logit prediction for each BNR configuration (Equations \ref{eq:nnnlp1}--\ref{eq:nnnlpfin}).

\begin{equation}\label{eq:nnnlp1}
    z^{(1)}_j = \sum\limits_{u\in N_0} w^{(0)}_{j,u} \bar{x}_{u} + b^{(0)}_j
\end{equation}

\begin{equation}
    a^{(1)}_j = \frac{1}{1+\exp\left( -z^{(1)}_j \right)}
\end{equation}

\begin{equation}
    z^{(2)}_j = \sum\limits_{j'\in N_{1}} w^{(1)}_{j,j'} a^{(1)}_{j'} + b^{(1)}_j
\end{equation}

\begin{equation}
    a^{(2)}_j = \frac{1}{1+\exp\left( -z^{(2)}_j \right)}
\end{equation}

\begin{equation}\label{eq:nnnlpfin}
    \hat{y}_{k,9} = \sum\limits_{j'\in N_2} w^{(2)}_{j,j'} a^{(2)}_{j'} + b^{(2)}_j
\end{equation}

The classification NNs were similarly embedded within an MINLP constraint including binary selection variables to ensure the the feasibility of the active BNR pathway. Specifically, a probability of feasibility was constrained greater than 0.5 by enforcing the logit prediction greater than or equal to 0.

\begin{equation}\label{eq:feas}
    \sum_k \gamma_k \hat{y}_{k,9} \geq 0
\end{equation}

Objective functions to maximise the recovered biogas, to maximise the recovered N and P nutrients within digestate, and to minimise the aeration requirement as a proxy to total operation cost were optimised individually to observe the trade-offs. These objective functions were of the form shown in Equation \ref{eq:objapp} for $k=6,7,8$. The objective function to maximise nutrient recovery modelled the quality of the recovered digestate in terms of the total nutrient (N and P) content in kg\,t$^{-1}$. The biogas recovery objective function modelled the total methane recovered, in m$^3$\,d$^{-1}$, from both the UASB and the sludge digester. Finally, the aeration requirement was modelled as the total oxygen flow, in m$^3$\,d$^{-1}$, required to reach the desired dissolved oxygen set point in the aerobic zones.

\begin{equation}\label{eq:objapp}
    \min \sum_{k} \gamma_k \left(\hat{\bar{y}}_{k,v}\sigma_{v} + \mu_{v}\right)
\end{equation}

The $\epsilon$-constraint method was also adopted to obtain additional non-dominated Pareto optimal solutions (Equation \ref{eq:epsilonapp}) where $\epsilon_v$ constraints output variable $v$ to be at least as good as a given upper bound.

\begin{equation}\label{eq:epsilonapp}
    \sum_{k}\gamma_k\left(\hat{\bar{y}}_{k,v}\sigma_v + \mu_v\right) \leq \epsilon_v
\end{equation}

Solutions to the MINLP problem thereby highlight optimal process configurations (optimised binary variables) as well as optimal process designs (optimised continuous variables). Initially, the influent COD was treated as an optimisation decision variable so as to demonstrate the multiple decision criteria trade-offs. Subsequently, the influent COD was treated as an uncertain parameter within a stochastic programming framework. All of the MINLP problems were solved using BARON version 22.9.30 \cite{sahinidis_baron_1996} with a maximum time limit of 500\,s using a CPU with 2.8\,GHz Quad-Core Intel Core i7 processor and 16\,GB of RAM.

\subsubsection{Stochastic programming formulation}

To optimise the superstructure configuration under the uncertainty in the influent COD concentration, the constraints and objective functions were reformulated to express the expected values of the relevant surrogate model outputs. For example the effluent quality constraints were reformulated to constrain the expected COD, TN, and TP concentrations below the specified quality limits whilst the expected biogas recovery, expected nutrient quality of the digestate, and expected aeration requirement were specified as the stochastic programming objective functions.

To this end, the influent COD was first redefined as an indexed parameter as opposed to an optimisation decision variable where the probability of each influent COD composition, $R_s$, being realised, $p_s$, was evaluated. Practically, the realised influent COD values were taken as a subset from the standardised sampled data $\bar{X}_2$. This simultaneously enables good sample coverage of the probability distribution and realised parameter values. The relevant constraints and objective functions were then expanded as a weighted sum-product between the probability of each realisation, $p_s$, and the realised surrogate model prediction of dependent variables, $\hat{\bar{y}}_{k,v,s}$. Equations \ref{eq:spcod}, \ref{eq:sptn}, and \ref{eq:sptn} show the stochastic programming reformulations of the effluent quality constraints for the expected COD, TN, and TP concentrations, respectively.

\begin{equation}\label{eq:spcod}
    \frac{1}{\sum p_s}\sum_{k} \gamma_k \left(\sum_s p_s \left(\hat{\bar{y}}_{k,3,s} \sigma_{3} + \mu_{3} \right)\right) \leq 50
\end{equation}
\begin{equation}\label{eq:sptn}
    \frac{1}{\sum p_s}\sum_{k} \gamma_k \left(\sum_s p_s \left(\hat{\bar{y}}_{k,4,s} \sigma_{4} + \mu_{4} \right)\right) \leq 10
\end{equation}
\begin{equation}\label{eq:sptp}
    \frac{1}{\sum p_s}\sum_{k} \gamma_k \left(\sum_s p_s \left(\hat{\bar{y}}_{k,5,s} \sigma_{5} + \mu_{5} \right)\right) \leq 5
\end{equation}

The surrogate models are thereby evaluated at $x_2 = R_s$ where $x_2$ is the input variable representing the influent COD (Table \ref{tab:vars and bounds}) and $R_s$ is the influent COD value realised for scenario $s$ equivalent to the standardised sampled values $\bar{X}_{s,2}$. The GPR constraint reformations are hereby shown by Equation \ref{eq:spgpr}.

\begin{equation}\label{eq:spgpr}
    \hat{\bar{y}}_{k,v,s} ={} \left.\sum\limits_{i \in n} \alpha_i \sigma_f^2 \exp\left( - \frac{1}{2 l^2} \left(\left( \bar{x}_1 - \bar{X}_{i,1} \right)^2 + \left( \bar{X}_{s,2} - \bar{X}_{i,2} \right)^2 \right)\right)\right|_{k,v}
\end{equation}

Similarly, the classification NNs were reformulated such that the second input dimension was parameterised using the sampled data as discrete scenario realisations to yield predictions denoted as $\hat{y}_{k,9,s}$. Whilst, the ``choose exactly one" constraint for the discrete BNR pathways remained unchanged (Equation \ref{eq:choose1}), the classification feasibility constraint was updated to ensure that solutions were robust to the uncertainties by ensuring the process was feasible for all influent COD scenarios (Equation \ref{eq:feas2}).

\begin{equation}\label{eq:feas2}
    \sum_k \gamma_k \hat{y}_{k,9,s} \geq 0
\end{equation}

Finally, the objective functions were updated to minimise the expected values over the possible uncertainties as shown by Equation \ref{eq:objappsto} for $v=6,7,8$.

\begin{equation}\label{eq:objappsto}
    \min \frac{1}{\sum p_s} \sum_{k} \gamma_k \left(\sum_s p_s\left(\hat{\bar{y}}_{k,v,s}\sigma_{v} + \mu_{v}\right)\right)
\end{equation}

\section{Results and discussion}

\subsection{Computer experiment results}

\begin{figure}[htb]
    \centering
    \includegraphics[width=5.5in]{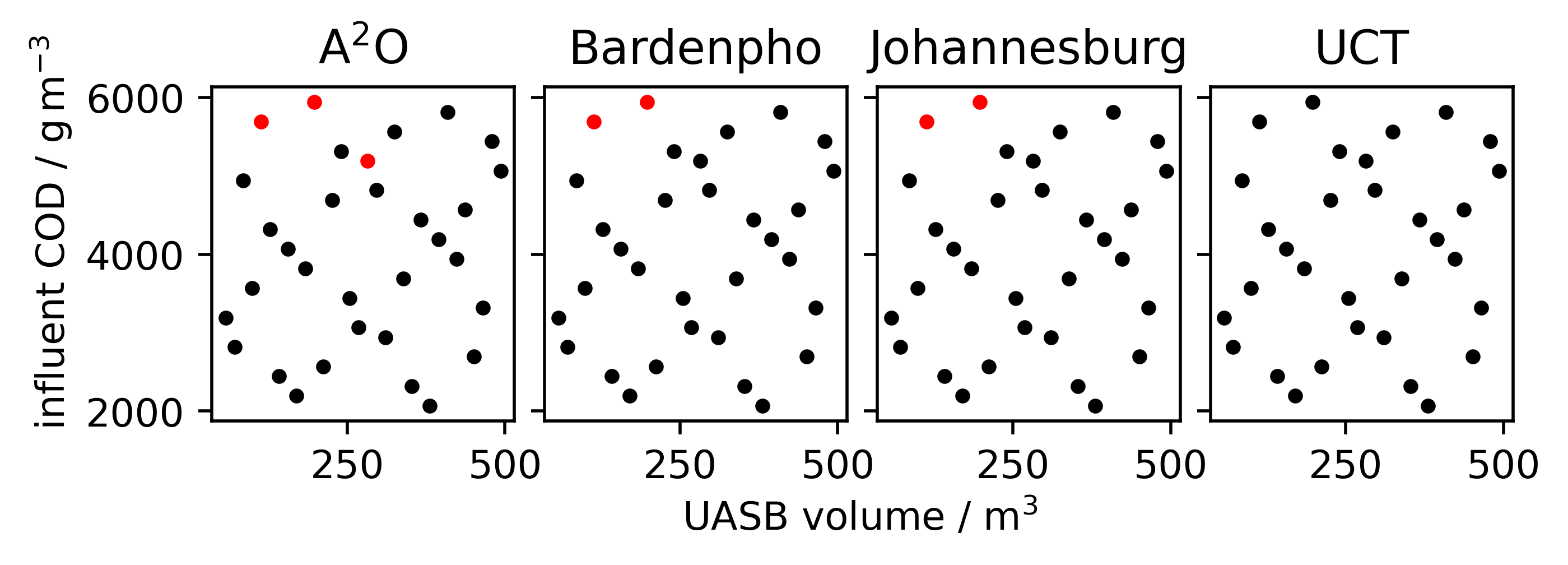}
    \caption[Computer experiments results]{Computer experiments results showing well-spaced Sobol samples within the input space. Converged samples are shown as black whilst non-converged samples are shown as red. Additionally, training samples are shown as filled whilst testing samples are shown as unfilled.}
    \label{fig:cer}
\end{figure}

Figure \ref{fig:cer} shows the samples, which were generated from the Python-GPS-X interface in 374\,s, distributed over the input space and the convergence results. It can be seen that 7 out of the total 128 samples failed to converge (about 5\,\%), primarily located at low UASB volumes at high COD loading for the A$^2$O, Bardenpho, and Johannesburg BNR pathways. The UCT process simulations converged for all 32 input samples. Overall, these results demonstrate the good sample space coverage exhibited by the static Sobol sampling over the 2 input dimensions. The total sampling time of 374\,s equates to just less than 3\,s per sample. Samples that failed to converge took roughly 3 times as long at just under 9\,s per sample which corresponds to the 3 allowed convergence attempts per sample. Although the total sampling time here is acceptable, for higher input dimensions necessitating greater numbers of input samples, and for more complex simulations with greater convergence failure rates, the total sampling time can contribute significant time costs to the surrogate modelling framework. The efficient sample generation exhibited here coupled with the high convergence rate indicates a well posed design space imposed by the input variable selection and their corresponding bounds.

Figure \ref{fig:cer} also shows the random split between training and testing data where 25\% of samples were reserved for testing, resulting in 24 training samples and 8 testing samples. The same train-test split was applied to each discrete BNR pathway so that the uncertainties introduced by this split were equivalent for each discrete scenario. Specifically, Figure \ref{fig:cer} shows how the random train-test split interferes with the spacing of the 32 samples resulting in regions of sparsely sampled training data within the 2-dimensional continuous search spaces. Additionally, the resulting testing set is not well spaced which can result in unrepresentative validation metrics for surrogate models trained to represent the entire search space. The implications of the random train-test split are highlighted in the subsequent sections.

\subsection{Surrogate modelling results}

\subsubsection{Validation results}

\begin{figure}[htb]
    \centering
    \includegraphics[width=4.5in]{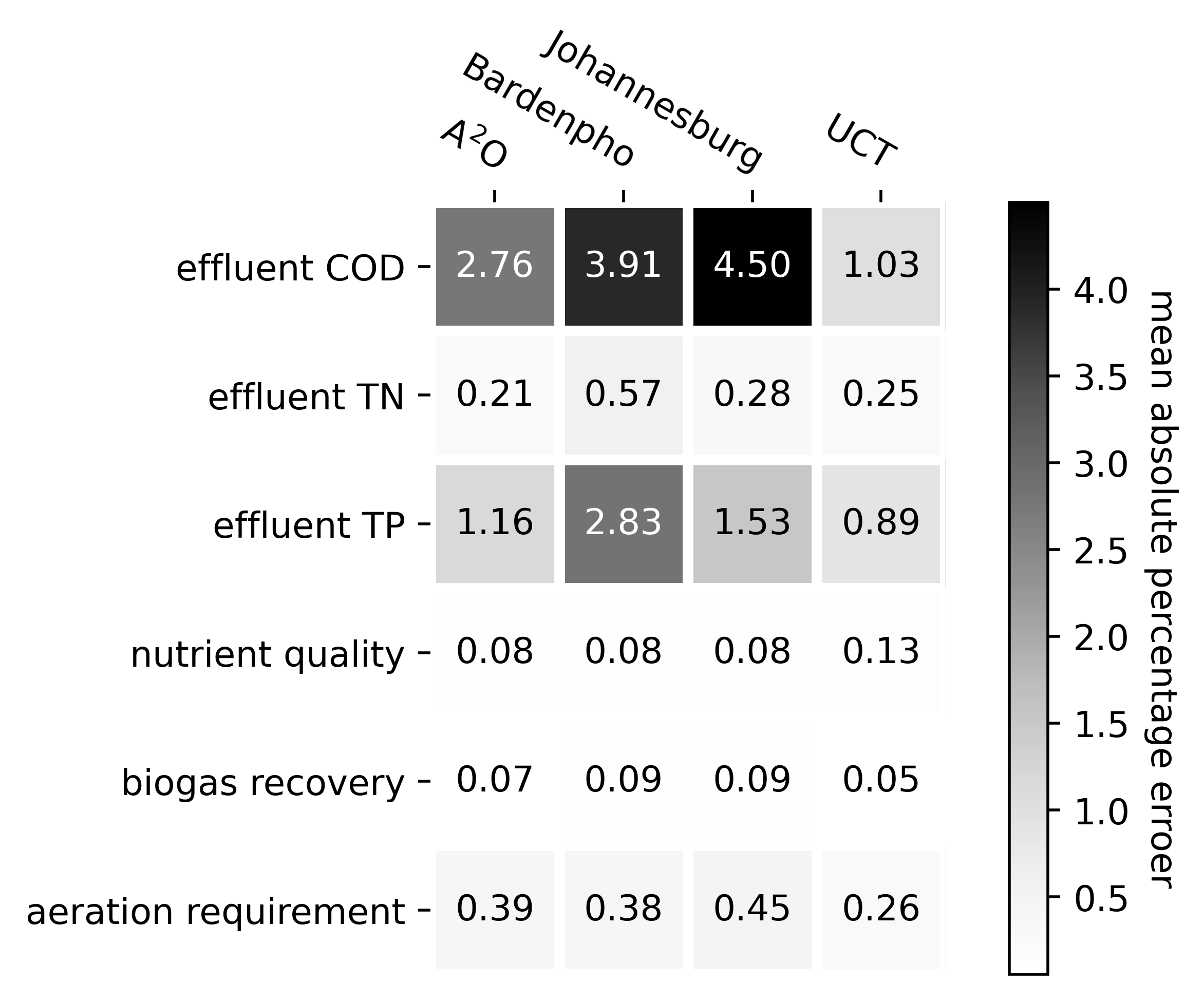}
    \caption[GPR mean absolute percentage errors]{GPR mean absolute percentage errors.}
    \label{fig:mapes}
\end{figure}

24 GPR models, each trained on 24 training samples in under 0.03\,s, were validated on the reserved testing data. The average MAEs across the 4 BNR pathways as well as the mean output values are shown in Table \ref{tab:maes}. Whilst the MAEs require knowledge of the magnitude of the underlying data, the MAPEs are normalised such that a value of 1 represents an error of the order of magnitude equivalent to the underlying data. In this way MAPEs can be directly compared across variables of different magnitude as shown in Figure \ref{fig:mapes}. It can be interpreted that the GPR errors for effluent COD and TP concentrations are high and that all other GPR errors are acceptable. It should be noted that high errors are expected due to the random train-test split negatively impacting the well-spaced properties of the training samples, and the resulting placement of the testing samples in the most sparsely populated regions.

\begin{table}[htb]
    \centering
    \begin{tabular}{l l c c}
        \hline
        Variable & Unit & MAE & $\mu$ \\
        \hline
        effluent COD & g\,m$^{-3}$ & 29.8 & 72.4 \\
        effluent TN & g\,m$^{-3}$ & 2.7 & 9.6 \\
        effluent TP & g\,m$^{-3}$ & 2.4 & 6.7 \\
        nutrient quality & kg\,t$^{-1}$ & 2.5 & 36.3 \\
        biogas recovery & m$^{3}$\,d$^{-1}$ & 67.8 & 707 \\
        aeration requirement & $,000$ m$^{3}$\,d$^{-1}$ & 7.5 & 38.5 \\
        \hline
    \end{tabular}
    \caption[Average GPR mean absolute errors (MAEs)]{Average mean absolute errors and mean values for output variable predictions across the 4 different BNR pathways.}
    \label{tab:maes}
\end{table}

The calculation of the MAPE also results in large values when the underlying data values are small. For example, effluent concentrations of COD, TN, and TP in the treated wastewater are very small for large part of the search space (depicted later in this chapter in Figure \ref{fig:effandfeas}). Additionally, the mean values are often skewed by high effluent concentrations at poor performing process design samples. The GPR validation results should therefore be used primarily for greater interpretation of the optimisation results in conjunction with the GPR predictive uncertainties.

\subsubsection{Response surface results}

\begin{figure}[htb]
    \centering
    \includegraphics[width=5.5in]{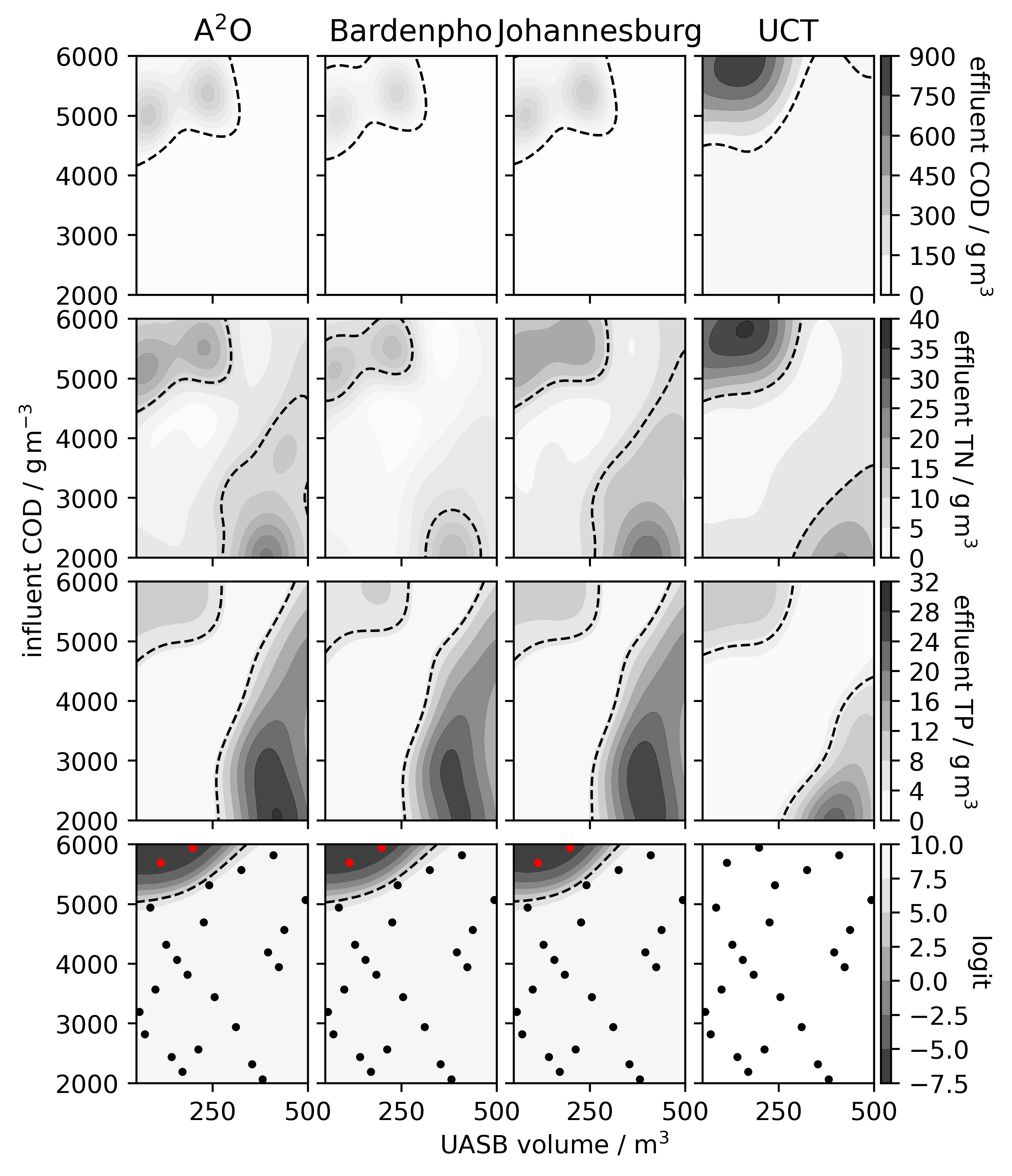}
    \caption[Surrogate modelling of the feasible region]{Surrogate modelling of the feasible region (shown by the dashed lines) constrained by an effluent COD limit of 50\,g\,m$^{-1}$, an effluent TN limit of 10\,g\,m$^{-1}$, and effluent TP limit of 5\,g\,m$^{-1}$, and a probability of simulation convergence greater than 0.5 (corresponding to a logit prediction of 0). Effluent COD, TN, and TP were modelled using GPR whilst the logit predictions were from classification NNs. Note that the UCT simulations converged for all samples therefore there is no region of infeasibility due to simulation convergence failures. Converged samples are shown as black whilst non-converged samples are shown as red. Additionally, training samples are shown as filled whilst testing samples are shown as unfilled.}
    \label{fig:effandfeas}
\end{figure}

The response surfaces generated by the GPR models predicting effluent concentrations of COD, TN, and TP, in addition to the feasible region predicted by the classification NN trained on convergence labels, are shown in Figure \ref{fig:effandfeas}. The effluent quality response surfaces also depict the feasible region separated by the contour placed at the maximum effluent concentration limits. Together, the agglomeration of these 4 distinct feasible regions defined the feasible search space for the subsequent superstructure optimisation.

Figure \ref{fig:effandfeas} shows that the effluent COD quality limit results in infeasible process designs at low UASB volumes for high influent COD concentrations. Additionally, it can be observed that about 75\% of the search space predicts effluent COD concentrations below the imposed limit of 50\,g\,m$^{-3}$ whilst the greatest predictions of effluent COD concentration were as high as 900\,g\,m$^{-3}$ and concentrated within 25\% of the search space. This may explain the high GPR errors for these variables due to these high magnitude outliers resulting in high MAPE errors relative to most of the training data.

The maximum limit for the effluent TN concentration (10\,g\,m$^{-3}$) results in infeasible process designs at low UASB volumes for high influent COD concentrations as well as at high UASB volumes for lower influent COD concentrations. The greatest effluent TN concentrations were observed for the UCT process with low UASB volumes and high influent COD concentrations. 

The TP effluent quality limit of 5\,g\,m$^{-3}$ enforced infeasible regions again at low UASB volumes and high influent COD concentrations as well as at high UASB volumes, particularly for low COD concentrations across all 4 BNR pathways, yet impacting feasible designs at high UASB volumes and high COD concentrations for A$^2$O, Bardenpho, and Johannesburg configurations. The greatest violations of the TP effluent quality limit were observable at high UASB volumes and low influent COD concentrations for the 3 aforementioned BNR process configurations.

The classification NN results are shown for the 4 BNR pathways in Figure \ref{fig:effandfeas}. The NN demonstrates a good separation between feasible and infeasible designs at logit predictions of 0 as shown by the separating hyperplane. Since all of the samples converged for the UCT pathway, the logit predictions are greater than 0 across the entire input space. The classification NNs were validated based on precision and recall scores which were both 1 across Bardenpho, Johannesburg, and UCT pathways, whilst A$^2$O exhibited a precision of 0.88 (due to the incorrectly classified non-converged sample within the testing set visible in Figure \ref{fig:effandfeas}) and a recall of 1.

\begin{figure}[htb]
    \centering
    \includegraphics[width=5.5in]{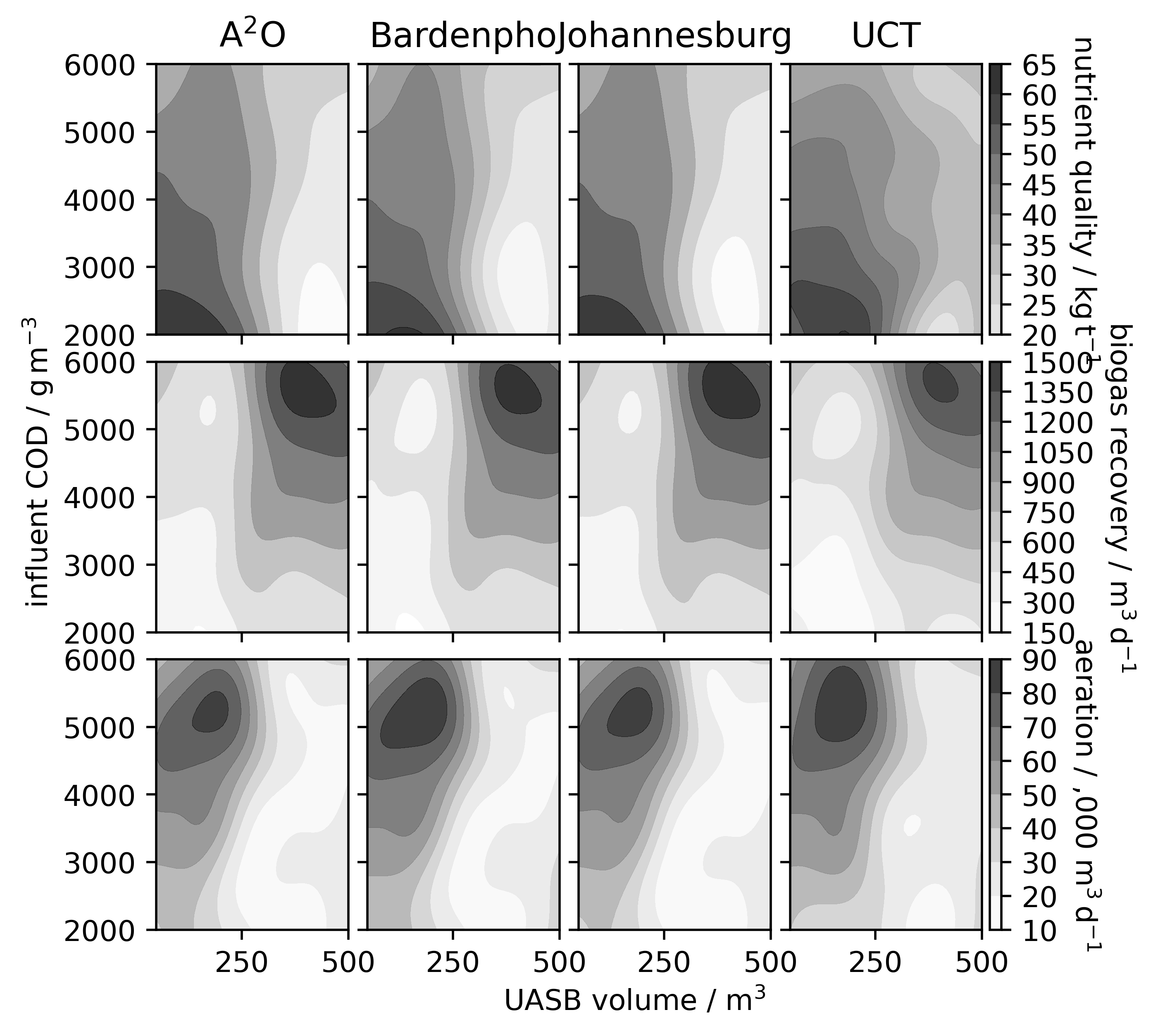}
    \caption[Objective function GPR response surfaces]{Objective function GPR response surfaces.}
    \label{fig:objsurfs}
\end{figure}

Figure \ref{fig:objsurfs} shows the GPR response surfaces for the objective functions enumerating the nutrient quality of the recovered digestate, the biogas recovery, and the aeration requirement of different process designs within the superstructure. Unconstrained by the feasibility constraints, the maximum nutrient quality was achieved at low UASB volumes and low influent COD concentrations for the Bardenpho process. Maximum biogas recovery was realised at high UASB volumes and high influent COD concentrations whilst the minimum aeration flow was required for the upper half of UASB volumes over a range of influent COD concentrations expect the greatest values.

\begin{figure}[htb]
    \centering
    \includegraphics[width=5.5in]{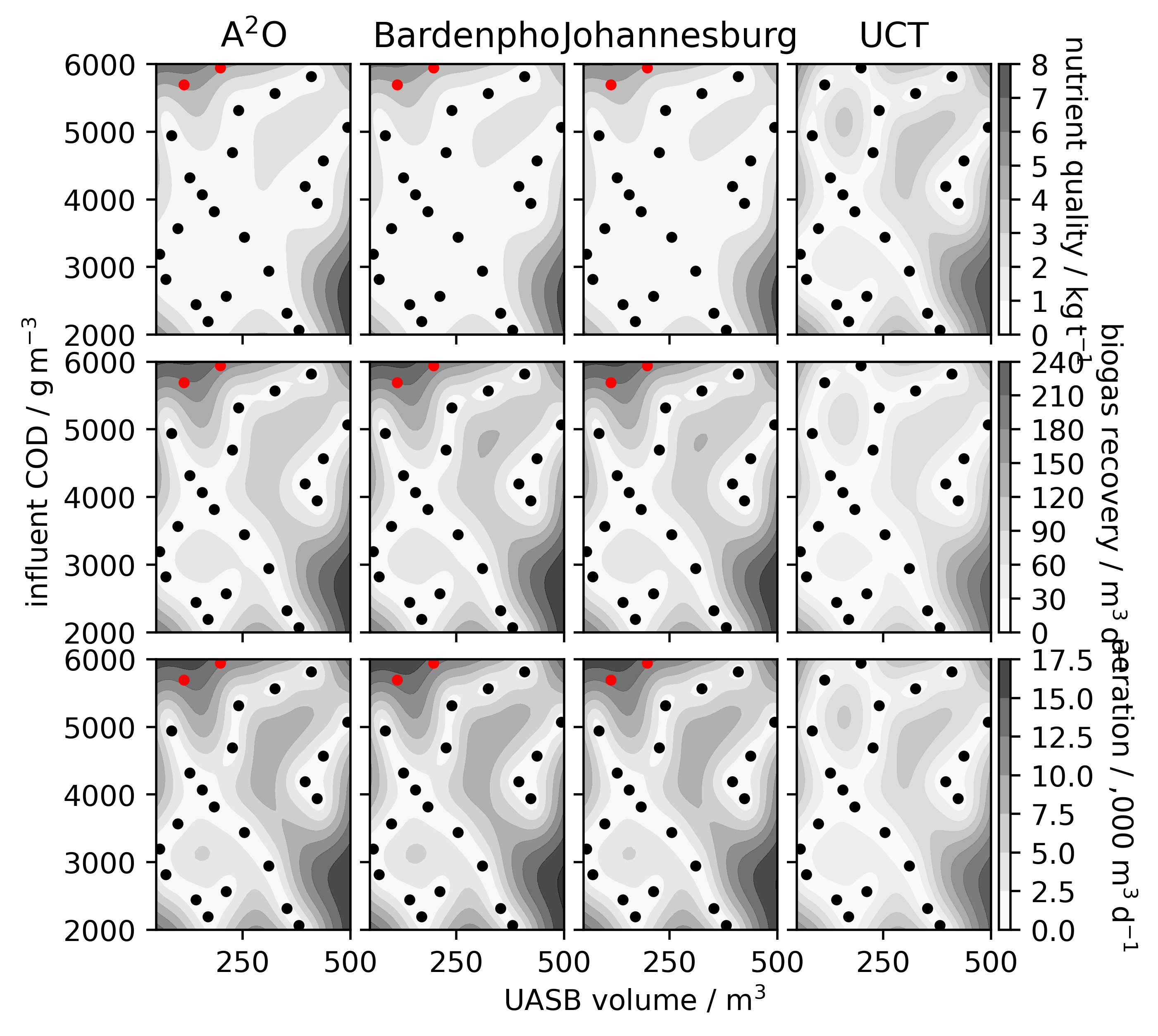}
    \caption[GPR standard deviation response surfaces]{GPR standard deviation response surfaces. The training data are superimposed where non-converged training samples (used for classification model training but not used to train the regression models) are shown in red.}
    \label{fig:objstds}
\end{figure}

Figure \ref{fig:objstds} shows the standard deviations in the GPR predictions from Figure \ref{fig:objsurfs}. The interpolation property of GPR models is observable by the negligible standard deviations close to the training samples with increasing uncertainties with increasing distance from samples. Note that the non-converged samples were not used to train the GPR models such that the uncertainty in the GPR predictions was unaffected by proximity to these samples. Because the same training data was used for each GPR model, the shape of the standard deviation response surfaces are all similar, only varying in the amplitude of uncertainties based on the magnitude of the corresponding variable predictions.

\subsection{Multi-objective optimisation results}

\begin{figure}[htb]
    \centering
    \includegraphics[width=5.5in]{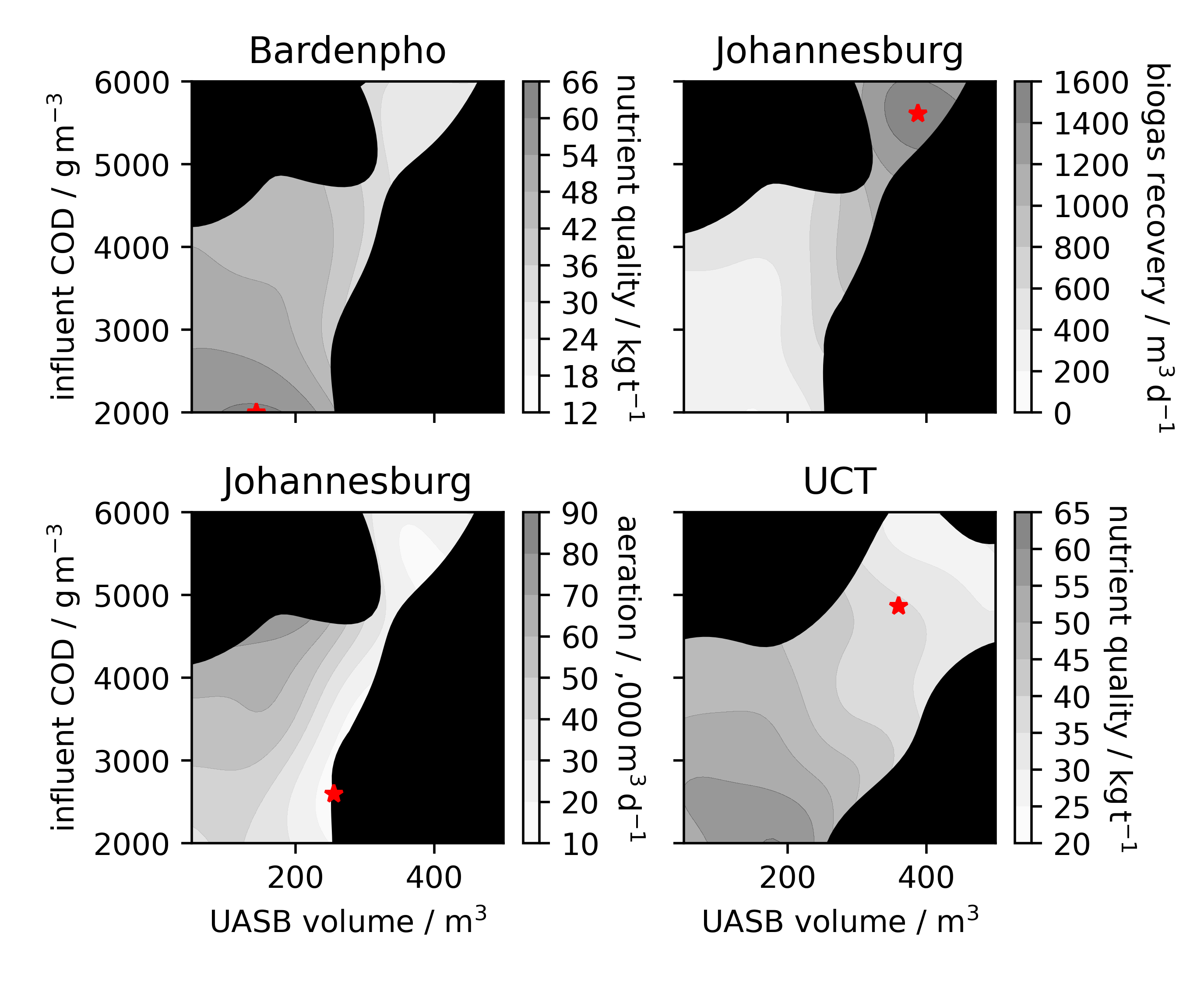}
    \caption[Optimisation solutions]{Optimisation solutions for 4 decision criteria: maximised nutrient quality (upper left); maximised biogas recovery (upper right); minimised aeration cost (lower left); and a Pareto optimal solution to maximise nutrient quality whilst ensuring a biogas recovery of at least 50\,\% the optimal (lower right). The optimised UASB volumes and corresponding influent COD concentrations are shown by the red starts whilst the optimal BNR pathway appears in the title above each response surface. The agglomeration of the effluent quality feasible regions and the simulation convergence-based feasible region is constrained within the black shaded regions.}
    \label{fig:appopt1}
\end{figure}

Figure \ref{fig:appopt1} shows the results of the superstructure optimisation to maximise the combined N and P nutrient quality of the recovered digestate at 61.2 $\pm$ 2.5\,kg\,t$^{-1}$ (where the uncertainty was obtained from the standard deviation in the GPR predictions based on the 95\% confidence interval). The optimal flowsheet for nutrient recovery implemented the Bardenpho BNR process with a 143\,m$^3$ UASB and was realised at an influent COD of 2,000\,g\,m$^{-3}$ as shown by the red star (Figure \ref{fig:appopt1}). The amount of recovered biogas with this flowsheet is 165 m$^3$d$^{-1}$ whilst the aeration requirement is 41,340 m$^3$d$^{-1}$. It can be seen from Figure \ref{fig:appopt1} that none of the constraints are active since the flowsheet produces an effluent with concentrations of COD, TN, and TP of 20.6\,g\,m$^{-3}$, 3.4\,g\,m$^{-3}$, and 1.6\,g\,m$^{-3}$, respectively.

Figure \ref{fig:appopt1} also shows the results of the superstructure optimisation to maximise the total biogas recovery as 1,500 $\pm$ 103\,m$^3$\,d$^{-1}$. The maximal biogas recovery was obtained using the Johannesburg BNR process with a 388\,m$^3$ UASB and was realised at an influent COD of 5,616\,g\,m$^{-3}$ as shown by the red star (Figure \ref{fig:appopt1}). The nutrient quality of the recovered digestate using this flowsheet is 26.5\,kg\,t$^{-1}$ whilst the aeration requirement is 18,920m$^3$d$^{-1}$. None of the effluent constraints were active for this solution since the flowsheet produced an effluent with concentrations of TN and TP of 4.9\,g\,m$^{-3}$, and 0.2\,g\,m$^{-3}$, respectively, whilst all of the COD was predicted to have been recovered.

Figure \ref{fig:appopt1} also shows the results of the superstructure optimisation to minimise the total aeration requirement as 15,790 $\pm$ 8,080\,m$^3$\,d$^{-1}$. This solution was obtained using the UCT BNR process with a 255\,m$^3$ UASB and was realised at an influent COD of 2,598\,g\,m$^{-3}$ as shown by the red star (Figure \ref{fig:appopt1}). The nutrient quality of the recovered digestate and the amount of recovered biogas using this flowsheet was 39.7\,kg\,t$^{-1}$ and 584\,m$^3$d$^{-1}$, respectively. The effluent constraints for this solution had concentrations of COD, TN, and TP of 12.8\,g\,m$^{-3}$, 9.8\,g\,m$^{-3}$, and only 5.0\,g\,m$^{-3}$, respectively. The TP effluent constraint was therefore active at this solution which can been seen in Figure \ref{fig:appopt1}.

A Pareto optimal flowsheet was determined to recover a digestate with an N and P nutrient quality of 35.9 $\pm$ 6.5\,kg\,t$^{-1}$ alongside 1,070\,m$^3$d$^{-1}$ of biogas, with an aeration requirement of 31,570\,m$^3$d$^{-1}$. This Pareto optimal flowsheet implemented to UCT process with a UASB volume of 360\,m$^3$d$^{-1}$, and operated optimally for an influent COD concentration of 4,871\,g\,m$^{-3}$ (Figure \ref{fig:appopt1}). The effluent from the Pareto optimal flowsheet had COD, TN, and TP concentrations of 0\,g\,m$^{-3}$, 2.4\,g\,m$^{-3}$, and 0\,g\,m$^{-3}$, respectively.

It is also worth noting the different influent COD concentrations realised for the optimal solutions in Figure \ref{fig:appopt1}. Specifically, the optimal nutrient recovery was realised at the lower bound of 2,000\,g\,m$^{-3}$ COD whilst the maximum biogas recovery was obtained for a COD concentration nearer the upper bound at 5,616\,g\,m$^{-3}$. Therefore, nutrient recovery could be the focus of the process system at lower influent COD concentrations, whilst biogas recovery is favoured at higher influent COD concentrations. To this end, a flexible resource recovery process, in which Bardenpho and Johannesburg BNR configurations were configured, could enable the operation of the resource recovery process to vary temporally based on the influent COD. However, this would require optimisation of the process system operations in response to real-time sensor data and future projections, for example harnessing model predictive control.

\begin{figure}[htb]
\centering
    \includegraphics[width=0.6\textwidth]{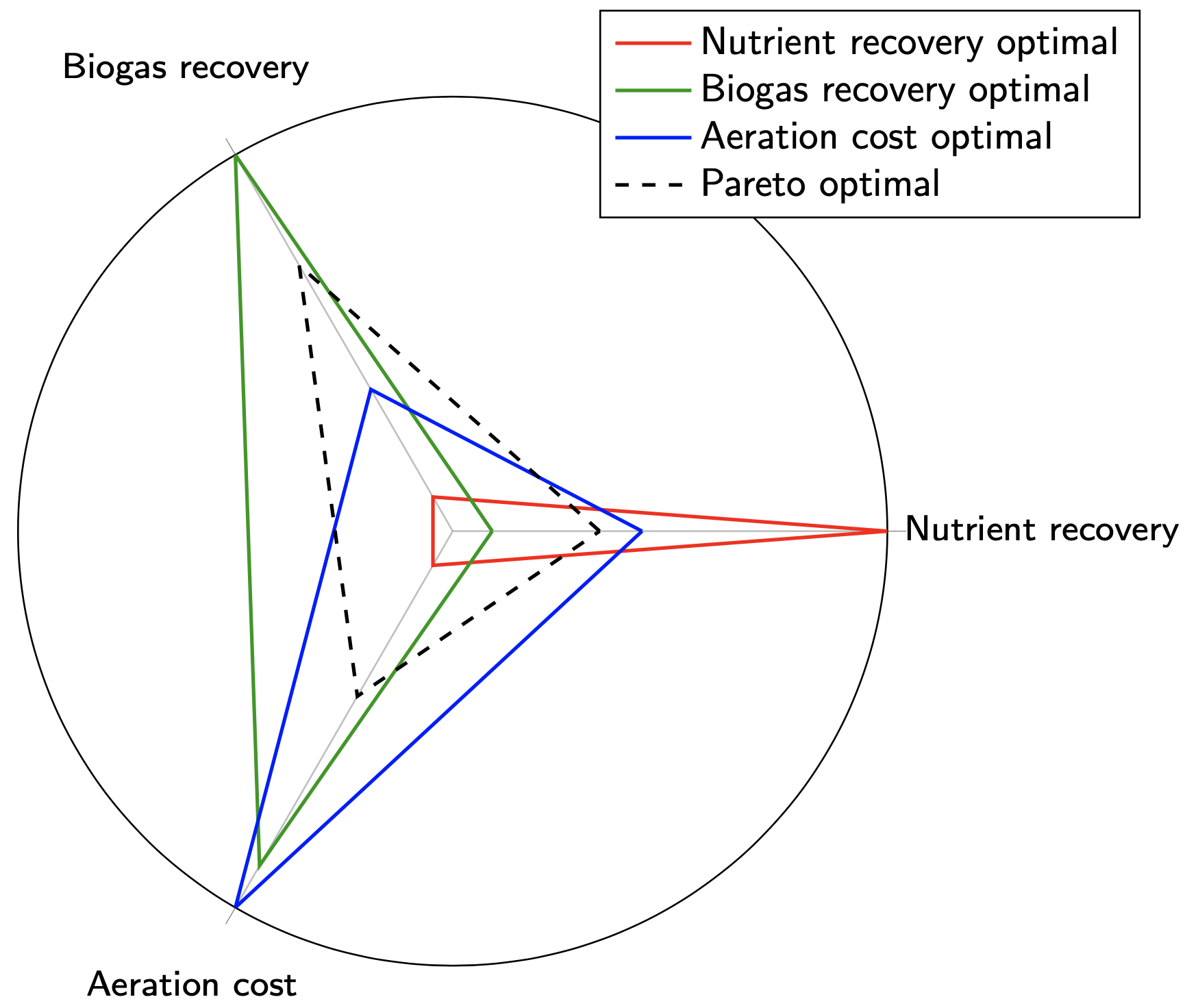}
    \caption[]{Multi-objective optimisation trade-offs. Values are normalised to the radial axes where solutions furthest from the origin are optimal and solutions nearest to the origin are the most sub-optimal trade-offs.}
    \label{fig:polar1}
\end{figure}

Figure \ref{fig:polar1} shows the trade-offs between the competing objective criteria where each single objective solution is visualised as well as the Pareto optimal solution. The maximised nutrient quality of the recovered digestate was 61.2\,kg\,t$^{-1}$ whilst the most sub-optimal nutrient quality existed for the solution to maximise biogas recovery at 26.5\,kg\,t$^{-1}$ (a reduction of over 55\,\%). Similarly, the maximum biogas recovery was 1,500\,m$^3$\,d$^{-1}$ whilst the lowest biogas recovery existed for the flowsheet to maximise the nutrient recovery at 164.5 m$^3$d$^{-1}$ or just 11\,\% of the maximum biogas recovery potential. Finally, the minimised aeration requirement was 15,790 \,m$^3$\,d$^{-1}$ whilst the greatest aeration requirement was almost 3 times as high and existed for the maximal nutrient recovery solution at 41,340\,m$^3$d$^{-1}$. The Pareto optimal solution achieved a trade-off between the 3 objectives with 59\,\% of the maximised nutrient recovery, 71\,\% of the maximum biogas recovery, necessitating 76\,\% of the aeration requirement to recover the maximum nutrient quality digestate.

\subsection{Optimisation under uncertainty results}

\begin{figure}[htb]
    \centering
    \includegraphics[width=5.5in]{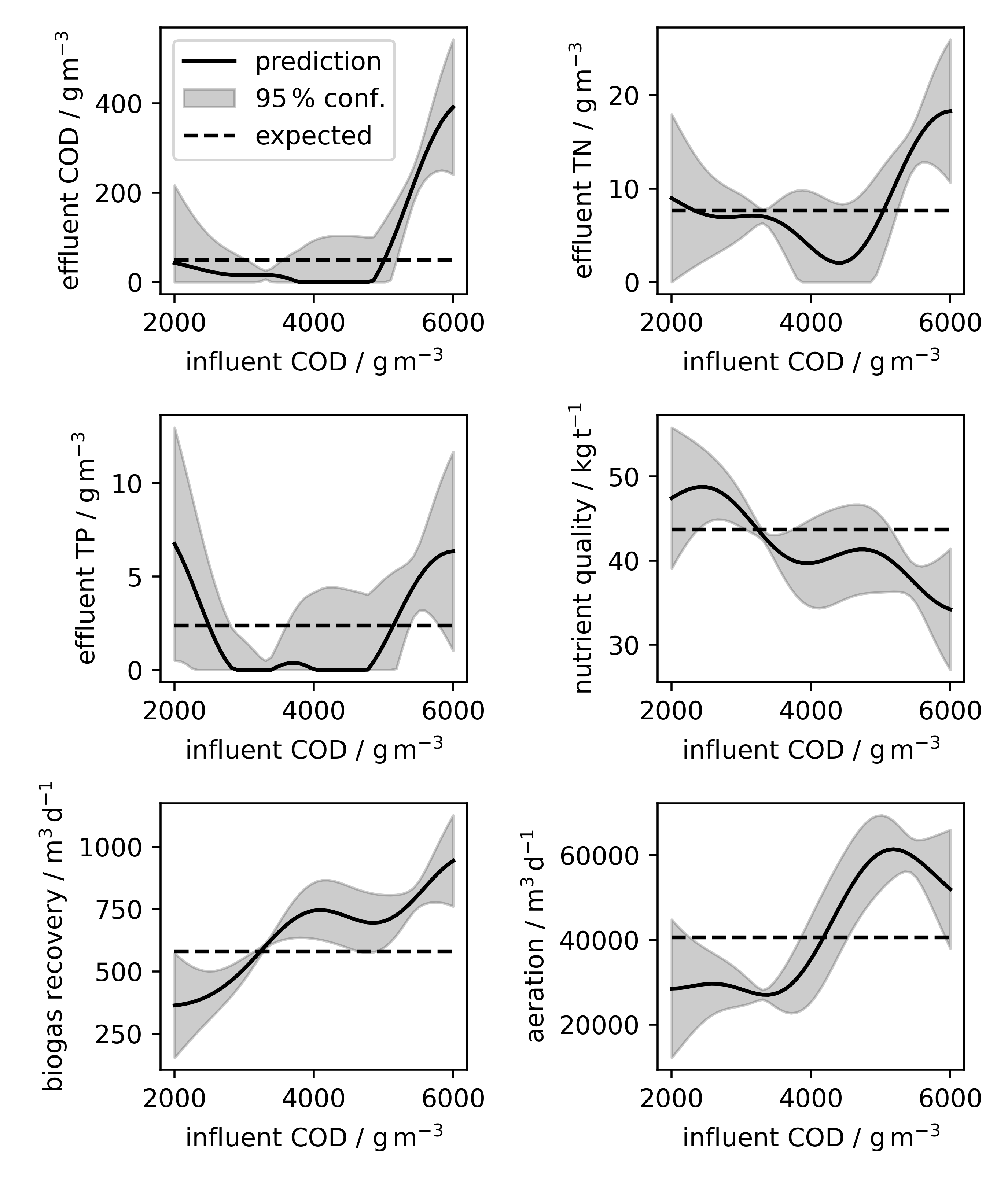}
    \caption[Stochastic programming optimised operational models]{Operational profiles of the solution to the optimisation problem accounting for uncertainty in the influent chemical oxygen demand (COD). Also shown are the predictive modelling uncertainties (as the 95\,\% confidence interval) from Gaussian process regression models. TN: total nitrogen, TP: total phosphorous.}
    \label{fig:sp1}
\end{figure}

Solutions to the optimisation under influent COD uncertainties determined the optimum BNR configuration and UASB volumes. The solution to the 3 objectives to maximise nutrient quality, biogas recovery, and minimise aeration cost converged on the same solution implementing a 274\,m$^3$ UASB followed by the UCT BNR process. Figure \ref{fig:sp1} thereby shows the GPR predictions for the optimal process which recovers a digestate with an expected nutrient quality of 38.6\,kg\,t$^{-1}$ (63\,\% of the maximised nutrient quality without any uncertainty considerations), 580\,m$^3$\,d$^{-1}$ expected biogas recovery (39\,\% of the maximum value found previously), and necessitating an expected aeration flow of 40,360\,m$^3$\,d$^{-1}$ which was over 150\,\% greater than the minimised value without any uncertainty considerations. Specifically, Figure \ref{fig:sp1} shows the responses of the optimal process design to the different realisations of influent COD between 2,000\,g\,m$^{-3}$ and 6,000\,g\,m$^{-3}$.

The expected effluent concentrations were constrained to be less than the effluent limits (50\,g\,m$^{-3}$, 10\,g\,m$^{-3}$, and 10\,g\,m$^{-3}$ for COD, TN, and TP, respectively). Figure \ref{fig:sp1} shows that the effluent constraint for the COD concentration was below its specified limit with an expected value of 48.1\,g\,m$^{-3}$. Additionally, it can be seen that the effluent COD is expected to transgress the limits at influent COD concentrations greater than 5,000\,g\,m$^{-3}$. 

Similarly, despite the expected effluent TN concentration of 7.6\,g\,m$^{-3}$ being below the effluent limit of 10\,g\,m$^{-3}$, the GPR predictive mean violates this limit at influent COD concentrations above 5,000\,g\,m$^{-3}$. Similarly the expected effluent TP concentration was 2.4\,g\,m$^{-3}$ whilst the predictive mean function over the realisations, despite not violating the effluent limit of 10\,g\,m$^{-3}$ over all realisations of the influent COD, the 95\,\% upper confidence bound violates this limit for extreme influent COD concentrations. In fact, infringement of the effluent limits by the 95\,\% upper confidence bound is common across COD, TN, and TP concentrations.

Figure \ref{fig:sp1} also shows the predictive GPR model for the recovered digestate nutrient quality and the uncertainty in these predictions dependent on realisations of the influent COD concentration. Also shown is the expected value across all realisations of the influent COD, accounting for the probability of these concentrations being realised. Generally, the nutrient quality appears to greatest for lower influent COD concentrations, whilst the nutrient quality at greater influent COD concentrations is reduced. In particular, the uncertainty in the nutrient quality peaks at extreme influent COD concentrations of 2,000\,g\,m$^{-3}$ and 6,000\,g\,m$^{-3}$. In fact, this characteristic of high uncertainty at the bounds of the influent COD realisations is common across all of the GPR models due to the dependence on the distribution of training data sampled from the process simulation software. The GPR modelling uncertainties could thereby be reduced by increasing the number of static samples generated in initial sampling strategies or by adopting an adaptive sampling approach wherein additional samples are iteratively evaluated from the process simulator so as to increase the modelling fidelity at sparsely sampled locations.

\clearpage

\section{Conclusions}
% pasted from chapter 4 conclusion

An object-orientated methodology was developed for surrogate modelling and DFO. GP and NN surrogate models were developed for both regression and classification applications. Explicit mathematical formulations of these surrogate models were validated against existing machine learning implementations and embedded within generalised optimisation formulations. Specifically, GP models were formulated as NLPs whilst NN models were formulated as MILPs or NLPs depending on whether the activation functions were piecewise linear or nonlinear, respectively. These abstracted optimisation formulations enabled plug-and-play surrogate-based optimisation modelling. A primary contribution of this work was the development of classification surrogate models to enable feasibility constraint formulations addressing uncertainties pertaining to simulation convergence failures.

Adaptive sampling algorithms based on GPs and Delaunay triangulation were developed to improve the exploration and exploitation of the search space. GP-based adaptive sampling methods enabled exploration via maximisation of the GP uncertainty or exploitation via the modified EI, both formulated as NLPs. Delaunay triangulation methods were formulated as MILPs to enable heuristic-based exploration and exploitation of the search space. The plug-and-play feature of the developed optimisation formulations enabled online feasibility constraints to be plugged-in to adaptive sampling formulations and increase the efficiency of sampling in the feasible region. Finally, the heuristic-based adaptive sampling MILP was adapted within a data-based direct-search optimisation algorithm to optimise black box problems. The adoption of the object-orientated paradigm means that these methods can be harnessed in a plug-and-play approach to increase the global optimisation and sampling efficiencies of a wide range of simulation-based optimisation applications.

% The results of an illustrative example demonstrate the surrogate modelling capabilities using GPs and NNs for both regression and classification models. Model-based DFO implementations embedding object-orientated GP and NN regression/classification models were formulated and solved. ReLU NN-based MILPs were presented as an effective approach to quickly optimise complex constrained BBO problems. This work also demonstrated the contribution of GP-based optimisation formulations enabling optimisation solution interpretability via prediction uncertainty quantification as well as the potential for incorporation of this uncertainty within decision-making frameworks with the developed mathematical programming formulations.

% A heuristic-based adaptive sampling algorithm based on Delaunay triangulated search space was introduced as an novel and effective approach to improve surrogate model accuracy with minimal sampling requirements. In fact, this heuristic-based approach was shown to achieve equivalent exploration of the feasible region as rigorous maximisation of GPR model uncertainty in a fraction of the computational time. This was shown to be due to the underlying MILP objective function used in this heuristic-based approach compared to the complex NLP used in GPR uncertainty or modified EI maximisation. Finally, the heuristic-based adaptive sampling MINLP was repurposed to enable data-based direct-search DFO of the underlying BBO problem. This approach was shown to be capable of obtaining high quality solutions with low computational times.

% pasted from chapter 5 conclusion
This work demonstrated the applicability of the methodology developed to the optimisation of process systems to recover resources from food and beverage processing wastewaters (FPWW). Specifically, a superstructure optimisation methodology was developed to optimise process systems to recover carbon, nitrogen, and phosphorous resources from a typical brewery wastewater. A superstructure embedding a high-rate anaerobic digestion reaction upstream of 4 different BNR configurations (A$^2$O, Bardenpho, Johannesburg, and UCT) was implemented in GPS-X process simulation software so as to harness high-fidelity data within the decision making framework. This work therefore contributed a practical simulation-based process synthesis application of a black box mixed-integer nonlinear programming framework embedding both GP and NN surrogate model formulations. To achieve this, GPR surrogate models were harnessed to represent resource recovery pathways with interpretable uncertainty quantification. Additionally, classification NN surrogate models were used to model the feasible region whilst addressing uncertainties from non-converged black box simulation data.

Multi-objective optimisation addressed the trade-offs between 3 competing decision criteria: one to maximise the total recovered biogas; one to maximise the nutrient quality of a recoverable digestate; and one to minimise the aeration requirement as a proxy to the total operational cost. As such, this work addressed the challenge of holistic process synthesis which currently constrains the deployment of resource recovery process to wastewater systems. Optimal nutrient recovery was realised at low COD compositions, whilst maximum biogas recovery was achieved at COD compositions near to the upper bound cited for typical brewery effluents. 

Uncertainties in the brewery wastewater COD composition were accounted for by implementing a stochastic programming reformulation of the optimisation problem, thereby enabling the process synthesis to account for the inherent uncertainties in wastewater characterisation. This work therefore demonstrates the applicability of optimisation under uncertainty methods to resource recovery from FPWW thereby contributing to the literature which focuses primarily on applications to organic solid wastes. The expected effluent compositions were enforced to be below maximum concentration limits accordingly. The optimisation also enforced robustness to process feasibility constraints derived from convergence of the process simulator. The solutions to the stochastic programming formulations maximised the expected nutrient recovery or biogas recovery, or minimised the expected aeration cost. The same solution, a 274\,m$^3$ UASB upstream of a UCT BNR configuration, was determined for all 3 objectives. This was suspected due to the difficulty of satisfying the expected effluent constraints over realisations of the influent COD uncertainties. The modelling framework enables machine learning models of the optimal process design to be harnessed for further operations optimisation. Specifically, GPR models provided predictions of process performance in response to variations in the influent COD composition in addition to the uncertainties in these predictions.

\subsection{Recommendations for future work}
Flexible process optimisation was recommended to determine flexible resource recovery processes in which the BNR pathway can be adjusted in response to temporal variations in the wastewater composition. Another recommendation was to integrate the uncertainties in the GPR surrogate models within the optimisation-based decision-making process. One strategy to achieve this is via minimising the upper confidence bound which can be calculated by adding factors of the GPR standard deviations to the GPR predictive mean. For example, the 95\,\% upper confidence bound can be obtained by adding 1.96 standard deviations to the mean predictions provided by the GPR modelling objects. This would ensure that obtained solutions simultaneously minimised the objective function and the uncertainty in solutions. The upper confidence bound could also be used within the constraint formulations to ensure the the optimisation solutions were robust to the GPR modelling uncertainties that arise due to interpolation between the training data.

There exist multiple research directions to improve the capabilities of the developed methodology. These include future work to increase the number of surrogate models available within the object-orientated framework. One example to achieve this is to consider enabling the selection of more kernel functions within GPC models, particularly a linear kernel which would reduce the complexity of GPC-based NLP formulations. Another example would be to enable recurrent NN structures to improve nonlinear deep learning capabilities. A third example is to incorporate other surrogate models into the framework for regression and classification: support vector machines, random forests, and decision trees have been highlighted as additional alternatives.

Future work could also focus on improving the implementations of the GP and NN surrogate models. This could be achieved by enabling optimisation of the order of the GPR polynomial kernel. Another option is to enable the optimisation of NN parameters such as number of layers, nodes in each layer, and activation functions \cite{diaz_effective_2017}. This could further be adapted to implement algorithms to optimise the selection of different surrogate models and their respective model structure as in \cite{boukouvala_argonaut_2017}. The surrogate modelling capabilities could be further improved by implementing further regularisation methods such as L1 and dropout regularisation for NNs as well as early termination prior to overfitting. Cross-validation methods are another method to validate and quantify overfitting of surrogate models. Additionally, analysis of the effect of multiple NN outputs compared to multiple single-output NNs on the resulting NN-based optimisation formulations should be explored. Other adaptive sampling formulations and acquisition functions could also improve the existing codebase's capabilities. \cite{bhosekar_advances_2018} provide a good discussion on the variants of GP models using different correlation and mean functions during specification of the GP prior. Additionally, \cite{bhosekar_advances_2018} provide a good discussion on the computational aspects on GP modelling pertaining to handling higher dimensional problems and the non-convexity of the maximum likelihood estimator.

Finally, future work should implement the open-source methodology on more applications. These could be programmatic testing applications to inform further programmable improvements of the codebase and its wider usability. In this regard, increased documentation and demonstrative applications would be beneficial. Application of the methodology to real world engineering problems is also particularly important to realise the impact of the developed methodology in practical operating environments. To this end, it would be beneficial to combine these data-driven surrogate models with well-established mechanistic models and expert knowledge. Such hybrid models would improve transparency, interpretability and extrapolation capabilities thereby gaining the trust of practical operational engineers with a lack of computer science skills. Finally, such hybrid models should leverage the increasing volumes of available sensor data. This would remove a layer of abstraction incurred from the reliance on simulation models and thereby further increase the transparency of black box modelling frameworks.

\bibliographystyle{unsrt}

\clearpage
\section*{Supplementary information}

\subsection*{Neural network training}

Training NNs consists of iterative algorithms to optimise the weights $W$, and biases $\textbf{b}$ for a given network structure. Training proceeds by making predictions using the initialised value for the weights and bias parameters. As such, these initial prediction are not very accurate. However, predictions are made iteratively during the training cycle referred to as the forward pass. One forward pass is completed by evaluating Equations \ref{eq:nn1}-\ref{eq:nn-out} for some provided inputs. Each forward pass yields thereby yields predictions from the output layer.

After each forward pass, the NN error is calculated by evaluating the loss function. The loss function is used as the objective function in an optimisation problem to minimise the error between NN predictions and training observations. The loss function used depends on the data and problem that the NN is being used to solve. Specifically, NNs for regression are trained using loss functions such as the mean absolute error or mean squared error (Equation \ref{eq:mseloss}) to minimise the loss between $n$-vectors for network predictions $\hat{\textbf{y}}$ and continuous regression observations $\textbf{y}$. Equation \ref{eq:nncost} shows an NN cost function incorporating a loss function as well as an L2 regularisation term which sums the elements of the weight matrices, where $\delta$ is a weight decay parameter and where the dependence on the weights and biases $C(W, \textbf{b})$ is incurred via the predictions in the loss function.

\begin{equation} \label{eq:mseloss}
    L(\hat{\textbf{y}}, \textbf{y}) = \frac{1}{n} \sum_{i=1}^n \left( \hat{y}_i - y_i \right)^2
\end{equation}

\begin{equation} \label{eq:nncost}
    C(W, \textbf{b}) = L(\hat{\textbf{y}}, \textbf{y}) + \frac{\delta}{2n} \sum_{i=1}^n w_i^2
\end{equation}

After calculating the error incurred in the forward pass, the backward pass proceeds to calculate the changes in weights that provide the best reduction in the cost function. This step is called backward propagation and typically uses a gradient decent optimisation algorithm to determine the best direction (changes in weights) to minimise the error. Backward propagation occurs in the opposite direction to the forward pass, proceeding from the output later to the input layer. The weights and biases are updated in the direction which minimises the NN cost function which is determined by calculating the gradient of the cost function with respect to the weights and biases. With respect to the weights, the gradient of the cost function equals the gradient of the loss function plus the gradient of the L2 weight regularisation term (Equation \ref{eq:bpw}). The gradient of the cost function is equal to the gradient of the loss function with respect to the biases (Equation \ref{eq:bpb}).

\begin{equation} \label{eq:bpw}
    \frac{\partial C}{\partial w^{(\lambda)}_{j,i}} ={} \frac{\partial L}{\partial w^{(\lambda)}_{j,i}} + \delta w^{(\lambda)}_{j,i}
\end{equation}

\begin{equation} \label{eq:bpb}
    \frac{\partial C}{\partial b^{(\lambda)}_{j}} ={} \frac{\partial L}{\partial b^{(\lambda)}_{j}}
\end{equation}

During back propagation, the gradients of the loss function with respect to the weights can be determined by using the chain rule to propagation back through the network (Equation \ref{eq:bpw2}).

\begin{equation} \label{eq:bpw2}
    \begin{aligned}
        \frac{\partial L}{\partial w^{(\lambda)}_{j,i}} ={} & \; \frac{\partial L}{\partial z^{(\lambda)}_{j}} \frac{\partial z^{(\lambda)}_{j}}{\partial w^{(\lambda)}_{j,i}} \\ 
        ={} & \; \frac{\partial L}{\partial a^{(\lambda)}_{j}}  \frac{\partial a^{(\lambda)}_{j}}{\partial z^{(\lambda)}_{j}} \frac{\partial z^{(\lambda)}_{j}}{\partial w^{(\lambda)}_{j,i}}
    \end{aligned}
\end{equation}

The derivative of the loss function with respect to a node output $a^{(\lambda)}_{j}$ can be written as shown by Equation \ref{eq:dlda} where the derivative of node input $z^{(\lambda+1)}_{k}$ with respect node output $a^{(\lambda)}_{j}$ can be determined as $w^{(\lambda)}_{k,j}$ by differentiating Equation \ref{eq:nn1} with respect to layer outputs.

\begin{equation}\label{eq:dlda}
    \begin{aligned}
        \frac{\partial L}{\partial a^{(\lambda)}_{j}} ={} & \; \sum_{k=1}^{N_{\lambda+1}} \frac{\partial L}{\partial z^{(\lambda+1)}_{k}} \frac{\partial z^{(\lambda+1)}_{k}}{\partial a^{(\lambda)}_{j}}  \\
        ={} & \; \sum_{k=1}^{N_{\lambda+1}} \frac{\partial L}{\partial z^{(\lambda+1)}_{k}} w^{(\lambda)}_{k,j}
    \end{aligned}
\end{equation}

Equation \ref{eq:dadz} shows the derivative of node output $a^{(\lambda)}_{j}$ with respect to node input $z^{(\lambda)}_{j}$ which can be determined as the derivative of the activation function $\xi^{(\lambda)'}$ by differentiating Equation \ref{eq:nn2} with respect to layer inputs. Some derivatives of common activation functions are shown in Table \ref{tab:activationfuncs}. Additionally, Equation \ref{eq:dzdw} shows how the derivative of node input $z^{(\lambda)}_{j}$ with respect to weight $w^{(\lambda)}_{j, i}$ can be written as node output $a^{(\lambda-1)}_{i}$ by differentiating Equation \ref{eq:nn1} with respect to weights.

\begin{equation}\label{eq:dadz}
    \frac{\partial a^{(\lambda)}_{j}}{\partial z^{(\lambda)}_{j}} = \xi^{(\lambda)'}\left( z^{(\lambda)}_{j} \right)
\end{equation}

\begin{equation}\label{eq:dzdw}
    \frac{\partial z^{(\lambda)}_{j}}{\partial w^{(\lambda)}_{j,i}} = a^{(\lambda-1)}_{i}
\end{equation}

Finally, an error signal of node $j$ in layer $\lambda$, $\epsilon_j^{(\lambda)}$ is defined by Equation \ref{eq:errorsig} to simplify Equation \ref{eq:bpw3} after substituting Equations \ref{eq:dlda}-\ref{eq:dzdw} into Equation \ref{eq:bpw2}.

\begin{equation}\label{eq:errorsig}
    \begin{aligned}
        \epsilon_j^{(\lambda)} ={} & \; \frac{\partial L}{\partial z^{(\lambda)}_{j}} \\
        ={} & \; \frac{\partial L}{\partial a^{(\lambda)}_{j}} \frac{\partial a^{(\lambda)}_{j}}{\partial z^{(\lambda)}_{j}} \\
        ={} & \; \sum_{k=1}^{N_{\lambda+1}} \left( \frac{\partial L}{\partial z^{(\lambda+1)}_{k}} w^{(\lambda)}_{k,j} \right) \xi^{(\lambda)'}\left( z^{(\lambda)}_{j} \right)
    \end{aligned}
\end{equation}

\begin{equation} \label{eq:bpw3}
    \begin{aligned}
        \frac{\partial L}{\partial w^{(\lambda)}_{j,i}} ={} & \; \sum_{k=1}^{N_{\lambda+1}} \left( \frac{\partial L}{\partial z^{(\lambda+1)}_{k}} w^{(\lambda)}_{k,j} \right) \xi^{(\lambda)'}\left( z^{(\lambda)}_{j} \right) a^{(\lambda-1)}_{i} \\
        ={} & \; \epsilon_j^{(\lambda)} a^{(\lambda-1)}_{i}
    \end{aligned}
\end{equation}

The gradient of the cost/loss function with respect to the biases is similarly approached by using the chain rule (Equation \ref{eq:bpb2}). The definition of the error signal (Equation \ref{eq:errorsig}) as well as the derivative of Equation \ref{eq:nn1} with respect to the biases (equal to 1), yields the simplified formulation of Equation \ref{eq:bpb2}

\begin{equation} \label{eq:bpb2}
    \begin{aligned}
        \frac{\partial C}{\partial b^{(\lambda)}_{j}} ={} & \; \frac{\partial L}{\partial z^{(\lambda)}_{j}} \frac{\partial z^{(\lambda)}_{j}}{\partial b^{(\lambda)}_{j}} \\
        ={} & \; \epsilon_j^{(\lambda)}
    \end{aligned}
\end{equation}

Once the direction of steepest decent has been determined by an optimisation algorithm, the NN parameters are updated by a given step size, $\alpha$, known as the learning rate (Equations \ref{eq:stepw} and \ref{eq:stepb}). The learning rate is another hyperparameter which can be tuned by the user between the values 0 and 1: too large and the optimiser might converge too quickly to a suboptimal local solution; too small and the optimiser could get stuck and fail to find an optimum within finite time. Optimisation algorithms available include Stochastic Gradient Descent (SGD), L-BFGS, and Adam \cite{kingma_adam_2017}.

\begin{equation} \label{eq:stepw}
    \begin{aligned} 
        w^{(\lambda)}_{j,i} ={} & w^{(\lambda)}_{j,i} - \alpha \frac{\partial C}{\partial w^{(\lambda)}_{j,i}} \\ 
        ={} & \; w^{(\lambda)}_{j,i} -\alpha \left( \epsilon_j^{(\lambda)} a^{(\lambda-1)}_{i} + \delta w^{(\lambda)}_{j,i} \right)
    \end{aligned}
\end{equation}

\begin{equation} \label{eq:stepb}
    \begin{aligned} 
        b^{(\lambda)}_{j} ={} & \; b^{(\lambda)}_{j} - \alpha \frac{\partial C}{\partial b^{(\lambda)}_{j}} \\
        ={} & \; b^{(\lambda)}_{j} - \alpha\epsilon_j^{(\lambda)}
    \end{aligned}
\end{equation}

Together, the forward and backward pass make one iteration which is typically carried out for a subset of the training data called a mini-batch (where the term batch is reserved for the whole data set). For a batch size equal to the size of the training data set, this is referred to as batch gradient decent, whereas for a batch size equal to 1 (i.e., 1 sample at a time) is what is used specifically during SGD. For any batch size greater than 1 but less than the size of the training set, the method is called mini-batch gradient decent. Batch gradient decent is useful for fitting to relatively smooth functions where the entire data can be used effectively to approximate a generalised smooth output. However, batch gradient decent is computationally expensive for very large data sets as gradients must be calculated for all samples at the same time. On the other hand, SGD only calculates one gradient at a time and takes an optimisation step accordingly, the result is quicker convergence for very large data sets, but the global optimum can be difficult to find as the objective function value fluctuates with every sample. Mini-batch gradient decent achieves a trade-off between the two extreme methods, using small batches enables better exploration of the solution space compared to SGD as well as better computational times compared to batch gradient decent. Table \ref{tab:nn-training} shows a pseudocode algorithm to train a NN by optimising the weights and biases $W^{(\lambda)}, \textbf{b}^{(\lambda)}$ via mini-batch gradient descent.

\begin{table}[htb]
    \centering
    \caption[Mini-batch gradient descent algorithm]{Mini-batch gradient descent algorithm.}
    \begin{tabular}{l l}
        \hline
        \multicolumn{2}{p{0.9\textwidth}}{
            \hangindent=15mm\textbf{input} : $X$ (training inputs), $Y$ (training observations), $n$ (number of training samples), $\Lambda$ (number of hidden layers), $N^{(\lambda)}$ (number of nodes in each layer), $\xi^{(\lambda)}$ (activation functions), \texttt{epochs} (number of epochs), $n_{\text{batch}}$ (batch size), $\alpha$ (learning rate), $\delta$ (weight decay), $L$ (loss function) 
        }\\
        \multicolumn{2}{p{0.9\textwidth}}{
            \hangindent=21mm\textbf{initialise} : $W^{(\lambda)}\leftarrow W^{\text{init}}$ (random weights), $\textbf{b}^{(\lambda)}\leftarrow \textbf{b}^{\text{init}}$ (random biases) 
        }\\
        \hline
        \textbf{for} \texttt{epoch} = 0 \textbf{to} \texttt{epochs} \textbf{do} \\
            \quad Randomly shuffle training inputs $X$ & \multirow{6}{*}{$\left. 
            \begin{array}{l} \\ \\ \\ \\ \\ \end{array} \right\} \text{Mini-batch setup}$} \\ 
            \quad Apply same permutation to training observations $Y$ \\
            \quad \textbf{for} $i=0$ \textbf{to} $n$ \textbf{step} $n_{\text{batch}}$ \textbf{do} \\
            \qquad $X^{\text{batch}} \leftarrow X[i:i+n_{\text{batch}}]$ \\
            \qquad $Y^{\text{batch}} \leftarrow Y[i:i+n_{\text{batch}}]$ \\
            \qquad Initialise empty array of predictions $\hat{Y}^{\text{batch}}$ \\
            \\
            \qquad \textbf{for} \textbf{x} \textbf{in} $X^{\text{batch}}$ \textbf{do} & \multirow{7}{*}{$\left. 
            \begin{array}{l} \\ \\ \\ \\ \\ \\ \\ \end{array} \right\} \text{Forward pass}$} \\
            \quad\qquad $\textbf{z}^{(1)} \leftarrow W^{(0)} \textbf{x} + \textbf{b}^{(0)}$ \\
            \quad\qquad \textbf{for} $\lambda=1$ \textbf{to} $\Lambda$ \textbf{do} \\
            \qquad\qquad $\textbf{a}^{(\lambda)} \leftarrow \xi^{(\lambda)} \left( \textbf{z}^{(\lambda)} \right)$ \\
            \qquad\qquad $\textbf{z}^{(\lambda+1)} \leftarrow W^{(\lambda)} \textbf{a}^{(\lambda)} + \textbf{b}^{(\lambda)}$ \\
            \quad\qquad $\hat{\textbf{y}} \leftarrow \textbf{z}^{(\Lambda+1)}$ \\
            \quad\qquad Update relevant row in $\hat{Y}^{\text{batch}}$ with $\hat{\textbf{y}}$ \\
            \\
            \qquad $C^{\text{batch}} \leftarrow L(\hat{Y}, Y) + \frac{\delta}{2n} \sum_{i=1}^n w_i^2$ & $\quad \; \triangleright$ Error evaluation \\
            \\
            \qquad $\epsilon_j^{(\lambda)} \leftarrow \sum_{k=1}^{N_{\lambda+1}} \left( \frac{\partial L}{\partial z^{(\lambda+1)}_{k}} w^{(\lambda)}_{k,j} \right) \xi^{(\lambda)'}\left( z^{(\lambda)}_{j} \right)$ & $\quad \; \triangleright$ Backward propagation \\
            \\
            \qquad \textbf{for} $\lambda=0$ \textbf{to} $\Lambda$ \textbf{do} & \multirow{5}{*}{$\left. 
            \begin{array}{l} \\ \\ \\ \\ \\ \end{array} \right\} \text{Optimisation step}$} \\
            \quad\qquad \textbf{for} $k=0$ \textbf{to} $N_{\lambda+1}$ \textbf{do} \\
            \qquad\qquad \textbf{for} $j=0$ \textbf{to} $N_{\lambda}$ \textbf{do} \\
            \quad\qquad\qquad $w_{k,j}^{(\lambda)} \leftarrow w^{(\lambda)}_{j,i} -\alpha \left( \epsilon_j^{(\lambda)} a^{(\lambda-1)}_{i} + \delta w^{(\lambda)}_{j,i} \right)$ \\
            \quad\qquad\qquad $b^{(\lambda)}_{j} \leftarrow b^{(\lambda)}_{j} - \alpha\epsilon_j^{(\lambda)}$ \\
        \hline
        \textbf{return} : $W^{(\lambda)}, \textbf{b}^{(\lambda)}$  \\
        \hline
    \end{tabular}
    \label{tab:nn-training}
\end{table}

An epoch refers to a forward and backward pass of the entire data set (batch) in multiple mini-batches, where an epoch contains a number of iterations equal to the size of the batch divided by the size of the mini-batch. The number of epochs should be tuned to ensure that the NN has enough exposure to the training data to achieve a good fit, but not so much that it succumbs to overfitting. To combat this, a validation data set can be used during training to test the trained NN on unseen data (similar to the testing data set but used during training to determine overfitting) and achieve regularisation. During training, the training error decreases as the NN is updated and fitted to the data. Initially, the validation error will also decrease as the NN goes from initially underfit to being fitted. Eventually, as overfitting starts to occur, the training error will continue to decrease but the validation error will start to increase. This phenomenon can be used to stop training and avoid overfitting, by specifying a large number of epochs along with a termination condition based on the validation error increasing.

\subsection*{OODX modelling objects}

\begin{longtable}[htb]{p{0.9\textwidth}}
\caption[\texttt{DataHandler} object]{\texttt{DataHandler} object.}
\label{tab:datahandler} \\

    \hline
    \endfirsthead

    \centerline{\tablename\ \thetable{} -- continued from previous page} \\
    \hline
    \endhead

    \hline \rightline{Table continued on next page} \\
    \endfoot

    \hline
    \endlastfoot

    \textbf{class} DataHandler() \\
    \quad Modelling object to sample, process, and store data. \\
    \hline
    \textbf{Attributes} \\
    \quad space : array-like of shape (n\_inputs, 2) \\
    \hangindent=2em \qquad Lower and upper bounds on each input feature in the original space. \\
    \quad x : array-like of shape (n\_samples, n\_inputs) \\
    \hangindent=2em \qquad Input samples in the original space. \\
    \quad y : array-like of shape (n\_samples, n\_outputs) \\
    \hangindent=2em \qquad Output samples in the original space. \\
    \quad t : array-like of shape (n\_samples, 1) \\
    \hangindent=2em \qquad Binary convergence targets. \\

    \quad x\_train : array-like of shape (n\_train\_samples, n\_inputs) \\
    \hangindent=2em \qquad Training input samples in the original space. \\
    \quad y\_train : array-like of shape (n\_train\_samples, n\_outputs) \\
    \hangindent=2em \qquad Training output samples in the original space. \\

    \quad t\_train : array-like of shape (n\_train\_samples, 1) \\
    \hangindent=2em \qquad Training binary convergence targets. \\

    \quad x\_test : array-like of shape (n\_test\_samples, n\_inputs) \\
    \hangindent=2em \qquad Testing input samples in the original space. \\

    \quad y\_test : array-like of shape (n\_test\_samples, n\_outputs) \\
    \hangindent=2em \qquad Testing output samples in the original space. \\

    \quad t\_test : array-like of shape (n\_test\_samples, 1) \\
    \hangindent=2em \qquad Testing binary convergence targets. \\

    \quad space\_ : array-like of shape (n\_inputs, 2) \\
    \hangindent=2em \qquad Lower and upper bounds on each input feature in the scaled modelling space. \\

    \quad x\_ : array-like of shape (n\_samples, n\_inputs) \\
    \hangindent=2em \qquad Input samples in the scaled space. \\

    \quad y\_ : array-like of shape (n\_samples, n\_outputs) \\
    \hangindent=2em \qquad Output samples in the scaled space. \\

    \quad x\_train\_ : array-like of shape (n\_train\_samples, n\_inputs) \\
    \hangindent=2em \qquad Training input samples in the scaled space. \\

    \quad y\_train\_ : array-like of shape (n\_train\_samples, n\_outputs) \\
    \hangindent=2em \qquad Training output samples in the scaled space. \\

    \quad x\_test\_ : array-like of shape (n\_test\_samples, n\_inputs) \\
    \hangindent=2em \qquad Testing input samples in the scaled space. \\

    \quad y\_test\_ : array-like of shape (n\_test\_samples, n\_outputs) \\
    \hangindent=2em \qquad Testing output samples in the scaled space. \\

    \quad x\_mean : array-like of shape (n\_inputs,) \\
    \hangindent=2em \qquad Mean over n\_samples of each input feature vector. \\

    \quad x\_std : array-like of shape (n\_inputs,) \\
    \hangindent=2em \qquad Standard deviation over n\_samples of each input feature vector. \\

    \quad y\_mean : array-like of shape (n\_outputs,) \\
    \hangindent=2em \qquad Mean over n\_samples of each output variable vector. \\

    \quad y\_std : array-like of shape (n\_outputs,) \\
    \hangindent=2em \qquad Standard deviation over n\_samples of each output variable vector. \\

    \quad x\_train\_mean : array-like of shape (n\_inputs,) \\
    \hangindent=2em \qquad Mean over n\_train\_samples of each input feature vector. \\

    \quad x\_train\_std : array-like of shape (n\_inputs,) \\
    \hangindent=2em \qquad Standard deviation over n\_train\_samples of each input feature vector. \\

    \quad y\_train\_mean : array-like of shape (n\_outputs,) \\
    \hangindent=2em \qquad Mean over n\_train\_samples of each output variable vector. \\

    \quad y\_train\_std : array-like of shape (n\_outputs,) \\
    \hangindent=2em \qquad Standard deviation over n\_train\_samples of each output variable vector. \\
    \hline
    \textbf{Methods} \\
    \quad init(n\_samples, space, n\_outputs=1, method=`lhs') \\
    \hangindent=2em \qquad Generate initial input samples using static sampling strategies and initialise empty arrays for outputs and binary convergence targets. \\
    \qquad \textbf{Parameters} \\
    \qquad\quad n\_samples : integer \\
    \qquad\qquad Number of input samples to generate. \\
    \qquad\quad space : array-like of shape (n\_inputs, 2) \\
    \qquad\qquad Lower and upper bounds on each input feature in the original space. \\
    \qquad\quad n\_outputs : integer, default=1 \\
    \qquad\qquad Number of output dimensions to initialise empty arrays. \\
    \qquad\quad method : string, default=`lhs` \\
    \qquad\qquad Static sampling strategy to generate initial input samples. \\
    \qquad \textbf{Returns} \\
    \qquad\quad self : object \\
    \qquad\qquad DataHandler class instance. \\
    \\
    \quad split(test\_size=0.3) \\
    \hangindent=2em \qquad Train-test split on data. \\
    \qquad \textbf{Parameters} \\
    \qquad\quad test\_size : float, default=0.3 \\
    \qquad\qquad Fraction of samples to be assigned to testing sets. \\
    \qquad \textbf{Returns} \\
    \qquad\quad self : object \\
    \qquad\qquad DataHandler class instance. \\
    \\
    \quad scale() \\
    \hangindent=2em \qquad Standardise input-output data. \texttt{x} is standardised and attributes the output as \texttt{x\_}. \texttt{y} is standardised to \texttt{y\_} using the mean and standard deviation of the subset of feasible outputs to ensure that any infeasible outputs do not skew the data. \texttt{x\_train} is standardised to \texttt{x\_train\_} using the mean and standard deviation of the input training set. \texttt{x\_test} is standardised to \texttt{x\_test\_} using the mean and standard deviation of the input training set to ensure that testing data is mapped into the same space in which the model was trained. \texttt{y\_train} is standardised to \texttt{y\_train\_} using the mean and standard deviation of the subset of feasible training outputs to ensure that any infeasible outputs do not skew the data. \texttt{y\_test} is standardised to \texttt{y\_test\_} using the mean and standard deviation of the subset of feasible output training samples to ensure that testing data is mapped into the same space in which the model was trained. \texttt{space} is standardised to \texttt{space\_} using the mean and standard deviation of the input training set is the data has been train-test split, otherwise using the mean and standard deviation of the original input data. \\
    \qquad \textbf{Returns} \\
    \qquad\quad self : object \\
    \qquad\qquad DataHandler class instance. \\
    \\
    \quad scale\_space(space) \\
    \hangindent=2em \qquad Map new lower, upper bounds on the features into the scaled modelling space. \\
    \qquad \textbf{Parameters} \\
    \qquad\quad space : array-like of shape (n\_inputs, 2) \\
    \hangindent=4em \qquad\qquad New lower and upper bounds on each input feature to be mapped into the scaled modelling space. \\
    \qquad \textbf{Returns} \\
    \qquad\quad new\_space : array-like of shape (n\_inputs, 2) \\
    \qquad\qquad Lower, upper feature bounds mapped into the scaled modelling space. \\
    \\
    \quad scale\_x(x) \\
    \hangindent=2em \qquad Map new input samples into the scaled modelling space. \\
    \qquad \textbf{Parameters} \\
    \qquad\quad x : array-like of shape (?, n\_inputs) \\
    \hangindent=4em \qquad\qquad New input samples to be mapped into the scaled modelling space. \\
    \qquad \textbf{Returns} \\
    \qquad\quad scaled\_x : array-like of shape (?, n\_inputs) \\
    \qquad\qquad New input samples mapped into the scaled modelling space. \\
    \\
    \quad inv\_scale\_x(x) \\
    \hangindent=2em \qquad Map scaled inputs back into the original space. \\
    \qquad \textbf{Parameters} \\
    \qquad\quad x : array-like of shape (?, n\_inputs) \\
    \hangindent=4em \qquad\qquad Input samples to be mapped back into the original space. \\
    \qquad \textbf{Returns} \\
    \qquad\quad inv\_scaled\_x : array-like of shape (?, n\_inputs) \\
    \qquad\qquad Input samples mapped back into the original space. \\
    \\
    \quad scale\_y(y) \\
    \hangindent=2em \qquad Map new output samples into the scaled modelling space. \\
    \qquad \textbf{Parameters} \\
    \qquad\quad y : array-like of shape (?, n\_outputs) \\
    \hangindent=4em \qquad\qquad New output samples to be mapped into the scaled modelling space. \\
    \qquad \textbf{Returns} \\
    \qquad\quad scaled\_y : array-like of shape (?, n\_outputs) \\
    \qquad\qquad New output samples mapped into the scaled modelling space. \\
    \\
    \quad inv\_scale\_y(y) \\
    \hangindent=2em \qquad Map scaled outputs back into the original space. \\
    \qquad \textbf{Parameters} \\
    \qquad\quad y : array-like of shape (?, n\_outputs) \\
    \hangindent=4em \qquad\qquad Output samples to be mapped back into the original space. \\
    \qquad \textbf{Returns} \\
    \qquad\quad inv\_scaled\_y : array-like of shape (?, n\_outputs) \\
    \qquad\qquad Output samples mapped back into the original space. \\
\end{longtable}

\begin{longtable}[htb]{p{0.9\textwidth}}
\caption[\texttt{NN} object]{\texttt{NN} object.}
\label{tab:nn} \\

    \hline
    \endfirsthead

    \centerline{\tablename\ \thetable{} -- continued from previous page} \\
    \hline
    \endhead

    \hline \rightline{Table continued on next page} \\
    \endfoot

    \hline
    \endlastfoot

    \textbf{class} NN(layers, activation='tanh', is\_classifier=False) \\
    \quad Neural network modelling object for regression and classification. \\
    \hline
    \textbf{Parameters} \\
    \quad layers : list \\
    \hangindent=2em \qquad List of the number of nodes in each layer. \\
    \quad activation : string, default=`tanh' \\
    \hangindent=2em \qquad Activation functions used throughout the network. \\
    \quad is\_classifier : Boolean, default=False \\
    \hangindent=2em \qquad Boolean variable equal to True if the NN is used for classification to change the loss function to binary cross entropy loss on the logits. \\
    \hline
    \textbf{Attributes} \\
    \quad name : string \\
    \hangindent=2em \qquad `NN' or `NNClf' depending on whether \texttt{is\_classifier} is False or True, repectively. \\
    \quad layers : list \\
    \hangindent=2em \qquad List of the number of nodes in each layer. \\
    \quad activation : string \\
    \hangindent=2em \qquad Activation functions used throughout the network. \\
    \quad weights : list \\
    \hangindent=2em \qquad List of weight matrices for each layer where the length of the list equals the number of layers in the network ($\Lambda + 1$). Each element of the list is array-like of shape ($N_{\lambda}$, $N_{\lambda - 1}$) where $N_\lambda$ is the number of nodes in layer $\lambda$, for layers $\lambda = 1, ..., \Lambda+1$. \\
    \quad biases : list \\
    \hangindent=2em \qquad List of bias vectors for each layer where the length of the list equals the number of layers in the network ($\Lambda + 1$). Each element of the list is array-like of shape ($N_{\lambda}$,) for layers $\lambda = 1, ..., \Lambda+1$. \\
    \hline
    \textbf{Methods} \\
    \hangindent=1em \quad fit(x, y, batch\_size=10, epochs=1000, learning\_rate=1e-2, weight\_decay=0.0, loss\_func=nn.MSELoss(), iprint=False) \\
    \hangindent=2em \qquad Fit NN object to input-output data. \\
    \qquad \textbf{Parameters} \\
    \qquad\quad x : array\_like of shape (n\_samples, n\_inputs) \\
    \qquad\qquad Input samples used to train the NN. \\
    \qquad\quad y : array\_like of shape (n\_samples, n\_outputs) \\
    \qquad\qquad Output samples used to train the NN. \\
    \qquad\quad batch\_size : integer, default=10 \\
    \qquad\qquad Batch size used in batch gradient descent. \\
    \qquad\quad epochs : integer, default=1000 \\
    \qquad\qquad The number of iterations of forward and backward propagation. \\
    \qquad\quad learning\_rate : float, default=0.01 \\
    \qquad\qquad The step size used during NN parameter optimisation. \\
    \qquad\quad weight\_decay : float, default=0.0 \\
    \hangindent=4em\qquad\qquad Weight decay parameter used to regularise the NN during parameter optimisation. \\
    \qquad\quad loss\_func : object, default=torch.nn.MSELoss() \\
    \hangindent=4em\qquad\qquad The loss function to be used during back propagation. Automatically switched to binary cross entropy loss on the logits if is\_classifier = True. \\
    \qquad\quad iprint : Boolean, default=False \\
    \qquad\qquad If True, prints the time taken to train the NN. \\
    \qquad \textbf{Returns} \\
    \qquad\quad self : object \\
    \qquad\qquad Trained NN class instance. \\
    \\
    \quad predict(x) \\
    \hangindent=2em \qquad Use a trained NN object to make predictions at new input data. \\
    \qquad \textbf{Parameters} \\
    \qquad\quad x : array\_like of shape (?, n\_inputs) \\
    \qquad\qquad New input samples used to make predictions. \\
    \qquad \textbf{Returns} \\
    \qquad\quad predictions : array\_like of shape (?, n\_outputs) \\
    \qquad\qquad Predictions from the NN made at new inputs. \\
    \\
    \quad formulation(x) \\
    \hangindent=2em \qquad Use a trained NN object to make predictions at new input data based on an explicit mathematical formulation. \\
    \qquad \textbf{Parameters} \\
    \qquad\quad x : array\_like of shape (?, n\_inputs) \\
    \qquad\qquad New input samples used to make predictions. \\
    \qquad \textbf{Returns} \\
    \qquad\quad predictions : array\_like of shape (?, n\_outputs) \\
    \hangindent=4em\qquad\qquad Predictions from the explicit mathematical formulation of the NN made at new inputs. \\
\end{longtable}

\begin{longtable}[htb]{p{0.9\textwidth}}
\caption[\texttt{GPR} object]{\texttt{GPR} object.}
\label{tab:gpr} \\

    \hline
    \endfirsthead

    \centerline{\tablename\ \thetable{} -- continued from previous page} \\
    \hline
    \endhead

    \hline \rightline{Table continued on next page} \\
    \endfoot

    \hline
    \endlastfoot

    \textbf{class} GPR(kernel=`rbf', noise=1e-10, porder=2) \\
    \quad Gaussian process regression model object. \\
    \hline
    \textbf{Parameters} \\
    \quad kernel : string, default=`rbf' \\
    \hangindent=2em \qquad Name of the kernel function to be used. \\
    \quad noise : float, default=1e-10' \\
    \hangindent=2em \qquad Nugget term for modelling noisy observations. \\
    \quad porder : integer, default=2 \\
    \hangindent=2em \qquad Order of polynomial kernel function. \\
    \hline
    \textbf{Attributes} \\
    \quad name : string \\
    \hangindent=2em \qquad Class attribute always equal to `GPR'. \\
    \quad kernel\_name : string \\
    \hangindent=2em \qquad Name of the kernel function implemented. \\
    \quad noise : float \\
    \hangindent=2em \qquad Nugget parameter value. \\
    \quad x\_train : array-like of shape (n\_samples, n\_inputs) \\
    \hangindent=2em \qquad Training inputs as GPR model parameter. \\
    \quad length\_scale : float \\
    \hangindent=2em \qquad Optimised parameter for `rbf' kernel. \\
    \quad constant\_value : float \\
    \hangindent=2em \qquad Optimised parameter for all kernel functions. \\
    \quad sigma\_0 : float \\
    \hangindent=2em \qquad Optimised parameter for linear/polynomial kernel functions. \\
    \quad inv\_K : array-like of shape (n\_samples, n\_samples) \\
    \hangindent=2em \qquad Inverse covariance matrix between training data. \\
    \quad porder : integer \\
    \hangindent=2em \qquad Order of polynomial kernel function. \\
    \hline
    \textbf{Methods} \\
    \hangindent=1em \quad fit(x, y, iprint=False) \\
    \hangindent=2em \qquad Fit GPR object to input-output data. \\
    \qquad \textbf{Parameters} \\
    \qquad\quad x : array\_like of shape (n\_samples, n\_inputs) \\
    \qquad\qquad Input training samples. \\
    \qquad\quad y : array\_like of shape (n\_samples, 1) \\
    \qquad\qquad Output training samples. \\
    \qquad\quad iprint : Boolean, default=False \\
    \qquad\qquad If True, prints the time taken to train the model. \\
    \qquad \textbf{Returns} \\
    \qquad\quad self : object \\
    \qquad\qquad Trained GPR object. \\
    \\
    \quad predict(x, return\_std=False, return\_cov=False) \\
    \hangindent=2em \qquad Use a trained GPR object to make predictions at new input data. \\
    \qquad \textbf{Parameters} \\
    \qquad\quad x : array\_like of shape (?, n\_inputs) \\
    \qquad\qquad New input samples used to make predictions. \\
    \qquad\quad return\_std : Boolean, default=False \\
    \qquad\qquad If True, returns the standard deviation of predictions. \\
    \qquad\quad return\_cov : Boolean, default=False \\
    \qquad\qquad If True, returns the covariance of predictions. \\
    \qquad \textbf{Returns} \\
    \qquad\quad predictions : array\_like of shape (?, 1) \\
    \qquad\qquad Predictions from the GPR made at new inputs. \\
    \qquad\quad std : array\_like of shape (?, 1), optional \\
    \qquad\qquad Standard deviation in predictions from the GPR made at new inputs. \\
    \qquad\quad cov : array\_like of shape (?, 1), optional \\
    \qquad\qquad Covariance in predictions from the GPR made at new inputs. \\
    \\
    \quad formulation(x, return\_std=False) \\
    \hangindent=2em \qquad Use an explicit mathematical formulation of a trained GPR object to make predictions at new input data. \\
    \qquad \textbf{Parameters} \\
    \qquad\quad x : array\_like of shape (?, n\_inputs) \\
    \qquad\qquad New input samples used to make predictions. \\
    \qquad\quad return\_std : Boolean, default=False \\
    \qquad\qquad If True, returns the standard deviation of predictions. \\
    \qquad \textbf{Returns} \\
    \qquad\quad predictions : array\_like of shape (?, 1) \\
    \qquad\qquad Predictions from the GPR made at new inputs. \\
    \qquad\quad std : array\_like of shape (?, 1), optional \\
    \qquad\qquad Standard deviation in predictions from the GPR made at new inputs. \\
\end{longtable}

\begin{longtable}[htb]{p{0.9\textwidth}}
\caption[\texttt{GPC} object]{\texttt{GPC} object.}
\label{tab:gpc} \\

    \hline
    \endfirsthead

    \centerline{\tablename\ \thetable{} -- continued from previous page} \\
    \hline
    \endhead

    \hline \rightline{Table continued on next page} \\
    \endfoot

    \hline
    \endlastfoot

    \textbf{class} GPC() \\
    \quad Gaussian process classification model object. \\
    \hline
    \textbf{Attributes} \\
    \quad name : string \\
    \hangindent=2em \qquad Class attribute always equal to `GPC'. \\
    \quad x\_train : array-like of shape (n\_samples, n\_inputs) \\
    \hangindent=2em \qquad Training inputs as GPC model parameter. \\
    \quad t\_train : array-like of shape (n\_samples, n\_inputs) \\
    \hangindent=2em \qquad Training targets as GPC model parameter. \\
    \quad l : float \\
    \hangindent=2em \qquad Optimised length scale parameter. \\
    \quad sigma\_f : float \\
    \hangindent=2em \qquad Optimised process variance parameter. \\
    \quad delta : array-like of shape (n\_samples, 1) \\
    \hangindent=2em \qquad Optimised GPC parameter. \\
    \quad inv\_P : array-like of shape (n\_samples, n\_samples) \\
    \hangindent=2em \qquad Optimised GPC parameter. \\
    \hline
    \textbf{Methods} \\
    \hangindent=1em \quad fit(x, t, iprint=False) \\
    \hangindent=2em \qquad Fit GPC object to input-output data. \\
    \qquad \textbf{Parameters} \\
    \qquad\quad x : array\_like of shape (n\_samples, n\_inputs) \\
    \qquad\qquad Input training samples. \\
    \qquad\quad t : array\_like of shape (n\_samples, 1) \\
    \qquad\qquad Training targets. \\
    \qquad\quad iprint : Boolean, default=False \\
    \qquad\qquad If True, prints the time taken to train the model. \\
    \qquad \textbf{Returns} \\
    \qquad\quad self : object \\
    \qquad\qquad Trained GPC object. \\
    \\
    \quad predict(x, return\_std=False, return\_class=False, threshold=0.5) \\
    \hangindent=2em \qquad Use a trained GPC object to make probability predictions that t=1 at new input data. \\
    \qquad \textbf{Parameters} \\
    \qquad\quad x : array\_like of shape (?, n\_inputs) \\
    \qquad\qquad New input samples used to make predictions. \\
    \qquad\quad return\_std : Boolean, default=False \\
    \hangindent=4em\qquad\qquad If True, returns the standard deviation of the latent variable at predictions. \\
    \qquad\quad return\_class : Boolean, default=False \\
    \hangindent=4em\qquad\qquad If True, returns class prediction as well as probability predictions. \\
    \qquad\quad threshold : float, default=0.5 \\
    \hangindent=4em\qquad\qquad Probability threshold for mapping probabilities onto class predictions. \\
    \qquad \textbf{Returns} \\
    \qquad\quad predictions : array\_like of shape (?, 1) \\
    \qquad\qquad Probability predictions that t=1 at new inputs. \\
    \qquad\quad std : array\_like of shape (?, 1), optional \\
    \qquad\qquad Standard deviation of the latent function. \\
    \\
    \quad formulation(x) \\
    \hangindent=2em \qquad Use an explicit mathematical formulation of a trained GPC object to make predictions at new input data. \\
    \qquad \textbf{Parameters} \\
    \qquad\quad x : array\_like of shape (?, n\_inputs) \\
    \qquad\qquad New input samples used to make predictions. \\
    \qquad \textbf{Returns} \\
    \qquad\quad predictions : array\_like of shape (?, 1) \\
    \qquad\qquad Probability predictions from the GPC made at new inputs. \\
\end{longtable}

\begin{longtable}[htb]{p{0.9\textwidth}}
\caption[\texttt{OODXBlock} object]{\texttt{OODXBlock} object.}
\label{tab:oodxblock} \\

    \hline
    \endfirsthead

    \centerline{\tablename\ \thetable{} -- continued from previous page} \\
    \hline
    \endhead

    \hline \rightline{Table continued on next page} \\
    \endfoot

    \hline
    \endlastfoot

    \textbf{class} OODXBlock(model) \\
    \quad Generate Pyomo Block formulations for machine learning model objects. \\
    \hline
    \textbf{Parameters} \\
    \quad model : object \\
    \hangindent=2em \qquad Machine learning model object i.e., NN, GPR, GPC. \\
    \hline
    \textbf{Attributes} \\
    \quad model : object \\
    \hangindent=2em \qquad Machine learning model object i.e., NN, GPR, GPC. \\
    \quad formulation : object \\
    \hangindent=2em \qquad Pyomo Block formulation. \\
    \hline
    \textbf{Methods} \\
    \hangindent=1em \quad get\_formulation(return\_std=False) \\
    \hangindent=2em \qquad Get Pyomo Block formulation for model attribute. \\
    \qquad \textbf{Parameters} \\
    \qquad\quad return\_std : Boolean, default=False \\
    \hangindent=4em\qquad\qquad If True, returns a mathematical program to model the GPR standard deviation. \\
    \qquad \textbf{Returns} \\
    \qquad\quad formulation : object \\
    \qquad\qquad Pyomo Block formulation. \\
\end{longtable}

\begin{longtable}[htb]{p{0.9\textwidth}}
\caption[\texttt{AdaptiveSampler} object]{\texttt{AdaptiveSampler} object.}
\label{tab:adaptivesampler} \\

    \hline
    \endfirsthead

    \centerline{\tablename\ \thetable{} -- continued from previous page} \\
    \hline
    \endhead

    \hline \rightline{Table continued on next page} \\
    \endfoot

    \hline
    \endlastfoot

    \textbf{class} AdaptiveSampler(space) \\
    \quad Object used to generate Pyomo-based adaptive sampling formulations. \\
    \hline
    \textbf{Parameters} \\
    \quad space : array-like of shape (n\_inputs, 2) \\
    \hangindent=2em \qquad Adaptive sampling search space. \\
    \hline
    \textbf{Attributes} \\
    \quad space : array-like of shape (n\_inputs, 2) \\
    \hangindent=2em \qquad Adaptive sampling search space. \\
    \quad delaunay : object \\
    \hangindent=2em \qquad SciPy Delauanay triangulation object. \\
    \hline
    \textbf{Methods} \\
    \hangindent=1em \quad max\_gp\_std(model) \\
    \hangindent=2em \qquad Generate Pyomo-based formulation to maximise the uncertainty of a GPR model for model-based exploration. \\
    \qquad \textbf{Parameters} \\
    \qquad\quad model : object \\
    \hangindent=4em\qquad\qquad GPR model object. \\
    \qquad \textbf{Returns} \\
    \qquad\quad m : object \\
    \qquad\qquad Pyomo formulation. \\
    \\
    \hangindent=1em \quad max\_triangle(x, include\_vertices=False) \\
    \hangindent=2em \qquad Generate Pyomo-based formulation to maximise the size of Delaunay triangulated regions for data-driven exploration. \\
    \qquad \textbf{Parameters} \\
    \qquad\quad x : array-like of shape (n\_samples, n\_inputs) \\
    \hangindent=4em\qquad\qquad Input data on which to triangulate. \\
    \qquad\quad include\_vertices : Boolean, default=False \\
    \hangindent=4em\qquad\qquad If True, vertices of the search space will be included in the triangulation. \\
    \qquad \textbf{Returns} \\
    \qquad\quad m : object \\
    \qquad\qquad Pyomo formulation. \\
    \\
    \hangindent=1em \quad modified\_expected\_improvement(model, y, sense) \\
    \hangindent=2em \qquad Generate Pyomo-based formulation to maximise the modified expected improvement for a GPR model for model-based exploitation. \\
    \qquad \textbf{Parameters} \\
    \qquad\quad model : object \\
    \hangindent=4em\qquad\qquad GPR model object. \\
    \qquad\quad y : array-like of shape (n\_samples, 1) \\
    \hangindent=4em\qquad\qquad Output data from which to determine the current best sample. \\
    \qquad\quad sense : string \\
    \hangindent=4em\qquad\qquad `max' or `min' depending on whether to exploit around the current maximal or minimal sample. \\
    \qquad \textbf{Returns} \\
    \qquad\quad m : object \\
    \qquad\qquad Pyomo formulation. \\
    \\
    \hangindent=1em \quad exploit\_triangle(x, y, sense, include\_vertices=False) \\
    \hangindent=2em \qquad Generate Pyomo-based formulation to maximise the size of Delaunay triangulated regions connected to the current best sample for data-driven exploitation. \\
    \qquad \textbf{Parameters} \\
    \qquad\quad x : array-like of shape (n\_samples, n\_inputs) \\
    \hangindent=4em\qquad\qquad Input data on which to triangulate. \\
    \qquad\quad y : array-like of shape (n\_samples, 1) \\
    \hangindent=4em\qquad\qquad Output data from which to determine the current best sample. \\
    \qquad\quad sense : string \\
    \hangindent=4em\qquad\qquad `max' or `min' depending on whether to exploit around the current maximal or minimal sample. \\
    \qquad\quad include\_vertices : Boolean, default=False \\
    \hangindent=4em\qquad\qquad If True, vertices of the search space will be included in the triangulation. \\
    \qquad \textbf{Returns} \\
    \qquad\quad m : object \\
    \qquad\qquad Pyomo formulation. \\
\end{longtable}

\end{document}